\documentstyle{article}
 
\begin{document}

\title{\bf Formas de transgresi\'on\\
como principio unificador\\
en Teor\'{\i}a de Campos}
\author{Pablo Mora \\
{\it Instituto de Fisica, Facultad de Ciencias}\\
{\it Igu\'a 4225, Montevideo, Uruguay}\\
}
\maketitle
\begin{abstract}
En este trabajo se consideran extensiones de las gravedades y
supergravedades de Chern-Simons asociadas al uso de formas de transgresi\'on
en las acciones correspondientes, en vez de formas de Chern-Simons.\newline
Se observa que las formas de transgresi\'on permiten:\newline
(i) hacer las teor\'{\i}as de Chern-Simons estrictamente invariantes gauge,%
\newline
(ii) tener un principio de acci\'on bien definido, de modo que la acci\'on
es un extremo cuando valen las ecuaciones del movimiento,\newline
(iii) calcular cargas conservadas covariantes de acuerdo con las calculadas
por m\'etodos hamiltonianos,\newline
(iv) y regularizar la acci\'on de modo que la entrop\'{\i}a calculada a
partir de la versi\'on eucl\'{\i}dea de esta acci\'on es finita y coincide
con la calculada de nuevo por m\'etodos hamiltonianos.\newline
Tambi\'en se introduce y estudia una clase de modelos para objetos
extendidos o branas con y sin supersimetr\'{\i}a con acciones definidas por
la suma de integrales de las formas de transgresi\'on para grupos de gauge
ordinarios, grupos espaciotemporales o sus extensiones supersim\'etricas.
Estos modelos son generalmente covariantes, independientes de background y
verdaderos sistemas de gauge.\newline
Un modelo de esta clase podr\'{\i}a proporcionar una formulaci\'on
independiente de background de la teor\'{\i}a M.
\end{abstract}

\newpage

\tableofcontents

\newpage
 
{\it "Los sabios de anta\~no no tem\'{\i}an estar solos en sus opiniones.}

{\it Sin grandes empresas. Sin planes.}

{\it Si fracasaban, sin pena.}

{\it Sin congratularse en el \'exito..."}

- Chuang-Tzu\newline

\section{Introducci\'on}

A pesar de muchos trabajos afirmando lo contrario, la Relatividad General,
actualmente la teor\'{\i}a aceptada de la gravitaci\'on, no es una
teor\'{\i}a de gauge, como lo son las teor\'{\i}as de las otras tres
interacciones fundamentales conocidas \cite{zanelli2}. La construcci\'on de una teor\'{\i}a
de gauge que incluya el campo gravitatorio plantea la dificultad de que,
mientras en el caso de las dem\'as interacciones se dispone de un \'ambito
espaciotemporal fijo con una m\'etrica de Minkowski, es este caso no hay tal
fondo fijo de referencia, al ser la m\'etrica (o el vielbein) parte de la
din\'amica. Cuando existe una m\'etrica de referencia fija la acci\'on usual
para los campos de gauge es la de Yang-Mills, en cuya construcci\'on se
utiliza la m\'etrica de Minkowski. Veremos que si no hay un fondo fijo dado
por una m\'etrica de referencia la acci\'on natural para una teor\'{\i}a de
gauge para los grupos espaciotemporales (Poincar\'e, los grupos de de Sitter
y sus extensiones supersim\'etricas) que pueda generalizar la relatividad
general est\'a dada por la forma de Chern-Simons.

\subsection{Panorama del desarrollo de las teor\'{\i}as de Chern-Simons}

Las teor\'{\i}as de gauge de Chern-Simons son modelos f\'{\i}sicos con una
lagrangiana dada por la forma de Chern-Simons para el grupo de gauge. 
Estos modelos fueron introducidos en el caso abeliano por A. Schwarz \cite
{aschwarz} y estudiados como 'modelos de juguete' en muchos trabajos
posteriormente (ver por ejemplo \cite{deser})hasta el reconocido
art\'{\i}culo de Witten \cite{witten-cs} en el que se muestra que estos
modelos son teor\'{\i}as cu\'anticas de campos exactamente solubles en 2+1
dimensiones, con observables dados por invariantes topol\'ogicos
(invariantes de nudos) de la variedad tridimensional de base.

Las gravedades y supergravedades de Chern-Simons son teor\'{\i}as de gauge
de Chern-Simons con grupo de gauge dado por uno de los grupos
espaciotemporales y alguna de sus extensiones supersim\'etricas
respectivamente. 
Estas teor\'{\i}as fueron introducidas en 
refs.\cite{vannieuwen,achu,witten1} para
espaciotiempos tridimensionales (2+1). Se observ\'o que la Relatividad
General en dimensi\'on 2+1 es equivalente {\it on shell} (cuando valen las
ecuaciones del movimiento) a la teor\'{\i}a de CS para el grupo de
Poincar\'e ISO(2,1), lo que fue explotado por Witten para mostrar que la
teor\'{\i}a es exactamente soluble a nivel cu\'antico \cite{witten1}.

M\'as tarde Chamseddine \cite{chamseddine} extendi\'o las supergravedades de
Chern-Simons a dimensiones mas altas y sugiri\'o que esta clase de modelos
pod\'{\i}an considerarse como la base de un enfoque para la unificaci\'on de
las interacciones fundamentales alternativo a la teor\'{\i}a de Supercuerdas 
\cite{superstring}. Las supergravedades de Chern-Simons (CS-SUGRA) en
dimensiones m\'as altas fueron extensivamente estudiadas en diferentes aspecto
por la 'Escuela Chilena' \cite
{banados1,zanelli1,troncoso1,troncoso2,banados2,zanelli2,hassaine}.

En uno de esos trabajos Troncoso y Zanelli \cite{troncoso1} sugirieron que
el l\'{\i}mite de bajas energ\'{\i}as de la teor\'{\i}a M \cite
{townsend1,hull1,witten2,townsend2} podr\'{\i}a ser una CS-SUGRA con grupo
de gauge $OSp(1\mid 32)$, contribuyendo desde otro \'angulo a la convergencia entre Teor\'{\i}as
CS y Supercuerdas, ya mostrada en las refs.\cite{witten1,kogan,moore1}. M\'as
recientemente Horava \cite{horava} propuso que una CS-SUGRA podr\'{\i}a ser en
realidad la Teor\'{\i}a M, la cual ser\'{\i}a en entonces una teor\'{\i}a de
campos ordinaria. La propuesta de Horava a sido considerada mas recientemente 
por Nastase \cite{nastase}. 

\subsection{Relaci\'on de la Relatividad General y la Supergravedad est\'andar con
las gravedades y  supergravedades de Chern-Simons}

La cuesti\'on de las relaciones entre las teor\'{\i}as CS y la Relatividad
General y/o supergravedad est\'andar en diversas dimensiones ha sido
discutido en las refs.\cite{witten1, chamseddine,zanelli2, banados1, troncoso1,
horava, banados2}. Un trabajo reciente que me parece importante, respecto 
al problema de hallar una soluci\'on (un 'vac\'{\i}o') de una supergravedad de 
Chern-Simons tal que la teor\'{\i}a linealizada alrededor de este vac\'{\i}o es la 
versi\'on linealizada de la supergravedad en 11D, es ref. \cite{hassaine}. 

Se puede entender la diferencia entre los dos enfoques considerando que hay 
 esencialmente dos clases de transformaciones locales en geometr\'{\i}a
diferencial.Estas son difeomerfismos (vistos como transformaciones generales
de coordenadas o deformaciones arbitrarias de la variedad, dependiendo de si
tomamos el punto de vista pasivo o activo) y rotaciones locales de una fibra
(transformaciones de gauge). Hay entonces tambi\'en dos maneras de hacer
local una simetr\'{\i}a global, que son realizarla como una clase de
transformaciones generales de coordenadas o realizarla como una
simetr\'{\i}a de gauge.

En lo que respecta a las simetr\'{\i}as espaciotemporales la primera via es
la que se toma en Relatividad General, mientras que la segunda se toma en
las gravedades de Chern-Simons. Ya se mencion\'o que estas teor\'{\i}as son
equivalentes en 2+1 dimensiones, pero esta equivalencia no vale en dimensiones 
m\'as altas.

En el caso supersim\'etrico nuevamente la primera opci\'on se toma en
supergravedad est\'andar. Los procedimientos principales para construir
estas teor\'{\i}as son el m\'etodo de Noether \cite{westbook} y los basados
en el 'Superespacio'\cite{gates}.

El m\'etodo de Noether involucra los pasos siguientes:

(i) Considerar representaciones de el \'algebra de la supersimetr\'{\i}a con
estados de esp\'{\i}n 2 ('gravitones') como m\'aximo.

(ii) Escribir una acci\'on incluyendo campos de esos espines con los
t\'erminos cin\'eticos est\'andar de segundo orden en las derivadas y sin
interacciones, contruida de modo tal que sea invariante bajo
transformaciones supersim\'etricas globales. Esta parte es bastante directa.

(iii) Hacer las transformaciones locales, entendidas como extensiones de
transformaciones generales de coordenadas, agregando tanto nuevos campos con
propiedades de transformaci\'on adecuadas como t\'erminos nuevos a las
reglas de transformaci\'on previas. Un punto importante es que se requiere
que la acci\'on resultante de este proceso sea de segundo orden como
m\'aximo en las derivadas, y que de lugar a ecuaciones del movimiento de
segundo orden

(iv) Iterar hasta que la acci\'on final sea invariante bajo las nuevas
transformaciones locales. No hay garant\'{\i}a de que este proceso termine
despues de un n\'umero finito de pasos, pero de hecho termina en la
mayor\'{\i}a de los casos interesantes.

En el m\'etodo del Superespacio el espaciotiempo con coordenadas $x^m$ se
extiende a un espacio con coordenadas adicionales que son n\'umeros de
Grassmann que anticonmutan $\theta ^{\alpha}$ y realizando las
transformaciones supersim\'etricas como transformaciones generales de
coordenadas del espacio extendido (superdifeomorfismos).

Como se mencion\'o, en las teor\'{\i}as de Chern-Simons la estrategia
seguida es tomar los grupos espaciotemporales o sus extensiones
supersim\'etricas como grupos de gauge. Las teor\'{\i}as CS contienen
t\'erminos de orden mayor que dos en las derivadas en su acci\'on, al
contrarioque la Relatividad General y las supergravedades est\'andar. Es
importante se\~nalar sin embargo que solo derivadas segundas de los campos
aparecen en las ecuaciones del movimiento \footnote{%
De hecho de primer orden, derivadas de segundo orden aparecen p.ej. si la
torsion es cero, porque la conexi\'on de esp\'{\i}n en ese caso es funci\'on
del vielbein involucrando derivadas primeras de este.}. Las teor\'{\i}as CS
no son entonces 'higher derivative theories', que es como se conoce a
aquellas teor\'{\i}as con ecuaciones del movimiento involucrando derivadas
de orden mayor al segundo de los campos, las cuales se sabe tienen muchas
dificultades que las hacen inconvenientes como modelos f\'{\i}sicos.

Como se se\~nal\'o desde varios puntos de vista Refs.\cite{witten1,
chamseddine, banados1, troncoso1, horava, banados2}. las teor\'{\i}as CS
pueden aproximarse para pequen\~as desviaciones respecto a ciertas
configuraci\'on de referencia o 'background' y para bajas energ\'{\i}as
(perturbaciones de peque\~a amplitud y longitud de onda larga alrededor de
esas configuraciones). La cuesti\'o de en que condiciones, para que
backgrounds y hasta que punto gravedades o supergravedades CS corresponden a
la Relatividad General o supergravedad est\'andar esta sin embargo lejos de
estar zanjada. Lo que est\'a claro es que las gravedades CS no pueden
considerarse en modo alguno descartadas como candidatos a teor\'{\i}as
f\'{\i}sicas de la gravitaci\'on, dando Relatividad General despues de
alguna compactificaci\'on din\'amica apropiada en alg\'un l\'{\i}mite de
longitudes de onda largas.

La presencia de t\'erminos de mayor orden en la curvatura en las gravedades
CS no deber\'{\i}a ser considerado problem\'atico, ya que tales t\'erminos
aparecen por ejemplo en correcciones de la teor\'{\i}a de cuerdas a la
Relatividad General, y no pueden descartarse, dada nuestra ignorancia sobre
el comportamiento del campo gravitacional en distancias cortas

En resumen hay  dos caminos  para entender estas
interacciones como consecuencia de principios de simetr\'{\i}a, el programa
de gauge desarrollado por Maxwell, Weyl, London, Yang, Mills y otros y el
propuesto por Riemann, Clifford, Einstein, Kaluza, Klein y otros de
'simetr\'{\i}as como transformaciones generales de coordenadas'. 
De las cuatro interacciones fundamentales conocidas, las
tres que conocemos a nivel microsc\'opico, como teor\'{\i}as cu\'anticas de
campos,  son teor\'{\i}as de gauge. 

Creo que no es irrazonable esperar que una completa descripci\'on cu\'antica de
todas las interacciones ser\'a realizada dentro del marco de las
teor\'{\i}as de gauge. Entonces el requerimiento de la independencia de
background lleva a alguna clase de teor\'{\i}a de Chern-Simons como la
\'unica posibilidad para tal teor\'{\i}a completa.

\subsection{Transgresiones y teor\'{\i}a de campos}

En este trabajo consideramos extensiones de las teor\'{\i}as de Chern-Simons
basadas en el uso de {\it formas de transgresi\'on} \cite
{stora,zumino,manes,alvarez,chern,naka}, las cuales son generalizaciones de las
formas de Chern-Simons involucrando dos campos de gauge. Rec\'{\i}procamente
las formas de Chern-Simons pueden pensarse como formas de transgresi\'on con
uno de los campos de gauge igual a cero.\newline
Se puede pensar el segundo campo de gauge en las formas de transgresi\'on
como un background de referencia fijo no din\'amico, o como un campo
din\'amico en pie de igualdad con el primero. En el segundo caso adem\'as
puede pensarse que ambos campos est\'an definidos en el mismo espaciotiempo,
o que est\'an definidos en variedades con un borde com\'un.\newline
A nivel de teor\'{\i}a de campos las formas de transgresi\'on poseen:\newline

{\bf (i) Invariancia Gauge}\newline

Las teor\'{\i}as de Chern-Simons no son estrictamente invariantes gauge,
sino cuasi-invariantes, en el sentido de que la acci\'on cambia por un
t\'ermino de borde bajo transformaciones de gauge. Las trangresiones en
cambio son invariantes gauge.\newline

{\bf (ii) Principio de Acci\'on}\newline

Para tener un principio de acci\'on bien definido, en el sentido de que la
acci\'on sea un extremo cuando valen las ecuaciones del movimiento es
necesario en general agregar t\'erminos de borde a la acci\'on, lo cual a
veces se hace caso por caso para configuraciones espec\'{\i}ficas. La
acci\'on de transgresi\'on permite dar una prescripci\'on general de los
t\'erminos de borde que hacen el principio de acci\'on bien definido, los
cuales son de hecho parte de su definici\'on (por lo que puede decirse que
'vienen con la acci\'on').\newline

{\bf (iii) Cargas Conservadas Covariantes}\newline

Las cargas conservadas que vienen de la acci\'on de Chern-Simons no son
covariantes, en el sentido de que su \'algebra de corchetes de Poisson
contiene t\'erminos centrales, como sucede siempre que se parte de una
acci\'on cuasi-invariante. Adem\'as en el caso de las masas de agujeros
negros en diversas dimensiones para gravedades de CS, los valores obtenidos
aplicando el Teorema de Noether a estas teor\'{\i}as no coinciden con los
obtenidos por m\'etodos hamiltonianos, que son los que tienen significado
f\'{\i}sico. Las cargas calculadas utilizando transgresiones como acciones
son covariantes, reflejando la invariancia estricta de la acci\'on, y dan
los mismos valores que los m\'etodos hamiltonianos.\newline

{\bf (iv) Termodin\'amica de Agujeros Negros}\newline

La entrop\'{\i}a de los agujeros negros en gravedades de Chern-Simons en
diversas dimensiones, calculada a partir de la versi\'on eucl\'{\i}dea de la
acci\'on, diverge, por lo que se debe 'regularizar' esta acci\'on con
'contrat\'erminos' apriopiados. La acci\'on de transgresi\'on
correspondiente da una entrop\'{\i}a finita y que coincide con la calculada
por m\'etodos hamiltonianos.\\

La parte arriba mencionada de este trabajo se basa en trabajo realizado 
en colaboraci\'on con Rodrigo Olea, Ricardo Troncoso y Jorge Zanelli, recogido 
en los art\'{\i}culos \cite{motz1,motz2}.

\subsection{Transgresiones y acciones para objetos extendidos}

Otra \'area interesante de aplicaci\'on de las formas de transgresi\'on
tiene que ver con el estudio de modelos de objetos extendidos de diversas
dimensionalidades, como cuerdas y membranas, llamados en general branas, el
cual ha recibido mucha atenci\'on en los \'ultimos a\~nos.\newline

Moore y Seiberg \cite{moore1} mostraron que muchas intrincadas propiedades
de una amplia clase de teor\'{\i}as bidimensionales (2D, signatura 1+1) con
invariancia conforme o 'Conformal Field Theories' (CFT) (las llamadas
'Rational Conformal Field Theories') se pueden entender de forma muy simple
si uno considera estas teor\'{\i}as como inducidas por una teor\'{\i}a de CS
en 3D en su borde bidimensional, como consecuencias de la invariancia gauge
y covariancia general de esta \'ultima. Las CFT en 2D son importantes porque
las Teor\'{\i}as de Cuerdas corresponden a teor\'{\i}as de este tipo.
Result\'o natural tratar de reescribir las acciones en 1+1 dimensiones de
las Supercuerdas como teor\'{\i}as de CS en 2+1 dimensiones \cite{mile} (ver
tambi\'en \cite{witten1,kogan,moore1}) por una especie de 'engrosamiento' de
la superficie de mundo, como modo de sacar ventajas de las propiedades
atractivas de las teor\'{\i}as de CS.\newline

Las formas de transgresi\'on se usar\'an en la construcci\'on de una clase
de modelos \cite{mn,mora} describiendo objetos extendidos con o sin
supersimetr\'{\i}a, como sistemas de gauge. La acci\'on de estos modelos es
la suma de las integrales de las formas de transgresi\'on para el grupo de
gauge en cuesti\'on, integradas sobre subvariedades de la variedad de base
(el volumen de mundo de la brana) y la variedad de base propiamente dicha.
El prop\'osito original era introducir objetos extendidos fundamentales en
supergravedades de Chern-Simons a trav\'es de la inmersi\'on de acciones de
CS de menor dimensi\'on en un background de CS. Si se permit\'{\i}a que
estas branas tuvieran bordes la acci\'on no ser\'{\i}a invariante gauge en
ese caso. Fijar el gauge en los bordes de las branas no era plausible, ya
que estos bordes pod\'{\i}an moverse. Pareci\'o natural entonces usar formas
de transgresi\'on en vez de CS en la construcci\'on, lo cual da acciones
invariantes gauge. El precio que se paga es la duplicaci\'on de los campos
de gauge. En esta clase de modelos confluyen varias l\'{\i}neas separadas de
trabajo, previamente no relacionadas. Sus principales ventajas son:\newline

{\bf (i) Invariancia Gauge} \newline

Los modelos de branas construidos con transgresiones tambi\'en son
invariantes gauge. Este es un punto importante que contrasta con lo que pasa
con la teor\'{\i}a de cuerdas est\'andar, la cual no es un sistema de gauge,
lo cual se critica como uno de sus puntos d\'ebiles \cite{woit}).\newline

{\bf (ii) Independencia de Background y Democracia Brana-Background}\newline

En estos modelos los objetos extendidos y el background est\'an descritos
por acciones de la misma forma, atractiva propiedad que podemos llamar
'democracia brana-background'. El background sin embargo no es fijo, sino
que es din\'amico e interact\'ua con las branas, por lo que el modelo es
independiente de background.\newline

{\bf (iii) Teor\'{\i}a de Cuerdas como una Teor\'{\i}a Topol\'ogica}\newline

Estos modelos avanzan el programa propuesto por Moore y Seiberg \cite{moore1}%
, Witten \cite{witten1}, Green \cite{mile} y Kogan \cite{kogan} de formular
la teor\'{\i}a de cuerdas como un teor\'{\i}a de CS en 2+1 dimensiones.
Incluso se puede conjeturar que uno de los modelos de la clase propuesta
podr\'{\i}a proporcionar una formulaci\'on independiente de background de la
teor\'{\i}a M.\newline

{\bf (iv) Branas Heter\'oticas Supersim\'etricas}\newline

Estos modelos proporcionan una extensi\'on supersim\'etrica de el modelo de
Dixon, Duff y Sezgin (DDS) \cite{dixon} para el acoplamiento de objetos
extendidos a campos de Yang-Mills. Sin embargo nuestros modelos difieren 
de estos en que los
modelos DDS contienen branas que se mueven en un background fijo dado por
campos de gauge $A$, la m\'etrica $g_{rs}$ y el campo-RR $B_d$, mientras que
nuestro modelo es independiente de background y todos los campos son campos
de gauge din\'amicos.\\

La parte de este trabajo que trata con objetos extendidos se 
basa en los trabajos \cite{mn,mora}, el primero de los cuales en colaboraci\'on con Hitoshi Nishino.

\subsection{Plan del trabajo}

El plan de este trabajo es el siguiente:

Las secciones 2 y 3 est\'an dedicadas a revisar material de otros autores
que se utilizar\'a  adelante.

En la Secci\'on 2 se repasar\'an  brevemente los elementos 
de la teor\'{\i}a de fibrados y clases caracter\'{\i}sticas que se
usan en la construcci\'on de nuestros modelos. 
En el ap\'endice A se incluyen detalles adicionales de este tema. 
 
En la secci\'on 3.1 se rev\'en los modelos f\'{\i}sicos de los que los
modelos propuestos ac\'a son extensiones. Estos son teor\'{\i}as de gauge y
gravedades de Chern-Simons (3.1), 
En la secci\'on 4 discutiremos las acciones de transgresi\'on en teor\'{\i}a
de campos, con la secci\'on 4.1 dedicada a las propiedades generales de esta
y una discusi\'on de las opciones disponibles en su formulaci\'on, la
secci\'on 4.2 a gravedad con formas de transgresi\'on, la secci\'on 4.3
dedicada a las cargas conservadas en general, la 4.4 a las cargas
conservadas para teor\'{\i}as con el grupo AdS como grupo de gauge.

La secci\'on 5 se dedica a la termodin\'amica de agujeros negros.

La secci\'on 6 comienza con una subsecci\'on en que 
se revisan los t\'opicos de la teor\'{\i}a de objetos
extendidos que conducen a la clase de modelos propuestos. 
Los temas revisados incluyen 
incluyendo las acciones de Green-Schwarz para supercuerdas
 y el acoplamiento de branas a campos de Yang-Mills  y
trabajos sobre la relaci\'on entre teor\'{\i}a de cuerdas y modelos de CS.

Luego se  introducen las acciones de branas basadas en formas de transgresi\'on, 
se discuten sus
simetr\'{\i}as, sus ecuaciones del movimiento y algunos aspectos de la
teor\'{\i}a cu\'antica, as\'{\i} como posibles relaciones con la teor\'{\i}a
de cuerdas.

La Discusi\'on y Conclusiones van en la secci\'on 7.

El ap\'endice A se dedica a la geometr\'{\i}a y topolog\'{\i}a de fibrados.
 
En el ap\'endice B se
repasan los grupos espaciotemporales y sus extensiones supersim\'etricas.

\newpage

 \section{Geometr\'{\i}a y Topolog\'{\i}a de los Campos de Gauge}

{\it That non-Abelian gauge fields are conceptually identical to ideas in
the beautiful theory of fiber bundles, developed by mathematicians {\bf $%
without$ $reference$ $to$ $the$ $physical$ $world$}, was a great marvel to
me. In 1975, I discussed my feelings with Chern and said "This is both
thrilling and puzzling, since you mathematicians dreamed up this concepts
out of nowhere." He immediately protested, "No, no, this concepts were not
dreamed up. They were natural and real."}

-C.N. Yang\newline

Los objetos conocidos en F\'{\i}sica como Campos de Gauge y en
Geometr\'{\i}a Diferencial como Fibrados desempe\~nan un rol central en
ambas disciplinas. En esta secci\'on se repasan las propiedades
geom\'etricas y topol\'ogicas b\'asicas de estos objetos que se utilizan m\'as
adelante en este trabajo. Debe consultarse el ap\'endice A por m\'as detalles.

La herramientas matem\'aticas que se requieren son esencialmente las
mismas usadas en el estudio de Anomal\'{\i}as en Teor\'{\i}a Cu\'antica de
Campos (TCC), por lo que las referencias en esta secci\'on son los
art\'{\i}culos en ese tema de Stora \cite{stora}, Zumino \cite{zumino},
Ma\~{n}\'es, Stora and Zumino \cite{manes}, y Alvarez-Gaum\'e y Ginsparg 
\cite{alvarez}, y el libro de Bertlmann \cite{bertlmann}. Un muy buen libro  
que contiene estos temas es ref.\cite{naka}.

En lo que respecta a
la literatura de matem\'aticas puras algunos de los resultados presentados
en esta secci\'on pueden encontrarse en el libro de Chern \cite{chern}. Por
una lista extensiva de referencias ver \cite
{stora,zumino,manes,alvarez,chern}. 

\subsection{Fibrados y campos de gauge}

Un fibrado diferenciable $(E,\pi ,M, F, G)$ consiste de los siguientes
elementos\cite{naka}:\\
(i) Una variedad diferenciable $E$ llamada el {\it espacio total}.\\
(ii) Una variedad diferenciable $M$ llamada el {\it espacio base}.\\
(iii) Una variedad diferenciable $F$ llamada la {\it fibra}.\\
(iv) Un mapa $\pi :E\rightarrow M$ llamado la {\it proyecci\'on}. La 
imagen inversa $\pi ^{-1}(p)\equiv F_p\approx F$ es llamada la fibra en p.\\
(v) Un grupo de Lie $G$ llamado {\it grupo de estructura}, que 
act\'ua en $F$ por la izquierda.\\
(vi) Un conjunto de abiertos $\{ U_i\}$ cubriendo $M$ con un difeomorfismo
$\phi _i:U_i\times F\rightarrow \pi ^{-1}(U_i)$ tal que $\pi\phi _i(p,f)=p$. El mapa
$\phi _i$ se llama una {\it trivializaci\'on local} dado que mapea 
$\pi ^{-1}U_i$ en $U_i\times F$.\\
(vii) Si escribimos $\phi _i(p,f)=\phi _{i,p}(f)$, el mapa
$\phi _{i,p}:F\rightarrow F_p$ es un difeomorfismo. Si la intersecci\'on de $U_i$ con $U_j$
es no vac\'{\i}a, se requiere que en la intersecci\'on
$t_{ij}(p)\equiv \phi_{i,p}^{-1}\phi _{j,p}:F\rightarrow F$ sea un elemento de $G$. Por lo tanto $\phi _i$
y $\phi _j$ est\'an relacionados por un mapa suave de la intersecci\'on de $U_i$ y $U_j$ a $G$,
$\phi _j(p,f)=\phi _i(p,t_{ij}(p)f)$. Las $\{t_{ij}\}$ se llaman {\it funciones de transici\'on}.\\

En f\'{\i}sica el espacio base $M$ es el espaciotiempo,con coordenadas denotadas por $x$.
La fibra $F$ es usualmente un espacio vectorial
(isomorfo a $R^n$) dado por los valores de un campo de materia (lo m\'as f\'acil es pensarlo 
como un conjunto de escalares, pero suelen ser espinores)
$\psi ^I$ con \'{\i}ndice en una 
representaci\'on del \'algebra de un grupo $G$. El grupo $G$ es el grupo de estructura, 
correspondiente en f\'{\i}sica al {\it grupo de gauge}. Los campos de materia se escriben usualmente 
$\psi (x) =\psi ^I(x)T^I$, donde $T^I$ son los generadores del \'algebra del grupo en alguna representaci\'on.
La acci\'on del grupo de estructura $G$ en la fibra se define por  $\psi\rightarrow\psi ^g=g^{-1}\psi$
donde $g(x)\equiv exp[\lambda ^I(x)T^I]$ es un elemento del grupo. Esta acci\'on del grupo corresponde 
en f\'{\i}sica a las {\it transformaciones de gauge}.\\

La derivada exterior usual no transforma covariantemente bajo transformaciones de gauge
$d\psi ^g=d(g^{-1}\psi )\neq g^{-1} d\psi $. Para definir una {\it derivada 
covariante} $D$ se introduce la {\it conexi\'on en el fibrado} dada por la
1-forma definida en $M$, $~A=A_m^I(x) T^I \, dx^m $. 
La conexi\'on corresponde en f\'{\i}sica al  {\it potencial de gauge}.
La derivada covariante se define por su 
acci\'on sobre una forma diferencial con \'{\i}ndices en el
grupo $\Omega =\Omega ^IT^I$ como
$$D\Omega=d\Omega +[A,\Omega ]$$
donde $d$ es la derivada exterior y el conmutador entre dos matrices de formas diferenciales 
de ordenes $p$ y $q$ se define por 
\begin{equation}
[ \Lambda _p,\Sigma _q ]= \Lambda _p\Sigma _q -(-1)^{pq}\Sigma _q\Lambda _p
\end{equation}

Definiendo la regla de transformaci\'on de $A$
bajo transformaciones de gauge
como  
$$A\rightarrow A^g=g^{-1}(A+d)g$$
se sigue que $D$ transforma covariantemente 
$$D^g=g^{-1}Dg$$
y tambi\'en
$$D^g\psi ^g=g^{-1}D\psi$$

El {\it tensor de campo} o {\it curvatura} se define como la
2-forma 
$$F=D^2=dA+A^2$$
De la definici\'on de la curvatura resulta que esta satisface id\'enticamente la 
{\it identidad de Bianchi } 
$$DF=0$$
La curvatura $F$ es covariante bajo transformaciones de gauge
$$F^g=(D^g)=g^{-1}Dg g^{-1}Dg=g^{-1}Fg$$
  
Claramente tanto $A$ como $F$ corresponden en realidad a una
matriz de formas diferenciales, para cualquier representaci\'on concreta de
los generadores $T^I$.  

Bajo transformaciones de gauge infinitesimales 
(con los elementos de $\lambda\rightarrow 0$)  se tiene
$$
\delta_{\lambda} A=d\lambda +[ A,\lambda ]~~=D(A)\lambda 
$$
de donde 
$$
\delta _{\lambda} F= [ F,\lambda ] 
$$

\subsection{Polinomios invariantes, formas de transgresi\'on y formas de Chern-Simons}
 
Un {\it polinomio invariante}  $~P(F)$~ se define como la suma formal 
\begin{equation}
P(F)=\sum_{n=0}^{N} \alpha _n \, \hbox{STr}\, \big( F^{n+1} \big) ~~,
\end{equation}
donde 
$$~\, \hbox{STr}\, \big( T^{I_1}\dots T^{I_{n+1}} \big)= g^{I_1\cdots I_{n+1}}$$
 corresponde a una {\it traza sim\'etrica invariante}\footnote{%
Ver ap\'endice B por el significado de 'sim\'etrica' en el caso de un
supergrupo, el cual tiene generadores fermi\'onicos. En ese caso debemos
hablar de una 'supertraza' en vez de una traza.}  en el \'algebra de $~G$. 
Esto es lo mismo que decir que $g^{I_1\cdots I_{n+1}}$ es un tensor
invariante sim\'etrico en el \'algebra del $G$, el cual por contrucci\'on tiene sus
\'{\i}ndices en la representaci\'on adjunta del grupo $G$.

En el ap\'endice A se prueba que los polinomios invariantes son cerrados
$$dP(F)=0$$
y por lo tanto {\bf localmente} exactos 
$$P(F)=d{\cal Q} _{2n+1}(A,F)$$
donde se introdujo la forma de Chern-Simons, definida por
$$
{\cal Q} _{2n+1}(A,F) \equiv (n+1)\int_0^1 d s~\, \hbox{STr} \left(AF_s^{n}\right)
$$
con $A_t=tA$ y $F_t=dA_t+A_t^2$.

Una relaci\'on similar pero que vale globalmente es la {\it f\'ormula de transgresi\'on}, que involucra 
dos potenciales de gauge $A_0$ y $A_1$ en la misma fibra, con curvaturas $F_0$ y $F_1$ respectivamente.
$$
\, \hbox{STr}\, \left(F_1^{n+1}\right)-\, \hbox{STr}\,
\left(F_0^{n+1}\right)=d{\cal T}_{2n+1}(A_1,A_0) 
$$
con la {\it forma de transgresi\'on} definida como
$$
{\cal T}_{2n+1}(A_1,A_0)\equiv  
(n+1)\int _0^1 dt~\, \hbox{STr}\, \left((A_1-A_0)F_t^{n}\right)
$$
con $A_t=tA_1+(1-t)A_0$ y $F_t=dA_t+A_t^2$.

La forma de transgresi\'on es invariante bajo transformaciones de gauge en las que $A_0$
y $A_1$ transforman con el mismo elemento $g$ del grupo $G$, 
debido a la covariancia de $J\equiv A_1-A_0$,$J^g=g^{-1}Jg$,  la 
covariancia de $F_t$,  $F_t^g=g^{-1}F_tg$, y la invariancia de la traza sim\'etrica.\\  

La invariancia 
bajo transformaciones de gauge de las transgresiones es de fundamental importancia para nosotros, 
ya que esa propiedad es la motivaci\'on para usar transgresiones en la construcci\'on de
acciones para sistemas f\'{\i}sicos,  que es de lo que  trata este trabajo.\\

Bajo variaciones infinitesimales gen\'ericas de $A_1$ y $A_0$ la variaci\'on 
de la transgresi\'on es   
$$
\delta {\cal  T}_{2n+1}=(n+1)<F_{1}^{n}\delta A_{1}>-(n+1)<F_{0}^{n}\delta
A_{0}>-n(n+1)~d~\int_{0}^{1}dt<JF_{t}^{n-1}\delta A_{t}>
$$
con $A_{t}=tJ+A_{0}=t A_1+(1-t)A_0$, $\delta A_{t}=t\delta  A_1+(1-t)\delta A_0$
y 
$F_{t}=dA_{t}+A_{t}^{2}$

Bajo transformaciones de gauge involucrando solo $A_1$ tenemos $\delta
A_1=D_1\lambda $, $\delta A_t=tD_1\lambda $ y entonces 
$$
\delta {\cal T}_{2n+1} =d[(n+1)<F_1^n\lambda>
-n(n+1)\int_0^1dt~t~<JF_t^{n-1}D_1\lambda >]
$$
Esto significa que la transgresi\'on var\'{\i}a por un t\'ermino de borde si solo uno 
de los campos se var\'{\i}a. Un resultado an\'alogo vale si se var\'{\i}a solo $A_0$ (ver ap\'endice A)

La expresi\'on previa con $A_1=A$ y $A_0=0$ da la variaci\'on de gauge de la
forma de Chern-Simons:
$$
\delta {\cal Q}_{2n+1} =d[(n+1)<F^n\lambda>
-n(n+1)\int_0^1dt~t~<AF_t^{n-1}D\lambda >]
$$
con $F_t=tF+(t^2-t)A^2$ y $D\lambda =d\lambda+[A,\lambda ]$ . Esto implica que la 
forma de Chern-Simons no es invariante gauge, sino que cambia por un t\'ermino de borde. 
Por esta raz\'on se dice que la forma de Chern-Simons es cuasi-invariante, al contrario que 
las transgresiones, que son invariantes. Como se ver\'a esto representa una gran diferencia 
cuando se consideran las cargas conservadas o la entrop\'{\i}a de agujeros negros en teor\'{\i}as 
de la gravitaci\'on con acciones de Chern-Simons o transgresiones.\\

Las consideraciones de esta secci\'on se extienden directamente a
supergrupos, que se definen en el ap\'endice B.

\newpage

\section{Repaso de la Construcci\'on de Acciones de Chern-Simons}

{\it If one may borrow a term used by the biologists, one would say that
there is gradually forming a "dogma" that all interactions are due to gauge
fields.}\newline
C.N. Yang\newline

\subsection{La Acci\'on}

Se considera un sistema f\'{\i}sico consistente de campos de gauge definidos
en cierta variedad de base como en la secci\'on 2.1 y se busca una acci\'on
que describa la din\'amica cl\'asica y cu\'antica del sistema. Se asume que
no hay una m\'etrica dada de antemano en la variedad que podamos usar en la
construcci\'n de la lagrangiana. La lagrangiana en dimensi\'on $D$ debe ser
una D-forma invariante gauge. El \'unico objeto local covariante gauge que
se puede construir con los potenciales de gauge $A$ es el tensor de campo $F$%
. Como no tenemos una m\'etrica, no podemos construir una acci\'on de
Yang-Mills. Una 2n-forma gen\'erica covariante gauge ser\'{\i}a $%
F^{I_1}...F^{I_n}$ y podemos construir un objeto invariante gauge
contrayendo esta con un tensor invariante $g_{I_1...I_n}$ (el cual debe ser
sim\'etrico en sus \'{\i}ndices debido a que al ser las $F$ 2-formas
conmutan, de modo que $F^{I_1}...F^{I_n}$ es sim\'etrico ) El invariante $%
g_{I_1...I_n}F^{I_1}...F^{I_n}$ no sirve como lagrangiana sin embargo,
debido a que es una derivada total (localmente) como ya vimos. Sin embargo
podemos tomar como nuestra acci\'on para dimensiones impares $D=2n+1$ 
\begin{equation}
S=k~\int_{S^{2n+1}}{\cal Q}_{2n+1} (F,A)=k~\int_{M^{2n+2}}\, \hbox{STr}\, (F^{n+1})
\end{equation}
donde $k$ es una constante. Esta lagrangiana, dada por la forma de
Chern-Simons, no es invariante gauge, cambia localmente por una derivada
total 
\[
\delta _\lambda {\cal Q} _{2n+1}(A,F)=-dQ^1_{2n}(A,F,\lambda) 
\]
lo que significa que la acci\'on dada es invariante gauge en una variedad
sin borde suponiendo que la topolog\'{\i}a de la fibra es trivial. La
variaci\'on de la forma de Chern-Simons es solo localmente exacta, entonces
si la topolog\'{\i}a de la fibra es no trivial las contribuciones de
diferentes cartas locales superpuestas no se cancelar\'an en la regi\'on de
intersecci\'on. La acci\'on tampoco es invariante bajo transformaciones de
gauge globalmente no triviales. Por construcci\'on la acci\'on es
generalmente covariante y no depende de ninguna m\'etrica definida en la
variedad.

\subsection{ Ecuaciones del movimiento}

Bajo variaciones del potencial de gauge se tiene 
\begin{equation}
\delta S= k~(n+1)\int_{M^{2n+2}}\, \hbox{STr}\, (D(\delta A)F^{n}) =
k~(n+1)\int_{M^{2n+2}}d\left[\, \hbox{STr}\, (\delta AF^{n})\right]
\end{equation}
donde usamos $\delta F=D(\delta A)$ y ec.(4). Del teorema de Stokes 
\begin{equation}
\delta S=k~ (n+1)\int_{S^{2n+1}}\, \hbox{STr}\, (\delta AF^{n})
\end{equation}
Entonces las ecuaciones del movimiento $\frac{\delta S}{\delta A}=0$ son 
\cite{chamseddine,banados1,troncoso1} 
\begin{equation}
\, \hbox{STr}\, (T^IF^{n})=0
\end{equation}
En 2+1 dimensiones ($n=1$) y si la 'm\'etrica del grupo' $d^{IJ}=\, %
\hbox{STr}\, (T^IT^J)$ es invertible ('no degenerada') las ecuaciones del
movimiento implican que $F^I=0$ y por lo tanto el campo de gauge es gauge
puro. Esto significa que no hay grados de libertad locales que se propaguen,
y el campo de gauge puede hacerce cero por una transformaci\'on de gauge en
cualquier carta local (pero no en todas simultaneamente) En dimensiones mas
altas esto no es verdad, las ecuaciones del movimiento aun tienen la
soluci\'on $F^I=0$, pero existen soluciones para las que esto no es cierto.

\subsection{Gravedad y Supergravedad de Chern-Simons}

Las gravedades de Chern-Simons son teor\'{\i}as de CS con un grupo
espaciotemporal como grupo de gauge 
\begin{equation}
A=e^r~P_r+\frac{1}{2}\omega ^{rs}J_{rs}
\end{equation}
donde $e^r$ es el vielbein (por una constante con dimensiones de longitud
inversa) y $\omega ^{rs}$ es la conexi\'on de spin. La dimensi\'on del
espaciotiempo es $D=2n+1$ y se toma $P_r=J_{r,D+1}$. La Torsi\'on $T^r$ y la
Curvatura $R^{rs}$ se definen como 
\begin{eqnarray}
T^r = de^r + \omega ^r_s e^s \\
R^{rs}= d\omega ^{rs}+ \omega ^{rp} \omega _p^{~s}
\end{eqnarray}
Bajo transformaciones de gauge infinitesimales 
\begin{equation}
\lambda=\lambda^rP_r+\frac{1}{2} \lambda^{rs}M_{rs}
\end{equation}
los campos cambian como 
\begin{equation}
\delta e^r = D \lambda^r~~~,~~~\delta\omega ^{rs} = 0
\end{equation}
para traslaciones de gauge y 
\begin{equation}
\delta e^r = \lambda^r_{~s} e^s ~~~, ~~~\delta\omega ^{rs} = -D\lambda^{rs}
\end{equation}
para transformaciones de Lorentz de gauge. Las expresiones para los grupos
de de Sitter son similares, con algunos t\'erminos adicionales que se
reducen a cero en el caso de Poincar\'e, y pueden calcularse a partir del
\'algebra (ver las referencias al comienzo de esta secci\'on).

Para especificar el modelo se debe dar el tensor invariante a usar. La
opci\'on mas com\'un es el pseudo-tensor de Levi-Civita 
\begin{equation}
<J_{r_1r_2}...J_{r_Dr_{D+1}}>=\epsilon _{r_1...r_{D+1}}
\end{equation}
El polinomio invariante obtenido en este caso se llama 'densidad de Euler'
En el caso $D=2+1$ la acci\'on es 
\begin{equation}
S_3=k~\epsilon _{rsp}\int _{S^3} e^r \left( d\omega ^{sp}+\omega ^{sq}\omega
_q^{~p} + \lambda \frac{1}{3}e^se^p\right)
\end{equation}
donde $\lambda = 0$ para el grupo de Poincar\'e $ISO(2,1)$, $\lambda = +1$
para el grupo AdS $SO(2,2)$ y $\lambda = -1$ para el grupo dS $SO(3,1)$. Las
ecuaciones del movimiento correspondientes a extremizar la acci\'on bajo
variaciones de la conexi\'on de spin $\omega ^{rs}$ son 
\begin{equation}
T^r = de^r + \omega ^r_s e^s=0
\end{equation}
Se sigue que si el vielbein es invertible (no degenerado) como matriz $%
det(e^r_m)\neq 0$, donde $m$ es un \'{\i}ndice espaciotemporal 'curvo', se
puede escribir la conexi\'on de spin como funci\'on del vielbein (on-shell) 
\begin{equation}
\omega ^{rs}_m =\omega_m ^{rs} (e^r_m)
\end{equation}
Las ecuaciones del movimiento correspondientes a $e^r$ son 
\begin{equation}
\epsilon_{rsp}(R^{sp}+\lambda e^se^p)=0
\end{equation}
La acci\'on de Einstein-Hilbert (EH) en cualquier dimensi\'on es 
\begin{equation}
S_{EH}=k~ \epsilon _{r_1....r_D}\int_{S^D} R^{r_1r_2}e^{r_3}...e^{r_D}
\end{equation}
donde $R^{rs}$ se define como antes, pero en t\'erminos de la conexi\'on de
spin correspondiente al caso de torsi\'on cero $\omega (e)$ of eq.(75).

Esta claro que la Relatividad General en 2+1 dimensiones es equivalente {\it %
on-shell} a la teor\'{\i}a de Chern-Simons para el grupo de Poincar\'e ($%
\lambda =0$), si el vielbein es no degenerado.

En dimensiones mas altas y para el tensor invariante $%
<M_{r_1r_2}...M_{r_Dr_{D+1}}>=\epsilon _{r_1...r_{D+1}}$ la acci\'on es 
\begin{equation}
S_{2n+1}=k~\int_{S^{2n+1}}{\cal Q} _{2n+1}
\end{equation}
y las ecuaciones del movimiento son 
\begin{eqnarray}
\epsilon _{r_1....r_{2n+1}} T^{r_1}(R^{r_2r_3}+\lambda e^{r_2}e^{r_3})
...(R^{r_{2n-2}r_{2n-1}}+\lambda e^{r_{2n-2}}e^{r_{2n-1}})=0 \\
\epsilon _{r_1....r_{2n+1}} (R^{r_1r_2}+\lambda e^{r_1}e^{r_2})
...(R^{r_{2n-1}r_{2n}}+\lambda e^{r_{2n-1}}e^{r_{2n}})=0
\end{eqnarray}
Esta teor\'{\i}a no es equivalente ni siquiera {\it on-shell} a la
Relatividad General, y el que la torsi\'on sea cero no es requerido por las
ecuaciones del movimiento, al contrario de lo que sucecede en 2+1. Sin
embargo se cree que las gravedades de Chern-Simons en dimensiones mas altas
son equivalentes a la Relatividad General en ciertos l\'{\i}mites, como se
menciona en la secci\'on 3.1.5.

Es posible tambi\'en construir acciones de CS para los grupos
espaciotemporales usando combinacines simetrizadas de trazas ordinarias como
tensores invariantes. Estas se conocen como 'acciones de gravedades
ex\'oticas de Chern-Simons'. Se diferencian de las mencionadas anteriormente
en que en general la torsi\'on aparece expl\'{\i}citamente en estas
acciones. Los invariantes $\, \hbox{STr}\, (F^k)$ se conocen como
'densidades de Pontryagin' y son nulas a menos que $k$ sea par para los
grupos $SO(d)$ con cualquier signatura, lo que implica que acciones de CS de
este tipo solo existen para $D=4n-1$.

Las supergravedades de Chern-Simons ('CS-sugra') son modelos con acciones de
Chern-Simons para extensiones supersim\'etricas de los grupos
espaciotemporales. Si los generadores fermi\'onicos son espinores de
Majorana, los potenciales de gauge son de la forma 
\begin{equation}
A=e^r~P_r+\frac{1}{2}\omega ^{rs}M_{rs}+\psi^{\alpha} Q_{\alpha} +\sum _k
b_{(k)}^{r_1...r_k} (Z_{(k)})_{r_1...r_k}
\end{equation}
donde $\psi _m^{\alpha}$ es el 'gravitino' y las $b_{(k)}^{r_1...r_k}$
1-formas son los potenciales de gauge asociados a las cargas bos\'onicas
adicionales que puedan requerirse para cerrar el \'algebra. Los tensores
invariantes pueden ser una extensi\'on supersim\'etrica apropiada de $%
\epsilon _{r_1...r_{k}}$ (que puede ser dif\'{\i}cil de construir), o
cualquier producto simetrizado de trazas de generadores (que puede
construirse de manera directa). 

\subsection{Teor\'{\i}a Cu\'antica}

La teor\'{\i}a cu\'antica se define formalmente a trav\'es de la integral de
caminos 
\begin{equation}
Z=\sum_{topologias}\int{\cal D}A~e^{iS/\hbar}
\end{equation}
donde se asume que sumamos sobre todas las configuraciones del campo de
gauge, y sobre todas las geometr\'{\i}as y topolog\'{\i}as de la variedad
deferencial de base. En principio deber\'{\i}an adem\'as utilizarse
procedimientos adecuados, analogos al m\'etodo de Fadeev-Popov, para evitar
redundancias en esta suma, como sumar configuraciones de gauge
correspondientes al mismo estado f\'{\i}sico.

Los observables naturales invariantes gauge son no locales, los llamados
'loops de Wilson' o 'lazos de Wilson' 
\begin{equation}
W[\gamma ] =Tr~\left[ {\cal P}~exp(\int _{\gamma}A)\right]
\end{equation}
donde $\gamma$ es una curva cerrada cualquiera y ${\cal P}~exp$ es la
exponencial ordenada de camino, definida como el producto de las $%
1+A_{\mu}dx^{\mu }$ en segmentos infinitesimales en que se divide el camino,
en el orden en que este es recorrido. Witten mostr\'o que para teor\'{\i}as
de gauge de CS en 2+1 los valores esperados cu\'anticos de los loops de
Wilson son invariantes topol\'ogicos de nudos 
\begin{equation}
K(\gamma )=<W[\gamma ]>=\int{\cal D}A~W[\gamma ]~e^{iS/\hbar}
\end{equation}
como cab\'{\i}a esperar de la ausencia de una m\'etrica de referencia que
proporcionara una noci\'on de distancia entre puntos de $\gamma$.

Un punto importante es que la constante de acoplamiento en la acci\'on esta
cuantizada, en el sentido de que la consistencia de la teor\'{\i}a a nivel
cu\'antico requiere que esta tome alguno de un conjunto discreto de valores.
La idea es que la acci\'on puede escribirse como 
\begin{equation}
S = k~\int_{M^{2n+2}}\, \hbox{STr}\, (F^{n+1})
\end{equation}
o 
\begin{equation}
\overline{S} =k~\int_{\overline{M}^{2n+2}}\, \hbox{STr}\, (F^{n+1})
\end{equation}
dependiendo de si extendemos $S^{2n+1}$ en $M^{2n+2}$ o en $\overline{M}%
^{2n+2}$ de modo que $S^{2n+1}=\partial M^{2n+2}= \partial\overline{M}%
^{2n+2} $. Es natural pedir que la f\'{\i}sica descrita por tal teor\'{\i}a
no pueda depender del modo en que uno extienda $S^{2n+1}$, ya que la
teor\'{\i}a est\'a definida en esta. Para que la integral de camino no
dependa de que extensi\'on elegimos debemos requerir 
\begin{equation}
S-\overline{S}=2\pi i~m
\end{equation}
donde $m$ es un entero. Pero 
\begin{equation}
S-\overline{S}= k~\int_{M_T^{2n+2}}\, \hbox{STr}\, (F^{n+1})
\end{equation}
donde $M_T^{2n+2}$ es la variedad cerrada formada por la uni\'on de $%
M^{2n+2} $ y $\overline{M}^{2n+2}$, esta \'ultima con la orientaci\'on
invertida, unidas en $S^{2n+1}$. Observamos que $\int_{M_T^{2n+2}}\, %
\hbox{STr}\, (F^{n+1})$ es un n\'umero de Chern, invariante topol\'ogico. La
condici\'on de cuantizaci\'on en $k$ queda 
\begin{equation}
k~\int_{ M_T^{2n+2} }\, \hbox{STr}\, (F^{n+1})=2\pi i~m
\end{equation}
Esta condici\'on de cuantizaci\'on es satisfecha por cualquier $M_T^{2n+2}$
posible solo si la integral $\int_{ M_T^{2n+2} }\, \hbox{STr}\, (F^{n+1})$
es proporcional a un n\'umero entero. Sabemos por el teorema de \'{\i}ndice
ec.(37) que esto es verdad para la traza simetrizada usual. Tambi\'en es
cierto en el caso del grupo $SO(D)$ para $D$ par y signatura arbitraria, con 
$\epsilon _{r_1...r_D}$ como tensor invariante, en cuyo caso la integral se
conoce como el 'n\'umero de Euler' de la variedad $\chi (M_T)$(el cual
tambien se relaciona con el \'{\i}ndice de un operador diferencial).

Una importante consecuencia del resultado anterior es que se espera que la
acci\'on cl\'asica sea ya la 'acci\'on efectiva cu\'antica'. Normalmente se
espera que despues de insertar la acci\'on cl\'asica en la integral de
caminos y calcular los efectos cu\'anticos usando t\'ecnicas apropiadas
(desarrollos porturbativos y regularizaci\'on, etc.) se obtiene la acci\'on
efectiva cu\'antica como la acci\'on cl\'asica mas correcciones cu\'anticas
('contrat\'erminos'), que pueden tener la misma dependencia funcional en los
campos din\'amicos que la acci\'on cl\'asica (dando lugar solo a la
renormalizaci\'on de los campos, masas y constantes de acoplamiento) o, como
sucede en general, una dependencia diferente.

Para teor\'{\i}as de CS los posibles contrat\'erminos consistentes con las
simetr\'{\i}as de la acci\'on (invariancia gauge y covariancia general) son
tambi\'en Chern-Simons. Esto fue se\~nalado por primera vez en el estudio de
las anomal\'{\i}as en TCC (y las matem\'aticas son esencialmente las mismas
aca) donde se conoce como 'teorema de Adler-Bardeen' \cite{Adler}. El
teorema de Adler-Bardeen para anomal\'{\i}as quirales establece que las
correcciones cu\'anticas de mayor orden (radiativas) a la anomal\'{\i}a solo
dan origen a la renormalizaci\'on de la funci\'on de ondas (redefinici\'on
de los campos) y de las cargas, pero la forma de la anomal\'{\i}a est\'a
determinada por la contribuci\'on de orden mas bajo. Este resultado fue
extendido a situaciones mas generales por Piguet y Sorella \cite{piguet}%
,quienes usaron m\'etodos BRST para independizarse de cualquier
procedimiento espec\'{\i}fico de regularizaci\'on. Podemos interpretar este
resultado mas la condici\'on de cuantizaci\'on como implicando que despues
de que todas las correcciones cu\'anticas y la regularizaci\'on se toman en
cuenta se debe finalizar con una acci\'on efectiva de la forma original con
una constante $k$ que satisface la condici\'on de cuantizaci\'on dada arriba.

En teor\'{\i}a cu\'antica de campos aquellas teor\'{\i}as que son bien
definidas para todas las escalas de energ\'{\i}a\footnote{%
por lo que no es necesario truncarlas a una escala determinada de distancias
o energ\'{\i}a, lo que se conoce como un 'cut-off'.} se dice que tienen un
'punto fijo ultravioleta'. Una clase de teor\'{\i}as que tienen esta
propiedad es la de las teor\'{\i}as 'asintoticamente libres', como por
ejemplo la Cromodin\'amica Cu\'antica(QCD) (ver p. ej. \cite{gross11}). Sin
embargo para estas teor\'{\i}as el punto fijo se dice 'trivial', porque en
el l\'{\i}mite ultravioleta la constante de acoplamiento se anula y la
teor\'{\i}a es libre. No se conocen teor\'{\i}as con puntos fijos
ultravioletas no triviales, pero hay teoremas en el sentido de que si esas
teor\'{\i}as existen el n\'umero de constantes de acoplamiento no nulas en
ese l\'{\i}mite debe ser finito (ver \cite{weinberg11}).

Es interesante observar que de los puntos previos se puede concluir que las
teor\'{i}as de Chern-Simons al nivel cu\'{a}ntico son TCC con un punto fijo
ultravioleta no  trivial.\newline

\newpage

\section{Transgresiones en Teor\'{\i}a de Campos}

{\it Santiago, y adelante.}

{\it Hern\'an Cort\'es}

\subsection{Transgresiones como acciones de teor\'{\i}as de gauge}

En esta secci\'on se consideran generalizaciones de las teor\'{\i}as con
acciones de Chern-Simons en las que la lagrangiana se toma como dada por
formas de Transgresi\'on, $L_{trans}={\cal T}_{2n+1}$ en dimensi\'on $D=2n+1$, en
vez de formas de Chern-Simons. Este tipo de generalizaci\'on fue considerada 
primero en \cite{postdam,mn,mora} y posteriormente en \cite{francaviglia1,francaviglia2}. 
Los resultados originales contenidos en esta secci\'on y la siguiente sobre termodin\'amica 
de agujeros negros fueron presentados en  nuestro trabajo en \cite{motz1,motz2}.\\ 

El uso de transgresiones \cite{motz1,motz2} tiene la ventaja inmediata de que la
acci\'on es ahora estrictamente invariante gauge, en vez de solo
cuasi-invariante, pero ademas tiene las importantes ventajas mencionadas en
la introducci\'on de:\newline
(i) dar un principio de acci\'on bien definido, en el sentido de que la
acci\'on es un extremo cuando valen las ecuaciones del movimiento con
condiciones de borde apropiadas, \newline
(ii) dar cargas conservadas covariantes a trav\'es del m\'etodo de Noether 
(debido a la invariancia de gauge estricta, como se observ\'o en general en \cite{brown}),
y en el caso de gravitaci\'on una masa que coincide con la obtenida por
m\'etodos hamiltonianos (al contrario de lo que pasa con la acci\'on de
Chern-Simons),\newline
(iii) regularizar la acci\'on a trav\'es de los t\'erminos de borde dictados
por la invariancia gauge de modo que la entrop\'{\i}a de los agujeros negros
en gravitaci\'on de AdS calculada a partir de la acci\'on eucl\'{\i}dea es
la correcta, comparada con la calculada con m\'etodos hamiltonianos (donde
estos t\'erminos de borde deben calcularse caso por caso).\newline
Se puede pensar que una vez que se logra la invariancia gauge de la
teor\'{\i}a, los dem\'as puntos se siguen como resultado de la 'magia del
principio de gauge', pero a\'un as\'{\i} resulta sorprendente que la
extensi\'on de las acciones de Chern-Simons requerida por la invariancia
gauge resuelva adem\'as estos otros problemas.\newline

Las ecuaciones del movimiento pueden determinarse a partir de la f\'ormula
general para variaciones de las transgresiones 
\begin{equation}
\delta {\cal T}_{2n+1}= (n+1)<F_1^n\delta A_1>- (n+1)<F_0^n\delta A_0>
-n(n+1)d\int_0^1<JF_t^{n-1}\delta A_t>
\end{equation}
donde las interpolaciones son entre $A_0$ y $A_1$.

Las ecuaciones del movimiento (E.d.M.)que se siguen de esta acci\'on son 
\begin{equation}
<F_1^nT^I>=0~~,~~<F_0^nT^I>=0
\end{equation}
las cuales deben suplirse con condiciones de borde apropiadas que anulen el
t\'ermino de borde 
\[
-n(n+1)d\int_0^1<JF_t^{n-1}\delta A_t>
\]
, de modo que sea $\delta {\cal T}_{2n+1}= 0$ cuando valen las E.d.M., con lo que
la acci\'on ser\'{\i}a realmente un extremo. El t\'ermino de borde en la
variaci\'on de la transgresi\'on se anula si las variaciones $\delta A_1$ y $%
\delta A_0$ se toman como cero en el borde, pero esto, que equivale a tomar
los potenciales de gauge fijos en el borde como condici\'on de borde, es
demasiado restrictivo, considerando que la lagrangiana contiene solo
derivadas primeras de los campos.

Una condici\'on de borde que parece natural dada la forma del t\'ermino de
borde es pedir que $J=0$ (o que tienda a cero lo suficientemente r\'apido)
en el borde, con lo que $A_0=A_1$ en el borde, y que $F_t$ sea finito en el
borde. Esto asegurar\'{\i}a que el t\'ermino de borde se anule. Se
considerar\'a otra posible condici\'on de borde en el caso de gravitaci\'on.

En esta discusi\'on de las ecuaciones del movimiento se asumi\'o que ambos
campos de gauge son din\'amicos. Sin embargo se pueden distinguir dos
posibilidades:\newline
(i) la ya mencionada de considerar tanto $A_1$ como $A_0$ campos din\'amicos
que satisfacen las E.d.M., o\newline
(ii) solo $A_1$ es din\'amico, mientras $A_0$ es un background fijo. En ese
caso Solo $A_1$ debe satisfacer las E.d.M..\newline

Observese que cualquiera sean las condiciones de borde, tanto en caso (i)
como (ii) la acci\'on es un extremo para variaciones que se reducen a
transformaciones de gauge en borde. En el caso (i) podemos tomar $\delta A_1$
y $\delta A_0$ como arbitrarios en el interior ('bulk') pero reduciendose a
transformaciones de gauge infinitesimales con el mismo par\'ametro de gauge $%
\lambda$ en el borde. Esto es as\'{\i} porque debido a la invariancia gauge
de la trangresi\'on se tiene 
\[
0=\delta _{\lambda}{\cal T}_{2n+1}=(n+1)<F_1^n D_1\lambda >- (n+1)<F_0^n
D_0\lambda >-n(n+1)d\int_0^1<JF_t^{n-1}\delta _{\lambda}A_t> = 
\]
\[
=d\left\{ (n+1)<F_1^n \lambda >- (n+1)<F_0^n \lambda
>-n(n+1)\int_0^1<JF_t^{n-1}D_t\lambda >\right\} 
\]
mientras que si miramos a la variaci\'on general de la transgresi\'on se
tiene que los t\'erminos de bulk son cero debido a las E.d.M. y el t\'ermino
de borde es el mismo de la expresi\'on previa para variaciones de gauge,
porque de todos modo $<F_1^n \lambda >$ y $<F_0^n \lambda >$ son cero
(E.d.M.), lo que prueba que la acci\'on es un extremoa\'un si variaciones de
gauge de los potenciales se permiten en el borde.

En el caso (ii) no asumimos que $<F_0^nT^I>=0$, y se toma $\delta A_1$
arbitrario en el bulk pero reduciendose a transformaciones de gauge con
par\'ametro $\lambda$ en el borde, mientras $\delta A_0$ es una variaci\'on
de gauge con par\'ametro $\lambda$ tanto en el bulk como en el borde (con un 
$\lambda $ que es el mismo que aparece en las variaciones de $\delta A_1$ en
el borde). Un argumento an\'alogo al usado en el caso (i) muestra que
tambi\'en en este caso la variaci\'on de la acci\'on ser\'a cero. La
situaci\'on es similar a la discutida por Regge y Teitelboim \cite{regge} en
Relatividad General, donde la adici\'on de t\'erminos de borde se
requer\'{\i}a para tener un principio de acci\'on bien definido, mientras se
permit\'{\i}an transformaciones de Poincar\'e en el infinito espacial. La
similitud ser\'a a\'un mas estrecha al considerar gravitaci\'on.\newline

Las formas expl\'{\i}citas de las formas de Chern-Simons y Transgresi\'on en
3D son 
\begin{equation}
{\cal Q} _3=<AdA+\frac{2}{3}A^3>=<AF-\frac{1}{3}A^3>
\end{equation}
y 
\begin{equation}
{\cal T}_3 =<A_1dA_1+\frac{2}{3}A_1^3>-<A_0dA_0+\frac{2}{3}A_0^3>-<A_1A_0>
\end{equation}
En 5D estas son 
\begin{equation}
{\cal Q_5} =<A(dA)^2+\frac{2}{3}A^3dA+\frac{3}{5}A^5>= <AF^2-\frac{1}{2}A^3F+\frac{%
1}{10}A^5>
\end{equation}
y 
\begin{equation}
{\cal T}_5(0,1)={\cal Q}_5 (1)-{\cal Q}_5 (0)-dC_4
\end{equation}
donde 
\begin{equation}
C_4=\frac{1}{2}<(A_1A_0-A_0A_1)(F_1+F_0)+A_0A_1^3+A_0^3A_1+ \frac{1}{2}%
A_1A_0A_1A_0>
\end{equation}
La notaci\'on ${\cal Q}_5 (1)$ (${\cal Q}_5 (0)$) significa que el argumento es $A_1$ (%
$A_0 $).

\subsection{Gravedad con Formas de Transgresi\'on}

\subsubsection{Generalidades}

Consideraremos teor\'{\i}as de la gravitaci\'on en las que la acci\'on
est\'a dada por formas transgresi\'on con el grupo G dado por el grupo de
Anti-de Siter $SO(d-2,2)$, $d=2n+1$ \cite{motz1,motz2}, con generadores $J_{AB}$ con el algebra

\begin{equation}
[ J_{AB} , J_{CD} ]= + \eta _{BC}J_{AD} -\eta _{AC}J_{BD} -\eta _{BD}J_{AC}
+\eta _{AD}J_{BC}
\end{equation}
y la traza sim\'etrica definida por 
\begin{equation}
<J_{A_1A_2}...J_{A_{d-1}A_d}> = \epsilon _{A_1....A_d}
\end{equation}
La traza sim\'etrica puede definirse a veces con otra normalizaci\'on, por
ejemplo mas adelante usaremos 
\begin{equation}
<J_{A_1A_2}...J_{A_{d-1}A_d}> = \kappa \frac{2^n}{(n+1)} \epsilon
_{A_1....A_d}
\end{equation}
donde $\kappa =[2(d-2)!\Omega_{d-2}G_k]^{-1}$ con $\Omega_{d-2}$ el volumen
de la la esfera en $d-2$ dimensiones y $G_d$ la 'constante de Newton' en
dimensi\'on $d$. Esto solo da el mismo que aparece en la traza sim\'etrica
factor de normalizaci\'on en frente de la transgresi\'on o el CS. Los
generadores se dividen en generadores del 'grupo de Lorentz' $J_{ab}$ with $%
a,b=0,...,d-2$ y los generadores de 'traslaciones' $P_a=J_{a,d-1}$. El
potencial de gauge $A$ es 
\begin{equation}
A=\frac{1}{2}\omega ^{ab}J_{ab}+e^aP_a
\end{equation}
En d=2+1 la forma de Chern-Simons es 
\begin{equation}
{\cal Q}_3 (e,\omega )= \epsilon _{abc} (R^{ab}e^c + \frac{1}{3}e^3) + \frac{1}{2}
\epsilon _{abc}d(e^a\omega ^{bc})
\end{equation}
y la forma de Transgresi\'on es 
\begin{equation}
{\cal T}_3(e_1,\omega _1 ;e_0,\omega _0)= \epsilon _{abc} (R_1^{ab}e_1^c + \frac{1%
}{3}e_1^3) -\epsilon _{abc} (R_0^{ab}e_0^c + \frac{1}{3}e_0^3) +\frac{1}{2}
\epsilon _{abc}d[(e_1^a+e_0^a)( \omega _1^{bc} -\omega _0^{bc})]
\end{equation}
En d=4+1 la forma de Chern-Simons es 
\begin{eqnarray}
{\cal Q}_5(e,\omega )= \frac{3}{4}\epsilon _{abcde} ( R^{ab}R^{cd}e^e +\frac{2}{3}
e^ae^be^cR^{de} +\frac{1}{5}e^ae^be^ce^de^e) + \\
\frac{1}{4}\epsilon _{abcde} d(-2\omega ^{ab}d\omega ^{cd}e^e +\frac{1}{2}%
e^ae^be^c\omega ^{de}-\frac{3}{2}\omega ^{af}\omega _f^{~b} \omega ^{cd} e^e)
\nonumber
\end{eqnarray}
En una notaci\'on mas compacta 
\begin{eqnarray}
{\cal Q}_5(e,\omega )= \frac{3}{4}\epsilon  (3 R^{2}e +2 e^3 R +\frac{1}{5}e^5) +
\\
\frac{\epsilon }{4} d(-2\omega d\omega e +\frac{1}{2}e^3\omega -\frac{3}{2}%
((\omega ^{2})) \omega e)  \nonumber
\end{eqnarray}
donde el par\'entesis doble implica contracciones, como por ejemplo $%
((\omega ^{2}))\equiv \omega ^{af}\omega _f^{~b}$, y $((\omega e))\equiv
\omega ^{af}e_f$. Para la transgresi\'on en d=4+1 tenemos 
\[
{\cal T}_5=\frac{3}{4}\epsilon (R^2e+\frac{2}{3}Re^{3}+\frac{1}{5}e^5) - \frac{3}{4}
\epsilon (\tilde{R}^2 \overline{e}+\frac{2}{3}\tilde{R}\overline{e}^{3}+%
\frac{1}{5}\overline{e}^5)-
\]
\[
\frac{1}{4} \epsilon ~d[ \theta (e+\overline{e}) (R-\frac{1}{4}\theta
^2+\frac{1}{2}e^2) + \theta (e+\overline{e})(\tilde{R}-\frac{1}{4}\theta ^2+%
\frac{1}{2}\overline{e}^2)+\theta Re+\theta \tilde{R}\overline{e}] 
\]
donde $\theta ^{ab}=\omega ^{ab}-\overline{\omega  }^{ab}$.
 
\subsubsection{Configuraci\'on de Variedad Cobordante (VC)}

Una elecci\'on particular de la configuraci\'on $A_0$ que permite apartarse
lo menos posible de las teor\'{\i}as de gravitaci\'on de Chern-Simons, en el
sentido de agregar un m\'{\i}nimo de estructura adicional, es la
configuraci\'on de variedad cobordante (VC) \cite{motz1,motz2}, donde si $A_1=A$ y $A_0=%
\overline{A}$, tenemos 
\begin{equation}
A=\frac{1}{2}\omega ^{ab}J_{ab}+e^aP_a~~,~~ \overline{A}=\frac{1}{2}%
\overline{\omega} ^{ab}J_{ab}+\overline{e}^aP_a
\end{equation}
con $\overline{A}$ definido solo en el borde por 
\begin{equation}
\overline{e}^a=0~~~,~~~\overline{\omega}^{1\underline{i}}=0~~~,~~~ \overline{%
\omega}^{\underline{i}\underline{j}} = \omega ^{\underline{i}\underline{j}}
\end{equation}
donde el \'{\i}ndice 1 corresponde a la direcci\'on normal al borde y los
\'{\i}ndices subrayados, como $\underline{i}$, pueden tomar cualquier valor
diferente de 1.\newline

Para esta elecci\'on de $A_0$ puede verse que la acci\'on puede escribirse
como la forma est\'andar del t\'ermino de bulk para la lagrangiana de
Chern-Simons con el tensor invariante 
\begin{equation}
<J_{A_1A_2}...J_{A_{d-1}A_d}> = \kappa \frac{2^n}{(n+1)} \epsilon
_{A_1....A_d}
\end{equation}
el cual puede escribirse notablemente solo en t\'erminos de $e$ y $R$ (y no
de $\omega$) en la forma conocida como de Lanczos-Lovelock- Chern-Simons 
\begin{equation}
L_{LCS}(R,e)=\kappa \int\limits_{0}^{1}dt\epsilon \left( R+t^{2}e^{2}\right)
^{n}e  \label{LCS}
\end{equation}
mas un t\'ermino de borde dado por 
\begin{eqnarray*}
\alpha &=&-\kappa n\int\limits_{0}^{1}dt\epsilon \theta
e\sum\limits_{k=0}^{n-1}\frac{C_{k}^{n-1}}{2k+1}\sum%
\limits_{l=0}^{n-1-k}C_{l}^{n-1-k}\tilde{R}^{n-1-k-l}t^{2l}\theta
^{2l}t^{2k}e^{2k} \\
&=&-\kappa n\int\limits_{0}^{1}dt\int\limits_{0}^{t}ds\epsilon \theta
e\sum\limits_{k=0}^{n-1}C_{k}^{n-1}\sum\limits_{l=0}^{n-1-k}C_{l}^{n-1-k}%
\tilde{R}^{n-1-k-l}t^{2l}\theta ^{2l}s^{2k}e^{2k} \\
&=&-\kappa n\int\limits_{0}^{1}dt\int\limits_{0}^{t}ds\epsilon \theta
e\left( \tilde{R}+t^{2}\theta ^{2}+s^{2}e^{2}\right)^{n-1} .
\end{eqnarray*}
Es decir que la lagrangiana de transgresi\'on en este caso esta dada por 
\begin{equation}
L_{trans}=L_{LCS}+d\alpha
\end{equation}
Este resultado es particularmente notable si se tiene en cuenta que el
t\'ermino de borde que debe adicionarse a la lagrangiana $L_{LCS}$ para
obtener la lagrangiana de Chern-Simons con el tensor invariante dado no se
conoce en general para cualquier dimensi\'on.\newline

La configuraci\'{o}n de variedad cobordante permite otra elecci\'{o}n de
condiciones de borde particularmente conveniente que tambi\'{e}n hace que la
acci\'{o}n sea un extremo cuando valen las ecuaciones del movimiento. La f%
\'{o}rmula de la variaci\'{o}n de la transgresi\'{o}n para la configuraci%
\'{o}n de variedad cobordante da 
\[
\delta {\cal  T}_{2n+1}^0=d[\kappa n\int_{0}^{1}dtt\epsilon (\delta \theta
e-\theta \delta e)(\tilde{R}+t^{2}\theta ^{2}+t^{2}e^{2})^{n-1}]
\]
Lo que sugiere la condici\'{o}n de borde natural 
\[
\epsilon _{abca_{3}.....a_{2n+1}}\delta \theta ^{ab}e^{c}=\epsilon
_{abca_{3}.....a\theta _{2n+1}^{ab}}\delta e^{c}
\]
Puede verse \cite{motz2} que esto implica que la curvatura extr\'{i}nseca
del borde $K_{ij}$ satisface 
\[
\delta K_{ij}=0
\]
para las variaciones permitidas en esta condici\'{o}n de borde, y 
\[
K_{ij}=\Omega g_{ij}
\]
donde $\Omega $ es una constante y $g_{ij}$ es la m\'{e}trica del borde.
Esta \'{u}ltima ecuaci\'{o}n implica que la normal es un vector de Killig
conforme, ya que la curvatura extr\'{i}nseca est\'{a} dada por la derivada
de Lie de la m\'{e}trica del borde seg\'{u}n la normal 
\[
K_{ij}={\cal L}_{n}g_{ij}
\]

\subsection{Cargas Conservadas}

En esta subsecci\'on se estudiar\'an las corrientes y cargas conservadas
para teor\'{\i}as de Chern-Simons y Transgresi\'on, calculadas a partir del
Teorema de Noether \cite{motz1,motz2} . Empezaremos repasando el teorema de Noether, entonces
discutiremos las corrientes conservadas asociadas a las transformaciones de
gauge y a los difeomorfismos. Otros trabajos que se han ocupado del c\'alculo de las cargas 
conservadas en teor\'{\i}as de gauge y gravitaci\'on de Chern-Simons, que no tienen sin embargo 
una superposici\'on de contenido significativa con nuestro trabajo, son las referencias \cite{francaviglia1,francaviglia2,sarda}. Un trabajo previo en el caso de 2+1 dimensiones es ref.\cite{postdam}. El prolema relacionado de la elecci\'on de t\'erminos de borde apropiados y de las cargas conservadas en teor\'{\i}as de la gravitaci\'on de Einstein-Hilbert con constante cosmol\'ogica en dimensiones pares se trat\'o en ref.\cite{aroscargas}. \newline

\subsubsection{Teorema de Noether}

La variaci\'{o}n de formas diferenciales bajo difeomorfismos en que las
coordenadas cambian como $\delta x^{\mu }=\xi ^{\mu }$ esta dada por 
\[
\delta \alpha (x)=\alpha ^{\prime }(x)-\alpha (x)=-{\cal L}_{\xi }\alpha 
\]
donde ${\cal L}_{\xi }$ es la derivada de Lie, que para formas diferenciales
puede escribirse como 
\[
{\cal L}_{\xi }\alpha =[dI_{\xi }+I_{\xi }d]\alpha 
\]
con $d$ la derivada exterior y el operador de contracci\'{o}n dado por 
\[
I_{\xi }\alpha _{p}=\frac{1}{(p-1)!}\xi ^{\nu }\alpha _{\nu \mu _{1}...\mu
_{p-1}}dx^{\mu _{1}}...dx^{\mu _{p-1}} 
\]
El operador $I_{\xi }$ es una antiderivaci\'{o}n, en el sentido de que
actuando en el el producto exterior de dos formas diferenciales $\alpha _{p}$
y $\beta _{q}$ de ordenes p y q respectivamente da $I_{\xi }(\alpha
_{p}\beta _{q})=I_{\xi }\alpha _{p}\beta _{q}+(-1)^{p}\alpha _{p}I_{\xi
}\beta _{q}$ . Un resultado \'{u}til es que la derivada de Lie actuando
sobre potenciales de gauge es 
\[
{\cal L}_{\xi }A=D(I_{\xi }A)+I_{\xi }F 
\]
donde $D$ es la derivada covariante y F el tensor de campo.\newline
Se considera una densidad de lagrangiana dada por una forma diferencial $%
L(\phi ,\partial\phi )$, donde $\phi $ representa todos los campos
din\'{a}micos. La variaci\'{o}n de la lagrangiana bajo difeomorfismos esta
dada por $\delta L=-d(I_{\xi }L)$, ya que $dL=0$ porque el orden de $L$ es
igual a la dimensi\'{o}n del espacio. Se considera una clase de
transformaciones bajo las que la lagrangiana sea cuasi-invariante,
combinadas con difeomorfismos. Bajo estas la variaci\'{o}n de la lagrangiana
se asume de la forma 
\[
\delta L=d\Omega -d(I_{\xi }L) 
\]
donde la primera derivada total viene de las transformaciones consideradas y
la segunda de los difeomorfismos. Por otro lado el procedimiento usual que
lleva a las ecuaciones del movimiento (E.d.M.) de Euler-Lagrange da la
variaci\'{o}n de la lagrangiana como las ecuaciones del movimiento mas un
t\'{e}rmino de borde 
\[
\delta L=(E.d.M.)\delta \phi +d\Theta 
\]
donde las variaciones $\delta \phi $son infinitesimales pero arbitrarias en
su forma. A partir de estas expresiones de la variaci\'{o}n obtenemos,
asumiendo las variaciones en ambas restringidas a transformaciones de la
clase considerada en la primera expresi\'{o}n de $\delta L$ e igualando, que
si valen las E.d.M. 
\[
d[\Omega -I_{\xi }L-\Theta ]=0 
\]
Se sigue que la llamada 'corriente de Noether' 
\[
\star j=\Omega -I_{\xi }L-\Theta 
\]
Se puede ver que si se agrega un t\'ermino de borde a $L$, como $%
L^{\prime}=L+dB$, con B funci\'on de los campos y sus drivadas, entonces la
corriente conservada asociada a la invariancia bajo difeomorfismos cambia
como $\star j^{\prime}=\star j+I_{\xi}B$.\newline

En las siguientes dos subsecciones deduciremos la forma general de las
cargas de gauge y difeomorfismos para teor\'{\i}as de gauge con acciones de
transgresi\'on y Chern-Simons.

\subsubsection{Cargas de Gauge}

La variaci\'on de la transgresi\'on es 
\begin{equation}
\delta {\cal T}_{2n+1} =(n+1)<F_1^n\delta A_1>-(n+1)<F_0^n\delta
A_0>-n(n+1)~d~\int_0^1dt<JF_t^{n-1}\delta A_t>
\end{equation}
Bajo transformaciones de gauge 
\begin{equation}
\delta _ {\lambda } A _1 =-D _1 \lambda~~,~~\delta _ {\lambda } A _0=-D _0
\lambda
\end{equation}
de donde 
\begin{equation}
\delta _ {\lambda } A _t=-D _t \lambda =-d \lambda -A _t\lambda +\lambda A _t
\end{equation}
Las E.d.M., que asumiremos son satisfechas por ambos campos $A_1$ y $A_0$,
son $<F_1^nT^a>=0$ y $<F_0^nT^a>=0$, de donde se sigue que podemos leer la
forma $\Theta$ que aparece en el teorema de Noether de la expresi\'on de la
variaci\'on 
\begin{equation}
\Theta =n(n+1)\int_0^1dt<JF_t^{n-1}D_t \lambda>
\end{equation}
La forma $\Omega $ es cero en este caso, ya que la transgresi\'on es
invariante gauge. Se sigue que la corriente conservada es 
\begin{equation}
\ast j _{\lambda } =-\Theta = -n(n+1)\int_0^1dt<JF_t^{n-1}D_t \lambda >
\end{equation}
Adem\'as $\ast j _{\lambda } =d Q _{\lambda}$ con 
\begin{equation}
Q _{\lambda } = n(n+1)\int_0^1dt<JF_t^{n-1}\lambda >
\end{equation}
ya que 
\begin{equation}
dQ _{\lambda } = n(n+1)\int_0^1dt<D _t[JF_t^{n-1}\lambda ] >
\end{equation}
o 
\begin{equation}
dQ _{\lambda } = n(n+1)\int_0^1dt< \frac{d~}{dt}F_t F_t^{n-1}\lambda
-JF_t^{n-1}D _t\lambda ] >
\end{equation}
y, usando $\frac{d~}{dt} F_t^{n-1} =\frac{1}{n}\frac{d~}{dt} F_t^{n} $ 
\begin{equation}
dQ _{\lambda } = (n+1) < ( F _1^n- F _0^n)\lambda
>-n(n+1)\int_0^1dt<JF_t^{n-1}D _t\lambda  >
\end{equation}
donde el primer t\'ermino del segundo miembro es cero debido a las E. de M.. 
\newline

Esta expresi\'on de las cargas es v\'alida para Chern-Simons, poniendo $A_1=A
$ y $A_0=0$, ya que la configuraci\'on $A_0=0$ satisface las E.d.M..

\subsubsection{Cargas de Difeomorfismos}

La variaci\'on de la Transgresi\'on es 
\begin{equation}
\delta {\cal T}_{2n+1} =(n+1)<F_1^n\delta A_1>-(n+1)<F_0^n\delta
A_0>-n(n+1)~d~\int_0^1dt<JF_t^{n-1}\delta A_t>
\end{equation}
La variaci\'on de los potenciales bajo difeomorfismos es 
\begin{eqnarray}
\delta _{\xi}A _1 =-{\cal L } _{\xi} A _1= D _1 [ I _{\xi }A _1] -I _{\xi}F
_1= -[I _{\xi}d+d I _{\xi}]A _1 \\
\delta _{\xi}A _0 =-{\cal L } _{\xi} A _0= D _0 [ I _{\xi }A _0] -I _{\xi}F
_0= -[I _{\xi}d+d I _{\xi}]A _0 \\
\delta _{\xi}A _t=-{\cal L } _{\xi} A _t= D _t [ I _{\xi }A _t] -I _{\xi}F
_t= -[I _{\xi}d+d I _{\xi}]A _t
\end{eqnarray}
Podemos leer el $\Theta$ que aparece en el teorema de Noether de la
variaci\'on de la transgresi\'on 
\begin{equation}
\Theta =-n(n+1)\int_0^1dt<JF_t^{n-1}\delta _{\xi}A _t>
\end{equation}
o 
\begin{equation}
\Theta =n(n+1)\int_0^1dt<JF_t^{n-1} D _t [ I _{\xi }A _t] + JF_t^{n-1} I
_{\xi}F _t >
\end{equation}
pero 
\[
D_t[JF_t^{n-1}I _{\xi} A_t] = D_tJ F_t^{n-1}I _{\xi} A_t -J F_t^{n-1} D_t[I
_{\xi} A_t]= \frac{d~}{dt}F_t F_t^{n-1} I _{\xi} A_t-J F_t^{n-1} D_t[I
_{\xi} A_t]
\]
entonces 
\begin{equation}
\Theta =n(n+1)\int_0^1dt<\frac{d~}{dt}F_t F_t^{n-1} I _{\xi }A _t +
JF_t^{n-1} I _{\xi}F _t > -n(n+1)~d~\int_0^1dt<JF_t^{n-1} I _{\xi }A _t>
\end{equation}
Para el t\'ermino $I _{\xi}L$ en la corriente de Noether tenemos 
\begin{equation}
I _{\xi}L = I _{\xi}{\cal T}_{2n+1} =(n+1)\int_0^1dt< I _{\xi}J F_t^n -nJ F_t^{n
-1}I _{\xi } F _t>
\end{equation}
La corriente es $\ast j=\Omega-[\Theta +I _{\xi }L] $, pero $\Omega =0$
debido a la invariancia de la acci\'on bajo difeomorfismos, entonces 
\[
\ast j=-[\Theta +I _{\xi }L] =-(n+1)\int_0^1dt<n \frac{d~}{dt}F_t F_t^{n-1}
I _{\xi }A _t + I _{\xi} JF_t^{n} > 
\]
\[
+n(n+1)~d~\int_0^1dt <JF_t^{n-1} I _{\xi }A _t>
\]
pero $I_{\xi}A_t=t I _{\xi}J+I _{\xi} A _0$, entonces $I _{\xi}J =\frac{d~}{%
dt} I_{\xi}A_t $ y por lo tanto 
\[
<n \frac{d~}{dt}F_t F_t^{n-1} I _{\xi }A _t + I _{\xi} JF_t^{n} > = \frac{d~%
}{dt} < F_t^{n} I _{\xi }A _t > 
\]
lo que permite integrar los primeros t\'erminos de la corriente dando 
\begin{equation}
\ast j = < F_1^{n} I _{\xi }A _1 > - < F_0^{n} I _{\xi }A _0 >
+n(n+1)~d~\int_0^1dt <JF_t^{n-1} I _{\xi }A _t>
\end{equation}
Los primeros dos t\'erminos del segundo miembro son cero debido a las E. de
M., entonces 
\begin{equation}
\ast j = ~d Q _{\xi}
\end{equation}
con 
\begin{equation}
Q _{\xi} = +n(n+1)\int_0^1dt <JF_t^{n-1} I _{\xi }A _t>
\end{equation}
Como en el caso de las cargas de gauge, esta expresi\'on es v\'alida para
Chern-Simons, poniendo $A_1=A$ y $A_0=0$, ya que la configuraci\'on $A_0=0$
satisface las E.d.M.. Otra forma de plantear estas cargas es como 
\begin{equation}
Q^{\prime}_{\xi}=Q_{\xi}+I_{\xi}B
\end{equation}
Entonces de las ecuaciones (22) y (23) resulta 
\begin{equation}
Q_{\xi}^{trans}(0,1) =Q_{\xi}^{CS}(1)-Q_{\xi}^{CS}(0)-I_{\xi}C_{2n}(0,1)
\end{equation}
con 
\begin{equation}
C_{2n}=-(n+1)n\int_0^1ds\int_0^1dt<tA_sJ(F_s)_t^{n-1}>
\end{equation}
donde $(F_s)_t=tF_s+(t^2-t)A_s^2$, $A_s=sA_1+(1-s)A_0$ and $F_s=dA_s+A_s^2$,
y de ah\'{\i} 
\begin{eqnarray}
I_{\xi}C_{2n}(0,1)=-n(n+1)\int_0^1ds\int_0^1dt<t I_{\xi}A_sJ(F_s)_t^{n-1}+t
A_s I_{\xi}J(F_s)_t^{n-1}+ \\
(n-1)t A_s (F_s)_t^{n-2}I_{\xi} (F_s )_t>  \nonumber
\end{eqnarray}

\subsection{C\'alculo de las Cargas en Casos Concretos para Chern-Simons y
Transgresiones}

En esta subsecci\'on calcularemos las cargas conservadas para agujeros
negros en gravedades de Chern-Simons y transgresiones (ver refs. \cite{dimensionally,scan} sobre estas soluciones), como se hizo en refs. \cite{motz1,motz2}. En el caso de
transgresiones haremos el c\'alculo en dimensi\'on arbitraria, usando como
referencia tanto otro agujero negro, en particular el de $M=0$ (masa cero) y
el de $M=-1$ (AdS), y la configuraci\'on de variedad cobordante (VC).

Veremos que el resultado de la masa es el que se esperaba dado el resultado
hamiltoniano para transgresiones, pero no para Chern-Simons.

\subsubsection{Chern-Simons en d=2+1}

Se considera la soluci\'on de las E.d.M. de gravedad de Chern-Simons para un
agujero negro en rotaci\'on en d=2+1, con vielbein \cite{dimensionally} 
\begin{equation}
e^0=\Delta dt~~~,~~~e^1=\frac{1}{\Delta}dr~~~,~~~e^2=rd\phi -\frac{J}{2r}dt
\end{equation}
con 
\[
\Delta = \sqrt{r^2-2G_3M+\frac{2G_3 J}{4r^2}} 
\]
y conexi\'on de esp\'{\i}n 
\begin{equation}
\omega ^{01}=r dt -\frac{J}{2r}d\phi ~~~,~~~\omega ^{02}=-\frac{J}{%
2r^2\Delta }dr ~~~,~~~\omega ^{12}=-\Delta d\phi
\end{equation}
y la soluci\'on de anti de Sitter (AdS) con vielbein 
\begin{equation}
\overline{e}^0=\overline{\Delta } dt~~~,~~~\overline{e}^1=\frac{1}{\overline{%
\Delta }}dr~~~,~~~\overline{e}^2=rd\phi
\end{equation}
con 
\[
\overline{\Delta} = \sqrt{r^2+2G_3} 
\]
y conexi\'on de esp\'{\i}n 
\begin{equation}
\overline{\omega }^{01}=r dt ~~~,~~~\overline{\omega } ^{02}=0 ~~~,~~~%
\overline{\omega } ^{12}=-\overline{\Delta } d\phi
\end{equation}

{\bf A Gauge}\newline

En d=2+1 la corriente de gauge es 
\begin{equation}
*j_{\lambda}^{CS}= -2d<A\lambda>
\end{equation}

O, para gravedad CS, si $\lambda =\frac{1}{2}\lambda ^{ab}J_{ab}+\lambda
^aP_a $ 
\begin{equation}
*j_{\lambda}^{CS}=\kappa d[\epsilon _{abc}\omega ^{ab}\lambda ^c +\epsilon
_{abc}e^a\lambda ^{bc}]
\end{equation}
con la traza sim\'etrica mencionada 
\begin{equation}
<J_{A_1A_2}...J_{A_{d-1}A_d}> = \kappa \frac{2^n}{(n+1)} \epsilon
_{A_1....A_d}
\end{equation}
en el caso $n=1$. La constante es $\kappa =[2(d-2)!\Omega_{d-2}G_d]^{-1}$
con $\Omega_{d-2}$ el volumen de la la esfera en $d-2$ dimensiones y $G_d$
la 'constante de Newton' en dimensi\'on $d$. En este caso $\kappa =[4\pi
G_2]^{-1}$

La carga $\int _{\Sigma}*j_{\lambda}^{CS}$, donde $\Sigma$ en uns secci\'on
espacial es $0$ o $\infty$ a menos que se elija un $\lambda$ que cumpla
determinadas condiciones. Esto parece corresponder a la elecci\'on de los
llamados 'par\'ametros de reducibilidad', discutidos por Barnich et al. 
\cite{barnich}. En el caso de los difeomorfismos lo que se requiere es que $\xi
^{\mu}$ sea un vector de Killing asint\'oticamente (lo cual desde luego es
una parte bien conocida del folcklore de la Relatividad General). Para
transformaciones de gauge lo que se requiere es que $\lambda$ sea un
par\'ametro covariantemente constante asint\'oticamente, esto es $D\lambda =0
$ asint\'oticamente(para transgresiones debe ser $D_1\lambda =D_0\lambda =0$
asint\'oticamente), lo que implica que $\delta _{\lambda}A=0$ en el borde
espacial, condici\'on an\'aloga a la condici\'on de ser vector de Killing
para vectores quegeneran difeomorfismos. La condici\'on $D\lambda =0$
implica, tanto para el agujero negro como AdS, 
\begin{equation}
\lambda ^0=\lambda ^{01}=c_1r~~,~~\lambda ^1=\lambda ^{02}=0~~,~~ \lambda
^2=-\lambda ^{12}=c_2r
\end{equation}
donde $c_1$ y $c_2$ son constantes.

Las cargas conservadas $\int _{\Sigma}*j_{\lambda}^{CS}$ en $d=2+1$ dan:%
\newline
(i) $M$ para el par\'ametro de gauge que corresponde a $c_1=1$ y $c_2=0$.%
\newline
(ii) $J$ para el par\'ametro de gauge que corresponde a $c_1=0$ y $c_2=1$,%
\newline
si la integral, que se reduce a una integral en el borde espacial, se toma
en el circulo de radio infinito.\newline

{\bf B Difeomorfismos}\newline

En d=2+1 
\begin{equation}
*j_{\xi}^{CS}= d<AI_{\xi}A>
\end{equation}
Las cargas conservadas $\int _{\Sigma}*j_{\xi}^{CS}$ en $d=2+1$ dan:\newline
(iii) $M$ para el vector de Killing $\xi =\frac{\partial ~}{\partial t}$.%
\newline
(iv) $-J$ para el vector de Killing $\xi =\frac{\partial ~}{\partial \phi}$,%
\newline
no importa que radio se tome la integraci\'on (al contrario de lo que sucede
para las cargas de gauge)

\subsubsection{Momento Angular para Transgresi\'on en 2+1}

Mas adelante se evaluar\'a la masa de gauge y difeomorfismos en cualquier
dimensi\'on para transgresiones, pero la \'unica soluci\'on que se conoce
con momento angular es en d=2+1, por lo que daremos el resultado del
c\'alculo del momento angular tomando $A_1$ y $A_0$ como agujeros negros con
la misma masa y momento angular $J$ y $\overline{J}$ respectivamente. Las
corrientes de gauge y difeomorfismos de las acciones de transgresi\'on son,
en d=2+1 
\begin{equation}
*j_{\lambda}^{trans}= -2d<(A_1-A_0)\lambda>
\end{equation}
para la de gauge y 
\begin{equation}
*j_{\xi}^{trans}= d<(A_1-A_0)(I_{\xi}A_1+I_{\xi}A_0)>
\end{equation}
para la de difeomorfismos.

El resultado para la carga de difeomorfismos $\int _{\Sigma}*j_{\xi}^{trans}$
en $d=2+1$ es $-(J-\overline{J})$ para el vector de Killing $\xi =\frac{%
\partial ~}{\partial \phi}$,\newline
no importa que radio se tome la integraci\'on. El resultado para la carga de
gauge $\int _{\Sigma}*j_{\lambda}^{trans}$ en $d=2+1$ es el mismo, pero la
integraci\'on debe tomarse para radio infinito de la secci\'on espacial. Es
importante notar que el hecho de que el resultado involucre las diferencias
de los valores de $J$ para cada configuraci\'on no es obvio que debiera
cumplirse a partir de la expresi\'on original de la corriente conservada.%
\newline

\subsubsection{Masa de agujeros negros en cualquier dimensi\'on con
variedades cobordantes}

La carga de Noether asociada a la invariancia bajo difeomorfismos es para $%
d=2n+1$ 
\begin{equation}
Q_{\xi}=n(n+1)\int_0^1<\Delta AF_t^{n-1}I_{\xi}A_t>
\end{equation}
donde $\Delta A=A-\overline{A}$.\newline
Calcularemos esta carga tomando $A$ como una soluci\'on de las E.d.M.
correspondiente a un agujero negro para gravedad de AdS en dimensi\'on
arbitraria $d=2n+1$ (ver \cite{dimensionally,scan}) y $\overline{A}$ como la configuraci\'on correspondiente
a una variedad cobordante 
\begin{equation}
A=\frac{1}{2}\omega ^{ab}J_{ab}+e^aP_a~~,~~ \overline{A}=\frac{1}{2}%
\overline{\omega} ^{ab}J_{ab}+\overline{e}^aP_a
\end{equation}
Donde 
\begin{eqnarray}
e^0=\Delta dt~~,~~e^1=\frac{1}{\Delta}dr~~,~~e^m=r\tilde{e}^m \\
\omega ^{01}=rdt~~,~~\omega ^{1m}=-\Delta\tilde{e}^m~~,~~\omega
^{0m}=0~~,~~\omega^{mn}
\end{eqnarray}
donde las coordenadas 0 y 1 corresponden a las direcciones temporal y
espacial radial y $\tilde{e}^m$ y $\omega ^{mn}$ son el velbein y la
conexi\'on de esp\'{\i}n de la esfera $S^{d-1}$, correspondiente a las
variables angulares. Tenemos 
\begin{equation}
\Delta =\sqrt{r^2-(2G_kM+1)^{\frac{1}{n}}+1}=\sqrt{r^2-\alpha +1}
\end{equation}
donde $\alpha = (2G_kM+1)^{\frac{1}{n}}$. Para $\overline{A}$ tenemos 
\begin{equation}
\overline{e}^a=0~~~,~~~\overline{\omega}^{1\underline{i}}=0~~~,~~~ \overline{%
\omega}^{\underline{i}\underline{j}} = \omega ^{\underline{i}\underline{j}}
\end{equation}
donde los \'{\i}ndices subrayados como $\underline{i}$ pueden tomar
cualquier valor posible diferente de 1.\newline
Si 
\begin{equation}
\Delta A=A-\overline{A}\equiv \frac{1}{2}\Theta ^{ab}J_{ab}+E^aP_a
\end{equation}
con $\Theta ^{ab} \equiv \omega ^{ab} -\overline{\omega}^{ab} $ y $E^a\equiv
e^a-\overline{e}^a$. Entonces 
\begin{equation}
E^a=e^a~~~,~~~\Theta ^{1\underline{i}} =\omega ^{1\underline{i}%
}~~~,~~~\Theta ^{\underline{i}\underline{j}} =0
\end{equation}
y para $\Delta A$ tenemos 
\begin{equation}
\Delta A= \frac{1}{2} \Theta ^{ab}J_{ab}+e^aP_a =\Theta ^{1\underline{i}}J_{1%
\underline{i}}+e^aP_a
\end{equation}
Tambi\'en 
\begin{equation}
A_t=t\Delta A+\overline{A}=\frac{1}{2}[t\Theta +\overline{\omega}]J+[tE+%
\overline{e}]P
\end{equation}
entonces para el vector de Killig temporal $\xi =\frac{\partial ~}{\partial t%
}$ obtenemos 
\begin{equation}
I_{\xi}A_t= t e^0_tP_0+t\Theta ^{01}_tJ_{01}
\end{equation}
donde se us\'o que $E=e$, $\overline{e}=0$ y $I_{\xi}\overline{\omega}=0$.

Los tensores de campo son 
\begin{equation}
F=\frac{1}{2}\overline{R}^{ab}J_{ab}+T^aP_a ~~,~~ \overline{F}=\frac{1}{2}%
\overline{\tilde{R}}^{ab}J_{ab}+\overline{T}^aP_a
\end{equation}
donde $\overline{R}^{ab}=R^{ab}+e^ae^b $ o en una notaci\'on mas simple 
$\overline{R}=R+e^2$ y $\overline{\tilde{R}}^{ab}=\tilde{R}^{ab}+\overline{e}^a%
\overline{e}^b$, con $R^{ab}=d\omega ^{ab} + \omega ^a_{~c} \omega ^{cb} $ y 
$\tilde{R}^{ab}=d\overline{\omega}^{ab} + \overline{\omega}^a_{~c} \overline{%
\omega}^{cb} $ . Para las soluciones de agujero negro 
\begin{equation}
T^a=0~~,~~ \overline{R}^{0a}=0~~ ,~~ \overline{R}^{1m}=0~~,~~\overline{R}%
^{mn}=\alpha \tilde{e}^m\tilde{e}^n
\end{equation}
y para la configuraci\'on de variedad cobordante 
\begin{equation}
\overline{T}^a=0~~,~~ \overline{\tilde{R}}^{1\underline{i}} = \tilde{R}^{1%
\underline{i}} = 0~~ ,~~ R ^{\underline{i}\underline{j}} =\tilde{R} ^{%
\underline{i}\underline{j}} +(\Theta ^2) ^{\underline{i}\underline{j}}
\end{equation}
de donde 
\begin{equation}
\overline{\tilde{R}} ^{\underline{i}\underline{j}} =\overline{R} ^{%
\underline{i}\underline{j}} -[ (\Theta ^2) ^{\underline{i}\underline{j}}+(e
^2) ^{\underline{i}\underline{j}}]
\end{equation}
La configuraci\'on $\overline{A}$ de variedad cobordante con el agujero
negro tambi\'en satisface las ecuaciones del movimiento, como el propio
agujero negro.\newline
Necesitaremos 
\begin{equation}
F_t=dA_t+A_t^2=\overline{F}+t\overline{D}\Delta A+t^2 \Delta A^2
\end{equation}
con $\overline{D}\Delta A =d\Delta A +\overline{A}\Delta A +\Delta A 
\overline{A}$, entonces 
\begin{equation}
F_t=\frac{1}{2}[\overline{\tilde{R}}+t(\overline{D}\Theta +\overline{e}E+E%
\overline{e})+t^2(\Theta ^2+E^2)]J+ [\overline{T}+t(\overline{D}E +((\Theta 
\overline{e})))+t^2((\Theta E))]P 
\end{equation}
donde los parentesis dobles indican contracciones, y $\overline{D}$ es la
derivada covariante con $\overline{\omega}$, por ejemplo $[\overline{D}%
\Theta ]^{ab} = d\Theta ^{ab}+ \overline{w}^a_{~c}\Theta ^{cb} +\overline{w}%
^b_{~c}\Theta ^{ac}$. Por definici\'on $F_t=\equiv \frac{1}{2}\overline{R}_t J+T_tP$.\\

Para las dos configuraciones consideradas 
\begin{equation}
F_t=\frac{1}{2}[\tilde{R}+t\overline{D}\Theta +t^2(\Theta ^2+e^2)]J+ [t%
\overline{D}e+t^2((\Theta e))]P
\end{equation}

Juntando todo, con el vector de Killing $\xi =\frac{\partial ~}{\partial t}$
y la traza sim\'etrica que lleva a la gravedad de Chern-Simons usual para el
grupo AdS 
\begin{equation}
<J_{a_1 a_2}J_{a_3 a_4}...J_{a_{2n-1} a_{2n}}P_{a_{2n+1}}>=\kappa \frac{2^n}{%
(n+1)} \epsilon_{ a_1 a_2 a_3...a_{2n-1} a_{2n} a_{2n+1}}
\end{equation}
donde $\kappa =[2(d-2)!\Omega_{d-2}G_k]^{-1}$ con $\Omega_{d-2}$ el volumen
de la la esfera en $d-2$ dimensiones y $G_d$ la 'constante de Newton' en
dimensi\'on $d$.\newline
De la forma de $\Delta A$ y $I_{\xi}A_t$ vemos que el \'{\i}ndice 1 debe
estar en $\Delta A$ or $I_{\xi}A_t$, y por lo tanto tambi\'en el generador $P
$. Por lo tanto los \'{\i}ndices e $F_t$ deben ser angulares $mn$.

Necesitamos 
\begin{eqnarray}
[\overline{D}\Theta ]^{mn}=0 \\
(\Theta ^{2}+e^2)^{mn}=(\alpha -1)\tilde{e}^m\tilde{e}^n \\
\tilde{R}^{mn}=\tilde{e}^m\tilde{e}^n
\end{eqnarray}

Reuniendo todo obtenemos 
\begin{equation}
Q(\frac{\partial ~}{\partial t})=\kappa n\int_0^1
dt~t\epsilon_{01m_1...m_{2n-2}} (2\Theta _{t} ^{01} e^{m _1} +2e ^0 _t
\Theta ^{1 m _1}) [1 + t^2 (\alpha -1)]^{n-1} \tilde{e}_{m_2}...\tilde{e}%
_{m_{2n-1}}
\end{equation}
pero $\Theta _{t} ^{01} =r$, $e^{m _1} =r\tilde{e}^{m _1}$, $e ^0 _t=\Delta $
y $\Theta ^{1 m _1}=-\Delta \tilde{e}^{m _1}$, entonces 
\begin{equation}
Q(\frac{\partial ~}{\partial t})=\kappa n\int_0^1
dt~\epsilon_{01m_1...m_{2n-2}} 2t(\alpha -1) [1 + t^2 (\alpha -1)]^{n-1} 
\tilde{e}_{m_1}...\tilde{e}_{m_{2n-1}}
\end{equation}
esta expresi\'on puede integrarse en $t$ tomando 
\[
u= [1 + t^2 (\alpha -1)]
\]
y el resultado es 
\begin{equation}
Q(\frac{\partial ~}{\partial t})= \kappa \epsilon_{01m_1...m_{2n-2}} (\alpha
^n-1) \tilde{e}_{m_1}...\tilde{e}_{m_{2n-1}}
\end{equation}
Integrado en la esfera $S^{d-2}$ esto da 
\begin{equation}
\int_{S^{d-2}}Q(\frac{\partial ~}{\partial t})= \kappa
(d-2)!\Omega_{d-2}(\alpha ^n- 1 )
\end{equation}
donde usamos $\int_{S^{d-2}}\epsilon_{01m_1...m_{2n-2}} \tilde{e}_{m_1}...%
\tilde{e}_{m_{2n-1}} =(d-2)!\Omega_{d-2}$. Finalmente 
\begin{equation}
\int_{S^{d-2}}Q(\frac{\partial ~}{\partial t})= M
\end{equation}

\subsubsection{Masa de agujeros negros en cualquier dimensi\'on respecto a
otro agujero negro}

La carga de Noether asociada a los difeomorfismos es para $d=2n+1$ 
\begin{equation}
Q_{\xi}=n(n+1)\int_0^1<\Delta AF_t^{n-1}I_{\xi}A_t>
\end{equation}
donde $\Delta A=A-\overline{A}$.\newline
Calcularemos esta carga tomando tanto $A$ como $\overline{A}$ como
soluciones de las E.d.M. de tipo agujero negro para gravedad de AdS en
dimensi\'on arbitraria $d=2n+1$ \cite{dimensionally,scan}. Si 
\begin{equation}
A=\frac{1}{2}\omega ^{ab}J_{ab}+e^aP_a~~,~~ \overline{A}=\frac{1}{2}%
\overline{\omega} ^{ab}J_{ab}+\overline{e}^aP_a
\end{equation}
tenemos 
\begin{eqnarray}
e^0=\Delta dt~~,~~e^1=\frac{1}{\Delta}dr~~,~~e^m=r\tilde{e}^m \\
\omega ^{01}=rdt~~,~~\omega ^{1m}=-\Delta\tilde{e}^m~~,~~\omega
^{0m}=0~~,~~\omega^{mn}
\end{eqnarray}
donde las coordenadas 0 y 1 corresponden a las direcciones temporal y
espacial y $\tilde{e}^m$ y $\omega ^{mn}$ son el vielbein y la conexi\'on de
esp\'{\i}n de la esfera $S^{d-2}$ correspondiente a las variables angulares.
Tenemos 
\begin{equation}
\Delta =\sqrt{r^2-(2G_kM+1)^{\frac{1}{n}}+1}=\sqrt{r^2-\alpha +1}
\end{equation}
donde $\alpha = (2G_kM+1)^{\frac{1}{n}}$. Para $\overline{A}$ tenemos
expresiones similares con 
\begin{equation}
\overline{\Delta} =\sqrt{r^2-(2G_k\overline{M}+1)^{\frac{1}{n}}+1}=\sqrt{r^2-%
\overline{\alpha }+1}
\end{equation}
Los tensores de campo son 
\begin{equation}
F=\frac{1}{2}\overline{R}^{ab}J_{ab}+T^aP_a ~~,~~ \overline{F}=\frac{1}{2}%
\overline{\tilde{R}}^{ab}J_{ab}+\overline{T}^aP_a
\end{equation}
donde $\overline{R}^{ab}=R^{ab}+e^ae^b $ o en una notaci\'on mas simple $%
\overline{R}=R+e^2$ y $\overline{\tilde{R}}^{ab}=\tilde{R}^{ab}+\overline{e}%
^a\overline{e}^b$, con $R^{ab}=d\omega ^{ab} + \omega ^a_{~c} \omega ^{cb} $
y $\tilde{R}^{ab}=d\overline{\omega}^{ab} + \overline{\omega}^a_{~c} 
\overline{\omega}^{cb} $ . Para las soluciones de agujero negro 
\begin{equation}
T^a=0~~,~~ \overline{R}^{0a}=0~~ ,~~ \overline{R}^{1m}=0~~,~~\overline{R}%
^{mn}=\alpha \tilde{e}^m\tilde{e}^n
\end{equation}
y 
\begin{equation}
\overline{T}^a=0~~,~~ \overline{\tilde{R}}^{0a}=0~~ ,~~ \overline{\tilde{R}}%
^{1m}=0~~,~~\overline{\tilde{R}}^{mn}=\overline{\alpha} \tilde{e}^m\tilde{e}%
^n
\end{equation}
Tenemos 
\begin{equation}
\Delta A=A-\overline{A}\equiv \frac{1}{2}\Theta ^{ab}J_{ab}+E^aP_a
\end{equation}
con $\Theta ^{ab} \equiv \omega ^{ab} -\overline{\omega}^{ab} $ y $E^a\equiv
e^a-\overline{e}^a$. Tambi\'en 
\begin{equation}
A_t=t\Delta A+\overline{A}=\frac{1}{2}[t\Theta +\overline{\omega}]J+[tE+%
\overline{e}]P
\end{equation}
y 
\begin{equation}
F_t=dA_t+A_t^2=\overline{F}+t\overline{D}\Delta A+t^2 \Delta A^2
\end{equation}
con $\overline{D}\Delta A =d\Delta A +\overline{A}\Delta A +\Delta A 
\overline{A}$ entonces 
\begin{equation}
F_t=\frac{1}{2}[\overline{\tilde{R}}+t(\overline{D}\Theta +\overline{e}E+E%
\overline{e})+t^2(\Theta ^2+E^2)]J+ [\overline{T}+t(\overline{D}E +((\Theta 
\overline{e})))+t^2((\Theta E))]P \equiv \frac{1}{2}\overline{R}_t J+T_tP
\end{equation}
donde el parentesis doble indica contracciones, y $\overline{D}$ es la
derivada covariante con $\overline{\omega}$, por ejemplo $[\overline{D}%
\Theta ]^{ab} = d\Theta ^{ab}+ \overline{w}^a_{~c}\Theta ^{cb} +\overline{w}%
^b_{~c}\Theta ^{ac}$.\newline
Para las soluciones de agujero negro consideradas las componentes de $E$ son 
\begin{equation}
E^0=(\Delta -\overline{\Delta})dt~~,~~E^1=\left( \frac{1}{\Delta}- \frac{1}{%
\overline{\Delta}}\right) dr~~,~~E^m=0
\end{equation}
mientras las componentes de $\Theta$ son 
\begin{equation}
\Theta ^{01}=0~~, ~~\Theta ^{0m}=0~~,~~\Theta ^{mn}=0~~, ~~\Theta ^{1m}=(%
\overline{\Delta}-\Delta)\tilde{e}^m
\end{equation}
Entonces, en este caso, 
\begin{equation}
\Delta A=\Theta ^{1m}J_{1m}+E^aP_a
\end{equation}
donde $E^aP_a$ es seg\'un $dr$ y $dt$ solamente. Adem\'as 
\begin{equation}
I_{\xi}A_t=[t(\Delta -\overline{\Delta})+\overline{\Delta}]P_0+rJ_{01}
\end{equation}
para el vector de Killing $\xi =\frac{\partial ~}{\partial t}$. Tomamos de
nuevo la traza sim\'etrica 
\begin{equation}
<J_{a_1 a_2}J_{a_3 a_4}...J_{a_{2n-1} a_{2n}}P_{a_{2n+1}}>=\kappa \frac{2^n}{%
(n+1)} \epsilon_{ a_1 a_2 a_3...a_{2n-1} a_{2n} a_{2n+1}}
\end{equation}
donde $\kappa =[2(d-2)!\Omega_{d-2}G_k]^{-1}$ con $\Omega_{d-2}$
correspondiendo al volumen de la esfera en $d-2$ dimensiones y $G_k$ la
'constante de Newton' en dimensi\'on $d$.\newline
Para calcular $Q_{\xi}$, para $\xi =\frac{\partial ~}{\partial t}$,
descartamos t\'erminos seg\'un $dr$ o $dt$ porque la integral se tomar\'a a
tiempo fijo sobre las variables angulares. Esto implica que el \'{\i}ndice 1
debe estar en la parte de $\Delta A$. Entonces solo el generador $P_0$
contribuir\'a de $I_{\xi}A_t$, y de ah\'{\i} solo t\'erminos seg\'un $J_{mn}$
contribuir\'an de $F_t^{n-1}$. Necesitamos 
\begin{eqnarray}
[\overline{D}\Theta ]^{mn}=2 \overline{\Delta} (\overline{\Delta}-\Delta)%
\tilde{e}^m\tilde{e}^n \\
E^m\overline{e}^n+\overline{e}^mE^n=0 \\
(\Theta ^{2}+E^2)^{mn}=-(\overline{\Delta}-\Delta)^2\tilde{e}^m\tilde{e}^n \\
\overline{\tilde{R}}^{mn}=\overline{\alpha}\tilde{e}^m\tilde{e}^n
\end{eqnarray}
Juntando todo 
\[
Q(\frac{\partial ~}{\partial t})=\kappa n\int_0^1
dt~2\epsilon_{01m_1...m_{2n-2}} (\overline{\Delta }-\Delta ) [t(\Delta -%
\overline{\Delta}) +\overline{\Delta}]
\]
\[
\left(\overline{\alpha}+2t\overline{\Delta} (\overline{\Delta}-\Delta)-t^2(%
\overline{\Delta}-\Delta)^2\right) ^{n-1}\tilde{e}_{m_1}...\tilde{e}%
_{m_{2n-1}}
\]
Esto puede integrarse en $t$ tomando 
\[
u=\left(\overline{\alpha}+2t\overline{\Delta} (\overline{\Delta}-\Delta)-t^2(%
\overline{\Delta}-\Delta)^2\right)
\]
y el resultado es 
\begin{equation}
Q(\frac{\partial ~}{\partial t})= \kappa \epsilon_{01m_1...m_{2n-2}} (\alpha
^n-\overline{\alpha}^n )\tilde{e}_{m_1}...\tilde{e}_{m_{2n-1}}
\end{equation}
Integrado en la esfera $S^{d-2}$ da 
\begin{equation}
\int_{S^{d-2}}Q(\frac{\partial ~}{\partial t})= \kappa
(d-2)!\Omega_{d-2}(\alpha ^n-\overline{\alpha}^n )
\end{equation}
donde usamos $\int_{S^{d-2}}\epsilon_{01m_1...m_{2n-2}} \tilde{e}_{m_1}...%
\tilde{e}_{m_{2n-1}} =(d-2)!\Omega_{d-2}$. Entonces 
\begin{equation}
\int_{S^{d-2}}Q(\frac{\partial ~}{\partial t})= M-\overline{M}
\end{equation}
Este es el resultado que se esperar\'{\i}a, pero es no trivial el modo en
que se obtiene. En particular para una teor\'{\i}a de CS pura no se obtiene $%
M$ como resultado para la masa, como vimos en el caso de 5D.

\subsubsection{ Masa de gauge en cualquier dimension para transgresiones}

La carga de Noether asociada a transformaciones de gauge es para $d=2n+1$ 
\begin{equation}
Q_{\lambda }=n(n+1)\int_0^1<\Delta AF_t^{n-1}\lambda >
\end{equation}
donde $\Delta A=A-\overline{A}$.\newline
Calcularemos esta carga tomando tanto $A$ como $\overline{A}$ como
soluciones de las E.d.M. de tipo agujero negro para AdS en dimensi\'on
arbitraria $d=2n+1$ \cite{dimensionally,scan} . No consideraremos un par\'ametro de gauge arbitrario $%
\lambda$, sino uno que satisfazga la condici\'on de ser covariantemente
constante 
\begin{equation}
D\lambda = d\lambda +A\lambda -\lambda A=0
\end{equation}
v\'alida asintoticamente. Por ejemplo para soluciones de tipo agujero negro
requeriremos que la condici\'on valga para $r\rightarrow \infty$. Esta
condici\'on implica que la variaci\'on de gauge del potencial es cero
asint\'oticamente, $\delta _{\lambda}A=0$ para $r\rightarrow \infty$. Para
soluciones de tipo agujero negro hay $d-1$ soluciones independientes a esta
condici\'on, marcadas por $d-1$ constantes arbitrarias $C^1$, $C^m$, con $%
m=2,...,d$. La condici\'on de covariancia constante da en ese caso 
\begin{eqnarray}
\lambda ^1=\lambda ^{0m}=0 \\
\lambda ^0=\lambda ^{01}=C^1 r \\
\lambda ^m=-\lambda ^{1m}=C^m r \\
\lambda ^m_n\tilde{e}^n=\omega ^{m}_n C^n
\end{eqnarray}
Elegimos el par\'ametro de gauge temporal, generador de 'boosts' de gauge 
\begin{equation}
\lambda_{(1)}=r P_0+r J_{01}
\end{equation}
que corresponde a $C^1=1$ y $C^m=0$.

Notese que la f\'ormula para las cargas de gauge es id\'entica a la
f\'ormula para cargas de difeomorfismos, con $\lambda$ en vez de $I_{\xi}A_t$.
Pero 
\begin{equation}
I_{\xi}A_t=[t(\Delta -\overline{\Delta})+\overline{\Delta}]P_0+rJ_{01}
\end{equation}
para el vector de Killing temporal $\xi =\frac{\partial ~}{\partial t}$.
Para $r\rightarrow \infty $, 
\[
\Delta -\overline{\Delta}\approx \frac{\alpha -\overline{\alpha}}{2r}={\cal O%
}(\frac{1}{r})\rightarrow 0
\]
y 
\[
\overline{\Delta}\approx r+{\cal O}(\frac{1}{r})\rightarrow r
\]
entonces 
\[
I_{\xi}A_t\rightarrow \lambda _{(1)}
\]
. Se sigue que 
\begin{equation}
\int_{S^{d-2}}Q_{\lambda}= M-\overline{M}
\end{equation}
si la integral se calcula sobre una esfera de radio infinito. Esto contrasta
con lo que pasa en el caso de la masa de difeomorfismo, donde la integral
puede calcularse en cualquier hipersuperficie espacial que rodee el origen.

\subsubsection{C\'alculo de la masa del agujero negro en d=4+1 para
Chern-Simons}

El c\'alculo de la masa de gauge o de difeomorfismos para la gravedad de
Chern-Simons pura en 5D muestra el problema antes mencionado de que el valor
obtenido no es el par\'ametro M en la soluci\'on de agujero negro, el cual
se sabe por m\'etodos hamiltonianos que corresponde a la masa f\'{\i}sica.

{\bf A Gauge}\newline

La corriente de gauge es 
\begin{equation}
*j_{\lambda}^{CS}= d<[-3AF+A_1^3]\lambda >
\end{equation}
la cual al ser evaluada para el pr\'ametro de gauge covariantemente
constante considerado arriba $\lambda _{(1)}$ que genera los 'boosts' de
gauge e integrada de una 'masa' 
\begin{equation}
\int _{\Sigma}*j_{\xi}^{CS}=\frac{2}{6G_5}(\alpha -1)^2
\end{equation}
con la constante $\alpha$ definida arriba y $G_5$ la constante de Newton en
5D. Esta expresi\'on claramente no da $M$\newline

{\bf B Difeomorfismos}\newline

\begin{equation}
*j_{\xi}^{trans}=<[\frac{3}{2}AF+\frac{1}{2}AdA]I_{\xi}A>
\end{equation}
Esta corriente, evaluada para el vector de Killing temporal $\xi =\frac{%
\partial ~}{\partial t}$ e integrada da una 'masa' 
\begin{equation}
\int _{\Sigma}*j_{\xi}^{CS}=\frac{1}{2G_5}(\alpha +\frac{2}{3})(\alpha -1)
\end{equation}
Este valor es nuevamente distinto de $M$.

\newpage

\section{Termodin\'amica de Agujeros Negros}

En esta secci\'on se discutir\'a la termodin\'amica de los agujeros negros
de Chern-Simons. Se ver\'a que estos tienen asociada una temperatura
definida y una entrop\'{\i}a \cite{bekenstein,hawking}, y que esta entrop\'{\i}a se puede calcular a
partir de la evaluaci\'on de la acci\'on eucl\'{\i}dea (con tiempo
imaginario y peri\'odico) en la configuraci\'on de agujero negro considerada \cite{gibbonshawking}.

El problema que resuelven las formas de transgresi\'on en este contexto es
la regularizaci\'on de la acci\'on \cite{motz1,motz2}. Sucede que la acci\'on de Chern-Simons
pura diverge en estas configuraciones, lo cual se resuelve a nivel de una
formulaci\'on hamiltoniana de mini-superespacio agregando los t\'erminos de
borde necesarios para que la acci\'on sea un extremo (tenga variaci\'on
cero) cuando valen las ecuaciones del movimiento y condiciones de borde
apropiadas (como se dijo antes, pedir que las variaciones sean cero en el
infinito espacial es demasiado restrictivo) \cite{dimensionally,scan}. Este procedimiento tiene el
inconveniente de que los t\'erminos de borde apropiados tienen que buscarse
caso por caso, para cada soluci\'on. La acci\'on de transgresi\'on contiene
los t\'erminos de borde apropiados por construcci\'on, resolviendo as\'{\i}
este problema en general. Esto se prueba para cualquier dimensi\'on para la
configuraci\'on de variedad cobordante, y en 3 y 5 dimensiones si se toma
como referencia el agujero negro de masa cero.

\subsection{Repaso de los fundamentos}

Hace unos treinta a\~nos Bekenstein \cite{bekenstein} y Hawking \cite{hawking} observaron que las soluciones
de agujero negro en Relatividad General tienen una entrop\'{\i}a y una
temperatura definidas. Estos resultados se
extienden a agujeros negros en gravedades de Chern-Simons. Se puede llegar a
la temperatura y entrop\'{\i}a por varios caminos, todos los cuales conducen
a las mismas conclusiones. Una de las formas mas concisas y elegantes, si
bien algo oscura y misteriosa en sus fundamentos, tiene que ver con la
formulaci\'on de integrales de camino con acci\'on eucl\'{\i}dea de la
teor\'{\i}a cu\'antica de campos\cite{gibbonshawking}. En esta subsecci\'on
repasaremos las ideas b\'asicas de este m\'etodo, siguiendo esencialmente 
\cite{gibbonshawking}.

\subsubsection{Teor\'{\i}a Cu\'antica de Campos y Mec\'anica Estad\'{\i}stica }

En la formulaci\'on de integrales de camino la amplitud de pasar de la
configuraci\'on $\Phi _ 1$ de los campos en el instante $t_1$ a la
configuraci\'on $\Phi _ 2$ en el instante $t_2$ esta dada por 
\begin{equation}
<\Phi _ 2, t_2\mid \Phi _ 1,t_1>=\int {\cal D}\Phi e^{iI[\Phi ]}
\end{equation}
donde la integral se toma sobre todas las configuraciones que interpolan
entre las configuraciones inicial y final dadas y $I[\Phi ]$ es la acci\'on.
Tambi\'en 
\begin{equation}
<\Phi _ 2, t_2\mid \Phi _ 1,t_1> = <\Phi _ 2\mid e^{-iH(t_2-t_1)}\mid \Phi _
1>
\end{equation}
donde $H$ es el hamiltoniano. 

Supongamos que se hace ahora $t_2-t_1=-i\beta$, lo que equivale a
pasar a un tiempo imaginario, y se hace $\Phi _2=\Phi _1$, sumando sobre
todos los $\Phi _1$ se obtiene 
\begin{equation}
Tr[exp(-\beta H)]=\int {\cal D}\Phi e^{iI[\Phi]}
\end{equation}
donde la integral se toma sobre todos los campos peri\'odicos con
per\'{\i}odo $\beta$ en tiempo imaginario y para la acci\'on eucl\'{\i}dea $\hat{I}=-iI$. 
Pero
\begin{equation}
Z=Tr[exp(-\beta H)] 
\end{equation}
es simplemente la funci\'on
de partici\'on $Z$ en el ensemble can\'onico para el campo $\Phi$ con
temperatura $T=\beta ^{-1}$ (tomando de ac\'a en mas la constante de
Boltzman $k_B=1$), la cual  puede calcularse entonces usando este procedimiento.\\ 

La evaluaci\'on concreta de la funci\'on de
partici\'on usualmente se basa en que la integral funcional que la define  
es dominada por 
la contribuci\'on de las configuraciones pr\'oximas a aquellas para
las cuales la acci\'on es un extremo (aproximaci\'on de punto de silla) con
las condiciones de periodicidad requeridas. Si estas configuraciones son  
 $\Phi _0$, y las configuraciones pr\'oximas a estas son de la forma  
 $\Phi =\Phi _0 +\overline{\Phi}$ mientras que la
acci\'on eucl\'{\i}dea desarrollada alrededor de esa configuraci\'on hasta segundo orden
en los campos tiene la forma 
\begin{equation}
\hat{I}[\Phi]= \hat{I}[\Phi _0] + I_2[\overline{\Phi}]
\end{equation}
donde $I_2$ es de segundo orden en las fluctuaciones. Se sigue que 
\begin{equation}
ln Z=- \hat{I}[\Phi _0] +ln\int{\cal D}\Phi e^{-I_2[ \overline{\Phi }]}
\end{equation}
donde el primer t\'ermino del segundo miembro representa el background y el
segundo las fluctuaciones.

En el ensemble can\'onico el logaritmo de $Z$ y la energ\'{\i}a libre se
relacionan entre si y con la entrop\'{\i}a $S$ y la masa (energ\'{\i}a
interna) $M$ como 
\begin{equation}
ln Z=\beta F=S-\beta M
\end{equation}
Esta ecuaci\'on permite calcular la entrop\'{\i}a, dadas M (la energ\'{\i}a total) y la 
funci\'on de partici\'on Z.

\subsubsection{Termodin\'amica de Agujeros Negros}

En el caso de gravitaci\'on, analogamente,  la funci\'on de partici\'on
en funci\'on de la m\'etrica $g$ (o el vielbein y la conexi\'on de
esp\'{\i}n) es 
\begin{equation}
Z=\int{\cal D}g{\cal D}\Phi e^{iI[g,\Phi]}
\end{equation}
Se $Z$ como la funci\'on de partici\'on mec\'anica
estad\'{\i}stica para configuracione peri\'odicas en tiempo imaginario 
$\beta $ y para la acci\'on eucl\'{\i}dea $\hat{I}=-iI$, por analog\'{\i}a con la 
discusi\'on para otros campos cu\'anticos, a pesar de que no se conocen 
los estados cu\'anticos del campo gravitatorio.
Otra vez la  funci\'on de
partici\'on es dominada por las configuraciones pr\'oximas a aquellas para
las cuales la acci\'on es un extremo (aproximaci\'on de punto de silla) con
las condiciones de periodicidad requeridas. Si estas configuraciones son $%
g_0 $ y $\Phi _0$, y las configuraciones pr\'oximas a estas son de la forma $%
g=g_0+\overline{g}$ y $\Phi =\Phi _0 +\overline{\Phi}$ mientras que la
acci\'on desarrollada alrededor de esa configuraci\'on hasta segundo orden
en los campos tiene la forma 
\begin{equation}
\hat{I}[g,\Phi]= \hat{I}[g_0,\Phi _0] + I_2[\overline{g},\overline{\Phi}]
\end{equation}
donde $I_2$ es de segundo orden en las fluctuaciones. Se sigue que 
\begin{equation}
ln Z=- \hat{I}[g_0,\Phi _0] +ln\int{\cal D}g{\cal D}\Phi e^{-I_2[\overline{g}%
, \overline{\Phi }]}
\end{equation}
donde el primer t\'ermino del segundo miembro representa el background y el
segundo las fluctuaciones.

Tambi\'en en el caso gravitatorio vale que en el ensemble can\'onico el 
logaritmo de $Z$ y la energ\'{\i}a libre se relacionan entre si y con la 
entrop\'{\i}a $S$ y la masa (energ\'{\i}a
interna) $M$ como 
\begin{equation}
ln Z=\beta F=S-\beta M
\end{equation}

A fines de la d\'ecada de 1960 se observ\'o que 
aparentemente se podr\'{\i}a violar la segunda ley de la 
termodin\'amica y reducir la entrop\'{\i}a 
del universo arrojando objetos con entrop\'{\i}a en un agujero negro. 
Esta observaci\'on, junto con la propiedad previamente 
conocida de que el \'area de un agujero negro (o mas bien de su horizonte de eventos) 
siempre se incrementa (cl\'asicamente), y en particular el agujero negro que resulta de unir dos 
agujeros negros tiene un \'area mayor que la suma de la de los dos originales, llevo a J. Bekenstein 
\cite{bekenstein} a proponer
que los agujeros tienen una entrop\'{\i}a proporcional a su \'area. Posteriormente Hawking 
\cite{hawking} observ\'o que 
si ten\'{\i}an entrop\'{\i}a deb\'{\i}an tener una temperatura, la cual calcul\'o en base a argumentos de
teor\'{\i}a cu\'antica de campos en espaciotiempos curvos. En retrospectiva resulta natural que los 
agujeros negros sean objetos susceptibles de una descripci\'on termodin\'amica, ya que debido 
a los teoremas llamados de {\it no hair} se puede caracterizar un agujero negro con un peque\~no 
n\'umero magnitudes macrosc\'opicas (masa, momento angular, carga el\'ectrica y otras cargas 
conservadas asociadas a interacciones de largo alcance), con independencia como se form\'o o 
los detalles del estado de su interior (cualquier estado interior con los mismos valores de las 
magnitudes microsc\'opicas mencionadas da lugar al mismo agujero negro en el exterior).\\

Si se consideran backgrounds correspondientes a agujeros negros se encuentra
que las soluciones eucl\'{\i}deas solo son no singulares para un valor
espec\'{\i}fico de la temperatura. Esto se ve al considerar las m\'etricas
de agujero negro, que suelen tener la forma gen\'erica 
\begin{equation}
ds^2=-f(r)dt^2+\frac{dr^2}{f(r)}+r^2d\Omega ^2
\end{equation}
donde la funci\'on $f(r)$ tiene una ra\'{\i}z en el horizonte $f(r_+)=0$ y $%
d\Omega ^2$ representa el elemento de l\'{\i}nea en la esfera de dimensi\'on
d-2. Cerca del horizonte se puede desarrollar $f(r)$ como $%
f(r)=(r-r_+)f^{\prime}(r_+)$ con lo que la m\'etrica queda 
\begin{equation}
ds^2=-(r-r_+)f^{\prime}(r_+)dt^2+\frac{dr^2}{(r-r_+)f^{\prime}(r_+)}
\end{equation}
ignorando la parte angular. Si elegimos una nueva variable $\rho$ tal que 
\begin{equation}
d\rho =\frac{dr}{\sqrt{f^{\prime}_+}\sqrt{r-r_+}}
\end{equation}
(donde $f^{\prime}_+=f^{\prime}(r_+)$) lo que implica $r>r_+$ si $\rho$ es
real, entonces 
\begin{equation}
\rho =\frac{2\sqrt{r-r_+} }{\sqrt{f^{\prime}_+} }
\end{equation}
y de ah\'{\i} 
\begin{equation}
f^{\prime}_{+} (r-r_+)= \frac{[ f^{\prime}_{+} \rho ]^2}{4}
\end{equation}
por lo que la m\'etrica, sin la parte angular y pasando al espacio
eucl\'{\i}deo $t\rightarrow i\tau$ queda 
\begin{equation}
ds^2=+\frac{[f^{\prime}_{+}\rho ]^2}{4}d\tau ^2+d\rho ^2= \rho ^2 d\phi
^2+d\rho ^2
\end{equation}
con $\phi =\frac{f^{\prime}_{+}}{2}\tau $. Para evitar una singularidad
c\'onica $\phi$ debe tener per\'{\i}odo $2\pi$, que corresponde a $\tau $
con per\'{\i}odo $\beta$ dado por 
\begin{equation}
\beta =\frac{4\pi}{f^{\prime}_{+}}
\end{equation}
La temperatura correspondiente es la 'Temperatura de Hawking' $T_H=1/\beta $

Notese que las coordenadas definidas arriba $(\tau ,\rho )$, para el $\beta$ correspondiente a la 
temperatura de Hawking, son coordenadas polares en un plano correspondiente 
a valores de $r\ge r_{+}$
y cuyo origen corresponde a $r=r_{+}$. Es importante se\~nalar que no hay borde en el origen 
(correspondiente al horizonte de eventos)
del plano eucl\'{\i}deo .\\

Por ejemplo, para la m\'etrica de Schwarszchild $f(r)=1-2M/r$
(en las unidades en que $c=G=\hbar =k_B=1$), entonces $r_{+}=2M$ y $\beta =8\pi M$

Para los agujeros negros de Chern-Simons de la secci\'on previa  
$f(r)=\Delta ^2(r)$, entonces $r_+^2=\alpha -1$, y $\beta =2\pi /r_+$ \cite{dimensionally,scan}.

Estas diferentes dependencias de la energ\'{\i}a M con la temperatura dan lugar a 
un comportamiento bastante diferente de los calores espec\'{\i}ficos en funci\'on de M 
para agujeros negros de Schwarzschild y Chern-Simons (ver p. ej. ref.\cite{scan}).\\

En lo que respecta a la funci\'on de partici\'on Z (y de ah\'{\i} la entrop\'{\i}a), 
el orden m\'as bajo de aproximaci\'on (llamada aproximaci\'on semicl\'asica) se 
obtiene, como se observ\'o antes,
 evaluando
la acci\'on en la configuraci\'on peri\'odica en tiempo imaginario con per\'{\i}odo $\beta$, 
correspondiente a los valores
dados de la energ\'{\i}a M y cualquier otra magnitud macrosc\'opica que caracterize 
el estado termodin\'amico.

En el caso de gravitaci\'on en el ensemble can\'onico la magnitud 
macrosc\'opica es M y la soluci\'on 
correspondiente es el agujero negro de masa M, el cual es peri\'odico en 
tiempo imaginario ya que es est\'atico. 
En el caso de momento angular, carga el\'ectrica y/o otras cargas 
macrosc\'opicas distintas de cero, la soluci\'on 
correspondiente ser\'a el agujero negro con esas cargas.

Es importante mencionar algunas sutilezas de la evaluaci\'on de la 
acci\'on de agujeros negros para calcular Z.
En primer lugar la acci\'on que se debe tomar debe tener t\'erminos 
de borde apropiados para que la 
acci\'on sea un extremo (tenga variaci\'on cero) cuando valen las 
ecuaciones del movimiento \cite{regge}. En segundo lugar 
al pasar a la acci\'on eucl\'{\i}dea el origen est\'a en $r=r_{+}$, por lo que la si 
la integraci\'on se hace en la variable $r$ el rango de integraci\'on debe ser 
de $r=r_{+}$ a $r=\infty$, y los t\'erminos de borde que existan deben evaluarse 
en el borde en infinito, pero no en el horizonte de eventos, ya que ah\'{\i} no hay borde
en coordenadas eucl\'{\i}deas.\\

Para el caso del agujero negro de Schwarzschild en relatividad general la acci\'on con el t\'ermino de
borde apropiado  en una regi\'on $\Omega $ con borde $\partial\Omega $es
$$
I=\frac{1}{16\pi }\int _{\Omega}R\sqrt{-g}d^4x+\frac{1}{8\pi } \int _{\partial\Omega}K\sqrt{-h}d^3x
$$
donde $g$ es el determinante de la m\'etrica, $R$ es el escalar de curvatura, $h$ es el 
determinante de la m\'etrica inducida en el borde y $K$ es la traza de la curvatura extr\'{\i}nseca \cite{regge}.
La acci\'on da 
$$
I=4\pi M^2
$$
y la entrop\'{\i}a da
$$
S=\frac{A}{4}
$$
con $A=4\pi r_{+}^2$ el \'area del horizonte de eventos, en unidades 
de Planck (recordar que tomamos $c=G=\hbar =k_B=1$ ). Este es el 
resultado justamente reconocido de Bekenstein y Hawking \cite{bekenstein,hawking,gibbonshawking}.

\subsection{Entrop\'{\i}a de agujeros negros en cualquier dimensi\'on para
la configuraci\'on de variedad cobordante}

En esta secci\'on se calcula la entrop\'{\i}a de agujeros negro de Chern-Simons usando la acci\'on
de variedad cobordante discutida en la secci\'on anterior. La raz\'on para elegir esta configuraci\'on 
es que el t\'ermino de borde es tal que la acci\'on es un extremo cuando valen las 
ecuaciones del movimiento, para la condici\'on de borde considerada.

La entrop\'{\i}a de agujeros negros de Chern-Simons fue calculada antes 
utilizando m\'etodos hamiltonianos de minisuperespacio \cite{dimensionally,scan}. 
El m\'erito del enfoque discutido ac\'a es que se da una prescripci\'on general de los 
t\'erminos de borde, mientras que en el enfoque de minisuperespacio 
deben buscarse caso por caso para cada soluci\'on.

Los c\'alculos de esta secci\'on est\'an incluidos en \cite{motz2}.
 
Al elegir $\overline{e}^{a}=0$ en la conexi\'on de AdS $\overline{A}$, la
acci\'on de transgresi\'on se reduce a la de Lanczos-Lovelock-Chern-Simons
mas un t\'ermino de borde 
\[
I_{2n+1}(A,\overline{A})=\int\limits_{M}L_{2n+1}^{LCS}(R,e)+d\alpha 
\]
con 
\[
L_{2n+1}^{LCS}(R,e)=\kappa \int\limits_{0}^{1}dt\epsilon \left(
R+t^{2}e^{2}\right) ^{n}e 
\]
y 
\[
\alpha =-\kappa n\int\limits_{0}^{1}dt\int\limits_{0}^{t}ds\epsilon \theta
e\left( \tilde{R}+t^{2}\theta ^{2}+s^{2}e^{2}\right) ^{n-1}. 
\]

La constante $\kappa $ se toma como antes igual a $\kappa =\frac{1}{%
2(d-2)!\Omega _{d-2}G_{n-1}}.$

y tambi\'en las configuraciones de agujero negro est\'atico son las dadas
arriba, con m\'etrica

\[
ds^{2}=-\Delta ^{2}(r)dt^{2}+\frac{dr^{2}}{\Delta ^{2}(r)}+r^{2}d\Omega
_{d-2}^{2} 
\]

donde la funci\'on $\Delta (r)$ es 
\[
\Delta ^{2}(r)=1-\sigma +r^{2} 
\]

y la constante $\sigma =\left( 2G_{n-1}M+1\right) ^{\frac{1}{n}}$ es la que
antes llamamos $\alpha$ (cambiamos la notaci\'on para evitar confusi\'on con
el t\'ermino de borde $\alpha$), con el par\'ametro $M$ correspondiendo a la
masa, como puede verse a partir del modelo de mini-superespacio en el
formalismo hamiltoniano.

Si evaluamos la forma eucl\'{\i}dea del t\'ermino de borde en esta familia
de soluciones se obtiene

\[
\int\limits_{\partial M}\alpha _{E}= 
\]
\[
2(d-2)!\Omega _{d-2}\beta \kappa n\left[ \frac{\left( \Delta ^{2}\right)
^{^{\prime }}}{2}r\int\limits_{0}^{1}dt\left( 1-\Delta
^{2}+t^{2}r^{2}\right) ^{n-1}+\left( \Delta ^{2}-\frac{\left( \Delta
^{2}\right) ^{^{\prime }}}{2}r\right) \int\limits_{0}^{1}dtt\left(
1-t^{2}\Delta ^{2}+t^{2}r^{2}\right) ^{n-1}\right] ^{r=\infty } 
\]
donde $\beta $ es el per\'{\i}odo del tiempo eucl\'{\i}deo $\tau $. Debido a
la forma particular de la funci\'on en la m\'etrica la segunda integral no
depende de $r$, y ser\'a proporcional a la masa $M$ 
\[
\frac{\beta }{G_{n}}n\left( 1-\sigma \right) \int\limits_{0}^{1}dtt\left(
1-t^{2}\left( 1-\sigma \right) \right) ^{n-1}=-\beta M 
\]
y el t\'ermino restante, que contiene potencias divergentes de $r$, se
combina con este de modo que el borde se escribe 
\[
\int\limits_{\partial M}\alpha _{E}=2(d-2)!\Omega _{d-2}\beta \kappa n\left[
r^{2}\int\limits_{0}^{1}dt\left( \sigma +\left( t^{2}-1\right) r^{2}\right)
^{n-1}\right] ^{r=\infty }-\beta M. 
\]

La acci\'on eucl\'{\i}dea en el bulk --{\em on-shell}-toma la forma
expl\'{\i}cita 
\[
I_{E}^{LCS}=(d-2)!\Omega _{d-2}\beta \kappa \left[ 2nr^{2}\int%
\limits_{0}^{1}dt\left( t^{2}-1\right) \left( \sigma +\left( t^{2}-1\right)
r^{2}\right) ^{n-1}+\int\limits_{0}^{1}dt\left( \sigma +\left(
t^{2}-1\right) r^{2}\right) ^{n}\right] _{r=r_{+}}^{r=\infty } 
\]

y, dado que est\'a claro que la potencia mas alta en $\sigma $ en el segundo
t\'ermino es cero porque es independiente de $r$, podemos poner la
expresi\'on anterior en la forma 
\[
I_{E}^{LCS}=2(d-2)!\Omega _{d-2}\beta \kappa n\left[ \sum_{k=0}^{n-1}\frac{%
\left( n-1\right) !}{\left( n-1-k\right) !k!}\sigma
^{n-1-k}\int\limits_{0}^{1}dt\frac{2k+3}{2k+2}\left( t^{2}-1\right)
^{k+1}r^{2\left( k+1\right) }\right] _{r=r_{+}}^{r=\infty }. 
\]
Sorprendentemente, los coeficientes que vienen de la integraci\'on en $t$ se
simplifican, debido a la relaci\'on

\[
\int\limits_{0}^{1}dt\frac{2k+3}{2k+2}\left( t^{2}-1\right)
^{k+1}=-\int\limits_{0}^{1}dt\left( t^{2}-1\right) ^{k} 
\]

que hace que la continuaci\'on eucl\'{\i}dea de la acci\'on de
Lovelock-Chern-Simons se reduzca a

\[
I_{E}^{LCS}=-2(d-2)!\Omega _{d-2}\beta \kappa n\left[ r^{2}\int%
\limits_{0}^{1}dt\left( \sigma +\left( t^{2}-1\right) r^{2}\right) ^{n-1}%
\right] _{r=r_{+}}^{r=\infty }. 
\]

Juntando todo la acci\'on eucl\'{\i}dea total $I_{E}=I_{E}^{LCS}+\int%
\limits_{\partial M}\alpha _{E}$ tiene las divergencias en el infinito
espacial canceladas. Como veremos abajo la contribuci\'on del horizonte de
eventos $r_{+}$ da exactamente la entrop\'{\i}a del agujero negro, ya que

\[
S=I_{E}+\beta M=\frac{\beta }{G_{n}}nr_{+}^{2}\int\limits_{0}^{1}dt\left(
\sigma +\left( t^{2}-1\right) r_{+}^{2}\right) ^{n-1}. 
\]

Por definici\'on el radio del horizonte $r_{+}$ satisface la relaci\'on 
\[
\Delta ^{2}(r_{+})=0=1-\sigma +r_{+}^{2} 
\]

y asumiendo que el per\'{\i}odo en el tiempo eucl\'{\i}deo $\beta $ se
calcula como se dijo arriba,

\[
\beta =\frac{1}{T} 
\]

donde $T$ representa la temperatura del agujero negro

\[
T=\frac{1}{4\pi }\left. \frac{d\Delta }{dr}\right| _{r_{+}}. 
\]

Por lo tanto, la expresi\'on para la entrop\'{\i}a de un agujero negro de
Chern-Simons es

\[
S=\frac{2\pi k_{B}}{G_{n}}nr_{+}\int\limits_{0}^{1}dt\left(
1+t^{2}r_{+}^{2}\right) ^{n-1} 
\]

la que por medio de un cambio de variables lleva a la f\'ormula

\[
S=\frac{2\pi k_{B}}{G_{n}}n\int\limits_{0}^{r_{+}}dr\left( 1+r^{2}\right)
^{n-1}. 
\]

\subsection{Entrop\'{\i}a de agujeros negros con respecto al agujero negro
de masa cero en 3D y 5D}

Se puede pensar en regularizar la acci\'on eucl\'{\i}dea en el
c\'alculo de la entrop\'{\i}a tomando la configuraci\'on $A$ correspondiente
a un agujero negro de masa $M$ y la configuraci\'on $\overline{A}$ como una
configuraci\'on especial correspondiente a una soluci\'on est\'atica de las
ecuaciones del movimiento. Dado que el per\'{\i}odo $\beta$ esta fijado para
el agujero negro, la configuraci\'on $\overline{A}$ debe ser tal que $\beta$
pueda ser arbitrario, o no este definido para esta configuraci\'on. Dos
configuraciones que satisfacen este requisito son el espacio AdS ($M=-1$) y
el agujero negro de masa cero($M=0$). Elegimos el agujero negro de masa
cero. El c\'alculo \cite{motz1} se realizar\'a para d=3 y d=5 porque solo en esos caso
conocemos la forma expl\'{\i}cita de la acci\'on en t\'erminos de los
vielbeins y conexiones de esp\'{\i}n como la diferencia de acciones de
Lovelock-Chern-Simons para cada campo mas un t\'ermino de borde. Se
comprueba que el resultado en estos casos es el esperado, como en el caso de 
variedad cobordante, si se aplican las siguientes prescripciones:\newline
(i) La integral en el bulk se toma entre los radios cero e infinito para el
agujero negro de masa cero pero entre $r_{+}$ e infinito para el agujero
negro de masa $M$.\newline
(ii) Los t\'erminos de borde se consideran en infinito, pero no en $r_{+}$,
donde en la representaci\'on eucl\'{\i}dea de hecho no hay borde.\newline

\subsubsection{Entrop\'{\i}a en 3D}

La acci\'{o}n de transgresi\'{o}n para el grupo AdS en 3D, con la traza
sim\'{e}trica considerada arriba 
\[
I_{3}^{0}=\kappa \epsilon (Re+\frac{1}{3}e^{3})-\kappa \epsilon (\tilde{R}%
\overline{e}+\frac{1}{3}\overline{e}^{3})+\frac{1}{2}\kappa \epsilon ~d[(e+%
\overline{e})\theta ] 
\]
donde la notaci\'{o}n es la de la secci\'{o}n anterior. Se toma la
configuraci\'{o}n $e$, $\omega $ correspondiente a un agujero negro de masa $%
M$ y la configuraci\'{o}n $\overline{e}$, $\overline{\omega }$
correspondiente a un agujero negro de masa cero. La acci\'{o}n eucl\'{i}dea $%
I$ se eval\'{u}a en los rangos de las coordenadas en que est\'{a} definida 
\[
I=\kappa \int_{\Omega _{+}}\epsilon (Re+\frac{1}{3}e^{3})-\kappa
\int_{\Omega }\epsilon (\tilde{R}\overline{e}+\frac{1}{3}\overline{e}^{3})+%
\frac{1}{2}\kappa \int_{\Sigma }\epsilon ~[(e+\overline{e})\theta ] 
\]
donde $\Omega _{+}$ es la regi\'{o}n comprendida entre $\tau =0$ y $\tau
=\beta $ y entre $r=r_{+}$ y $r=+\infty $, $\Omega $ es la regi\'{o}n
comprendida entre $\tau =0$ y $\tau =\beta $ y entre $r=0$ y $r=+\infty $, y 
$\Sigma $ es la superficie con $r=+\infty $ y $\tau $ entre $0$ y $\beta $.
Utilizando la forma expl\'{\i}cita del vielbein, la conexi\'{o}n de esp\'{i}n
y la curvatura para cada configuraci\'{o}n resulta 
\[
I=\kappa \int_{0}^{r_{+}}dr~r~6\frac{2}{3}2\pi \beta -2\kappa \pi \beta
2G_{3}M 
\]
donde el primer t\'{e}rmino del segundo miembro viene del bulk y el segundo
del borde. Se tiene 
\[
I=\kappa (4\pi r_{+}^{2}\beta -2\pi \beta 2G_{3}M)=2\pi \beta \kappa 2G_{3}M 
\]
donde se uso que $r_{+}^{2}=2G_{3}M$. Usando que en 3D 
\[
\kappa =\frac{1}{2G_{3}2\pi } 
\]
resulta $I=\beta M$ pero $I=\beta F=S-\beta M$ de donde la entrop\'{i}a $S$
es 
\[
S=2\beta M 
\]
Usando $\beta =2\pi /r_{+}$ y $r_{+}^{2}=2G_{3}M$ resulta tambi\'{e}n 
\[
S=\frac{2\pi }{G_{3}}r_{+} 
\]
que coincide con el resultado de la subsecci\'{o}n anterior para 3D.

\subsubsection{Entrop\'{\i}a en 5D}

La acci\'{o}n de transgresi\'{o}n para el grupo AdS en 5D, con la traza
sim\'{e}trica antes considerada es 
\[
{\cal T}_{5}=\kappa \epsilon (R^2e+\frac{2}{3}Re^{3}+\frac{1}{5}e^5) - \kappa
\epsilon (\tilde{R}^2 \overline{e}+\frac{2}{3}\tilde{R}\overline{e}^{3}+%
\frac{1}{5}\overline{e}^5)-
\]
\[
\frac{1}{3}\kappa \epsilon ~d[ \theta (e+\overline{e}) (R-\frac{1}{4}\theta
^2+\frac{1}{2}e^2) + \theta (e+\overline{e})(\tilde{R}-\frac{1}{4}\theta ^2+%
\frac{1}{2}\overline{e}^2)+\theta Re+\theta \tilde{R}\overline{e}] 
\]
con la notaci\'{o}n de la secci\'{o}n anterior. Se toma de nuevo la
configuraci\'{o}n $e$, $\omega $ correspondiente a un agujero negro de masa $%
M$ y la configuraci\'{o}n $\overline{e}$, $\overline{\omega }$
correspondiente a un agujero negro de masa cero. La acci\'{o}n eucl\'{i}dea $%
I$ se eval\'{u}a en los rangos de las coordenadas en que est\'{a} definida 
\[
I =\kappa \int_{\Omega _{+}} \epsilon (R^2e+\frac{2}{3}Re^{3}+\frac{1}{5}%
e^5) - \kappa \int_{\Omega }\epsilon (\tilde{R}^2 \overline{e}+\frac{2}{3}%
\tilde{R}\overline{e}^{3}+\frac{1}{5}\overline{e}^5)-
\]
\[
\frac{1}{3}\kappa \int _{\Sigma } \epsilon ~[ \theta (e+\overline{e}) (R-%
\frac{1}{4}\theta ^2+\frac{1}{2}e^2) + \theta (e+\overline{e})(\tilde{R}-%
\frac{1}{4}\theta ^2+\frac{1}{2}\overline{e}^2)+\theta Re+\theta \tilde{R}%
\overline{e}] 
\]
donde otra vez $\Omega _{+}$ es la regi\'{o}n comprendida entre $\tau =0$ y $%
\tau =\beta $ y entre $r=r_{+}$ y $r=+\infty $, $\Omega $ es la regi\'{o}n
comprendida entre $\tau =0$ y $\tau =\beta $ y entre $r=0$ y $r=+\infty $, y 
$\Sigma $ es la superficie con $r=+\infty $ y $\tau $ entre $0$ y $\beta $.
Utilizando la forma expl\'{i}cita del vielbein, la conexi\'{o}n de esp\'{i}n
y la curvatura para cada configuraci\'{o}n resulta que el t\'ermino de borde
divergente cancela la divergencia que viene del bulk (regularizando asi la
acci\'on) y se obtiene 
\[
I=\frac{\beta }{2G_5}[-\frac{8}{3}r_{+}^4+4\alpha r_{+}^2 -2G_5M] 
\]
donde $\alpha =\sqrt{2G_5 M+1}$, y $G _5$ es la constante de Newton en 5D.
Usando como antes que $I=\beta F=S-\beta M$, la entrop\'{i}a $S$ es 
\[
S=\frac{\beta }{G_5}[-\frac{4}{3}r_{+}^4+2\alpha r_{+}^2] 
\]
Usando $\beta =2\pi /r_{+}$ y $r_{+}^{2}=\alpha -1$ resulta tambi\'{e}n 
\[
S=\frac{4\pi }{G_{5}}[r_{+}+\frac{r_{+}^3}{3}] 
\]
que coincide con el resultado de la subsecci\'{o}n anterior para 5D.

\newpage

\section{Acoplamiento de Branas con Gravedades de Chern-Simons}

~
\subsection{Repaso de Modelos de Objetos Extendidos}

~
{\it However, I do not believe that scientific progress is always best
advanced by keeping an altogether open mind. It is often necessary to forget
one's doubts and to follow the consecuences of one's assumptions wherever
they may lead-the great thing is not to be free of theoretical prejudices,
but to have the right theoretical prejudices. And always, the test of any
theoretical preconception is in where it leads}

-S. Weinberg\newline

La Teor\'{\i}a de Supercuerdas \cite{superstring} es actualmente el \'unico
candidato para una teor\'{\i}a cu\'antica de toda la materia y las
interacciones. Como tal es destacable que incluye una teor\'{\i}a cu\'antica
de la gravitaci\'on. La teor\'{\i}a se defini\'o originalmente trav\'es de
desarrollos perturbativos involucrando todas las superficies bidimensionales
interpolando entre configuraciones iniciales y finales dadas de lazos
cerrados y/o abiertos (dependiendo de la teor\'{\i}a). Esta definici\'on
perturbativa de la teor\'{\i}a es su principal desventaja, dado que la mayor
parte de las cuestiones interesantes que se plantean (incluyendo hacer
predicciones experimentalmente verificables) son intrinsecamente no
perturbativas.

La consistencia de la teor\'{\i}a cu\'antica resulta ser una condici\'on muy
restrictiva, que es pasada por solo cinco modelos definidos en un
espaciotiempo de diez dimensiones, llamados Tipo IIA, IIB and I y los dos
modelos de Cuerdas Heter\'oticas con grupos de gauge $SO(32)$ y $E_8\times
E_8$. Estas teor\'{\i}as tienen como l\'{\i}mite de bajas energ\'{\i}as
diferentes teor\'{\i}as de supergravedad est\'andar.

Durante los \'ultimos a\~nos se progres\'o mucho en la comprensi\'on de los
aspectos no perturbativos de la teor\'{\i}a, lo que llev\'o a establecer una
red de 'Dualidades' entre las diferentes teor\'{\i}as. Estas dualidades
implican que el r\'egimen de acoplamiento fuerte de alguno de los cinco
modelos consistentes corresponde al r\'egimen de acoplamiento d\'ebil de
otra y que las compactificaciones de diferentes teor\'{\i}as en diferentes
variedades llevan a la misma teor\'{\i}a. Algunas de las dualidades, sin
embargo, relacionan teor\'{\i}as de cuerdas en diez dimensiones con una
teor\'{\i}a desconocida en once dimensiones. Tambi\'en result\'o que al
nivel no perturbativo aparecen objetos extendidos de dimensi\'on mayor que
dos, conocidos como 'branes'. Estos resultados llevaron a postular que
existe una \'unica teor\'{\i}a cu\'antica supersim\'etrica que tiene como
l\'{\i}mites especiales las cinco teor\'{\i}as de cuerdas consistentes
conocidas, la cual ha sido llamada teor\'{\i}a M \cite
{townsend1,hull1,witten2,townsend2}. Se sabe que esta teor\'{\i}a tiene
supergravedad est\'andar en once dimensiones como su l\'{\i}mite de bajas
energ\'{\i}as, y que incluye objetos extendidos con volumen de mundo de
dimensi\'on 2+1 y 5+1 (membranas y 5-branas respectivamente)

En esta secci\'on se revisar\'an aspectos de la teor\'{\i}a de objetos
extendidos que se utilizar\'an en lo que sigue. Estos son (1) la
formulaci\'on de Green-Schwarz de las supercuerdas y superbranas con
supersimetr\'{\i}a espaciotemporal, (2) las acciones de cuerdas y branas
heter\'oticas y (3) trabajos orientados a reescribir la teor\'{\i}a de
cuerdas como un modelo de Chern-Simons.\newline

\subsubsection{La Acci\'on de Green-Schwarz}

La acci\'on de cuerdas con supersimetr\'{\i}a espaciotemporal de propuesta
por Green y Schwarz \cite{greenschwarz} es 
\begin{equation}
S= T~\int d\tau\int_0^{\pi}d\sigma (L_1+L_2)
\end{equation}
donde 
\begin{eqnarray}
L_1=-\frac{1}{2}\sqrt{-\gamma}\gamma ^{ij}\Pi _i^r\Pi_{jr} \\
L_2=-i~\epsilon ^{ij}\partial _i X^r \left[ \overline{\theta}^1\gamma
_r\partial _j\theta ^1-\overline{\theta}^2\gamma _r\partial _j\theta ^2%
\right] +\epsilon ^{ij}\overline{\theta}^1\gamma ^r\partial _i\theta ^1%
\overline{\theta}^2\gamma _r\partial _j\theta ^2
\end{eqnarray}
los \'{\i}ndices $i,j=0,1$ son \'{\i}ndices curvos de la superficie de
mundo, las coordenadas de la superficie de mundo son $\zeta ^i=(\zeta
^0,\zeta ^1)=(\tau ,\sigma )$, $r,s,p=0,...,10$ son \'{\i}ndices planos de
Lorentz, las variables din\'amicas son las coordenadas espaciotemporales $%
X^r(\sigma ,\tau )$, los espinores de Majorana-Weyl en 10D $\theta
^{A\alpha}(\sigma ,\tau )$, con $A=1,2$ y $\alpha =1,...32$, la m\'etrica
auxiliar en la superficie de mundo $\gamma ^{ij}$, y 
\begin{equation}
\Pi ^r= dX^r-i \overline{\theta}^A\gamma ^rd\theta ^A = \left(\partial
_iX^r-i \overline{\theta}^A\gamma ^r\partial _i \theta ^A\right) d\zeta ^i =
\Pi ^r_i d\zeta ^i
\end{equation}
La acci\'on es invariante bajo reparametrizaciones de la superficie de mundo
(o transformaciones generales de coordenadas de esta) y transformaciones
globales de super-Poincar\'e 
\begin{eqnarray}
\delta \theta ^A=\frac{1}{4}\Lambda _{rs}\gamma ^{rs}\theta ^A+ \epsilon ^A
\\
\delta X^r=\Lambda ^r_{~s}X^s+a^r+ \overline{\epsilon}^A\gamma ^r\theta ^A \\
\delta \gamma ^{ij}=0
\end{eqnarray}
con par\'ametros ($\Lambda ^r_s$, $a^r$, $\epsilon ^A$). La acci\'on
tambi\'en es invariante bajo la simetr\'{\i}a local fermi\'onica conocida
como 'simetr\'{\i}a kappa' 
\begin{eqnarray}
\delta _{\kappa} \theta ^A=2i\gamma {\bf .}\Pi _i \kappa ^{Ai}~~,~~ \delta
_{\kappa} X^r =i\overline{\theta}^A \gamma ^r\delta _{\kappa }\theta ^A \\
\delta _{\kappa}(\sqrt{-\gamma}\gamma ^{ij})= -16\sqrt{-\gamma}~[ P_{-}^{ik}%
\overline{\kappa}^{1j}\partial _k\theta ^1 P_{+}^{ik}\overline{\kappa}%
^{2j}\partial _k\theta ^2]
\end{eqnarray}
donde los $\kappa ^{Ai}$ son variables de Grassman, bi-vectores de la
superficie de mundo con un \'{\i}ndice espinorial suprimido de Majorana-Weyl
y satisfacen las condiciones de autodualidad y anti-autodualidad 
\begin{equation}
P_+^{ij}\kappa ^1_j=0 ~~,~~ P_-^{ij}\kappa ^2_j=0
\end{equation}
con los proyectores 
\begin{equation}
P^{ij}_{\pm} = \frac{1}{2}(\gamma ^{ij}\pm \epsilon ^{ij}/\sqrt{-\gamma})
\end{equation}
la simetr\'{\i}a-$\kappa$ permite elegir el gauge del cono de luz 
\begin{equation}
\gamma ^+\theta ^A=0~~,~~ X^+(\sigma ,\tau )= x^+ + \frac{p^+}{\pi T}\tau
\end{equation}
Las componentes de cono de luz de un vector $A^r$ se definen como $A^{\pm}=%
\frac{1}{\sqrt{2}}(A^0\pm A^9)$. Debido a $(\gamma ^+)^2=0$ la condici\'on
del cono de luz implica que la mitad de las componentes de $\theta ^A$ son
cero. Es destacable que la teor\'{\i}a descrita por la acci\'on de las
ecs.(90-92), la cual parece tener t\'erminos de interacci\'on complicados,
es en realidad una teor\'{\i}a libre en el gauge del cono de luz. El
requerimiento de la simetr\'{\i}a-$\kappa$ determina los coeficientes
relativos de $L_1$ y $L_2$.

En la ref.\cite{henneaux} se observ\'o que el t\'ermino $S_2=\int d\tau
d\sigma L_2$ puede interpretarse como la integral del la 3-forma de
Wess-Zumino-Witten (como en la ec.(24)) para el grupo de Super-Poincar\'e.
Se considera la forma de Maurer-Cartan $U=g^{-1}dg$ con $g=e^{i\hat{X}%
^aP_a+\theta ^{A\alpha}Q^A_{\alpha}}$. Entonces esencialmente $U=[dX^a+i%
\overline{\theta}\gamma ^ad\theta] +d\theta ^{A\alpha}Q^A_{\alpha}$ y 
\begin{equation}
S_2= \int _{M^3}\, \hbox{STr}\, [U^3] =\int _{M^3}\, \hbox{STr}\,
[(g^{-1}dg)^3]
\end{equation}
donde $M^3$ es una variedad tridimensional con borde en la superficie de
mundo de la cuerda. Las trazas de productos de generadores requeridas son $%
\, \hbox{STr}\, (P P P)$, $\, \hbox{STr}\, (P P Q)$, $\, \hbox{STr}\, (Q Q
P) $~ y $\, \hbox{STr}\, (Q Q Q)$. De estas, la \'unica no nula es $\, %
\hbox{STr}\, (P_a Q _{\alpha}Q_{\beta} ) = \left(\gamma _a~C^{-1}\right)
_{\alpha\beta} $. Notar que los \'{\i}ndices espinoriales del tensor
sim\'etrico invariante son en realidad \'{\i}ndices en la representaci\'on
adjunta del grupo desuper-Poincar\'e, a\'un cuando vistos desde el punto de
vista del subgrupo de Lorentz son \'{\i}ndices en la representaci\'on
fundamental espinorial de aquel grupo.

Las configuraciones de la supercuerdapueden pensarse como una inmersi\'on de
su superficie de mundo en un superespacio plano en 10D con $N=2$. Desde este
punto de vista la generalizaci\'on natural ser\'{\i}a considerar inmersiones
en superespacios curvos \cite{superstring}, lo que podr\'{\i}a interpretarse
como una cuerda propagandose en un background no trivial. Requerimientos de
consistencia imponen la condici\'on de que el background debe satisfacer las
ecuaciones de la supergravedad est\'andar que corresponde al l\'{\i}mite de
bajas energ\'{\i}as del modelo de supercuerdas considerado, y aquellos
backgrounds incluir\'an en general campos adicionales, como campos de gauge
y campos dados por p-formas (algunos de los cuales son llamasos 'campos de
Ramond-Ramond' o 'campos-RR'). Esto puede interpretarse pensando que el
background consistente para la cuerda esta formado por un condensado de
cuerdas.

Es posible generalizar la acci\'on de Green-Schwarz a objetos extendidos de
mayor dimensionalidad, llamados 'super p-branas' \cite{supermembrane}, en
backgrounds planos o curvos. Estos modelos son $\kappa$-sim\'etricos, pero
al contrario de lo que ocurre en el caso de las cuerdas esto no alcanza para
permitir la elecci\'on de un gauge en el que la teor\'{\i}a sea libre. Esto
hace que el problema de la cuantizaci\'on de los modelos de p-brana sea muy
dif\'{\i}cil, por lo que no est\'a resuelto hasta hoy.\newline

\subsubsection{Cuerdas y Branas Heter\'oticas}

El acoplamiento de objetos extendidos bos\'onicos de diversas dimensiones a
campos de gauge fue estudiado por Dixon, Duff y Sezgin (DDS)\cite{dixon}
siguiendo el modelo de la cuerda heter\'otica \cite{gross}. La acci\'on de
el modelo DDS para objetos extendidos con volumen de mundo de dimensi\'on
par $d$ en un espaciotiempo de dimensi\'on $D$ es 
\begin{equation}
S_{DDS}=S^{K}_d+S^{WZW}_d
\end{equation}
donde el t\'ermino cin\'etico es 
\begin{equation}
S_{d}^K=\int d^d\zeta\left\{ -\frac{1}{2}\sqrt{-\gamma}~\gamma ^{ij}\left[%
\partial _iX^r\partial _jX^sg_{rs}(X^p)+J_i^aJ^a_j\right]+\frac{1}{2}(d-2)%
\sqrt{-\gamma}\right\}
\end{equation}
aca las coordenadas del volumen de mundo son $\zeta ^i$, $i,j,k=0,...,d-1$,
las coordenadas espaciotemporales son $X^r(\zeta )$, $r,s,p=0,...,D-1$, la
m\'etrica del volumen de mundo es $\gamma ^{ij}(\zeta )$, y la m\'etrica del
background espaciotemporal es $g_{rs}(X)$. Las $J_i^a$'s se definen como $%
J^a_i=\partial _iX^rA^a_r-\partial _iY^lK_l^a$ donde los $A^a_r(X)$ son
campos de gauge para alg\'un grupo de gauge $G$, las $Y^l(\zeta )$ son
coordenadas en la variedad de grupo $G$ y los $K^a_l(Y)$ son formas de
Maurer-Cartan (MC) invariantes a la izquierda $K_l=K_l^aT^a=g^{-1}(Y)\frac{%
\partial ~}{\partial Y^l}g(Y)$. En el lenguaje de formas diferenciales $%
A=A^a_rT^adx^r$ y $K=K^a_lT^ady^l$, y sus pull-backs al volumen de mundo
(los cuales denotaremos con el mismo s\'{\i}mbolo, ya que ser\'a claro por
el contexto cual de los dos estamos usando) $A=A^a_rT^a\partial _iX^r d\zeta
^i$ y $K=K^a_lT^a\partial _iY^ld\zeta ^i$. Entonces $J=A-K$, an\'alogo al $J$
de la secci\'on 2.1.2. con $A_1=A$ y $A_0=K$, $K$ gauge puro. Las formas de
MC satisfacen la ec. de MC $dK+K^2=0$.

El t\'ermino de WZW es 
\begin{equation}
S^{WZW}_d=\int \left[B_d+C_d-b_d\right]=\int {\cal B}_d
\end{equation}
con 
\begin{eqnarray}
B_d=\frac{1}{d!}B_{r_1...r_d}\partial _{i_1}X^{r_1}... \partial
_{i_d}X^{r_d}d\zeta ^{i_1}...d\zeta ^{i_d} \\
b_d=\frac{1}{d!}b_{r_1...r_d}\partial _{i_1}X^{r_1}... \partial
_{i_d}X^{r_d}d\zeta ^{i_1}...d\zeta ^{i_d}
\end{eqnarray}
donde $B_d$ es un campo de Ramond-Ramond (campo RR) y $b_d$ es una forma de
WZW que satisface 
\begin{equation}
db_d=-{\cal Q} _{d+1}(K,0)
\end{equation}
localmente (recordar $d{\cal Q} _{d+1}(K,0)=\, \hbox{STr}\,(F^{\frac{d+2}{2}}=0)=0$%
). La d-forma $C_d(A,K,F)$ se define como antes 
\begin{equation}
C_d(A,K,F)=k_{01}{\cal Q} (A_t,F_t)
\end{equation}
con $A_t$ interpolando entre $A$ y $K$ como $A_t=tA+(1-t)K$.

La acci\'on DDS ec.(103) es invariante bajo transformaciones generales de
coordenadas del background y (independientemente) del volumen de mundo por
construcci\'on. Si la m\'etrica del background es plana $g_{rs}(X)=\eta
_{rs} $ entonces la acci\'on es invariante bajo transformaciones de
Poincar\'e globales.

La acci\'on DDS tambi\'en es invariante gauge. Para ver que el t\'ermino
cin\'etico es invariante notamos que $J^a_iJ^a_j=\, \hbox{STr}\, (J_iJ_j)$,
el cual es invariante porque como sabemos $J$ transforma covariantemente.
Para probar que el t\'ermino de WZW es invariante gauge se necesitan las
propiedades de transformaci\'on de $b_d$, $C_d$ y $B_d$. Se tiene $\delta
_v{\cal Q} _{d+1}(A,F)=-dQ^1_d(A,F,v)$, entonces m\'odulo formas exactas 
\begin{equation}
\delta _vb_d=Q^1_d(K,0,v)
\end{equation}
La f\'ormula de homotop\'{\i}a de Cartan con ${\cal P}( A_t,F_t )=
{\cal Q} _{d+1}(A_t,F_t ) $ y $A_t=tA+(1-t)K$ da 
\begin{equation}
{\cal Q} _{d+1}(A,F) -{\cal Q} _{d+1}(K,0)=dC_d(A,K,F) + k_{01}\, \hbox{STr}\,(F^{\frac{%
d+2}{2}})
\end{equation}
entonces, considerando que $k_{01}\, \hbox{STr}\,(F^{\frac{d+2}{2}})$ es
invariante gauge se tiene, m\'odulo formas exactas 
\begin{equation}
\delta _vC_d(A,K,F) = -Q^1_{d}(A,F,v) +Q^1_{d}(K,0,v)
\end{equation}

Finalmente se postula la regla de transformaci\'on para el campo RR 
\footnote{%
En teor\'{\i}a de cuerdas el campo RR es parte del espectro de modos cero y
su regla de transformaci\'on se deduce de la expansi\'on perturbativa.} 
\begin{equation}
\delta _vB_d=Q^1_d(A,F,v)
\end{equation}
que implica que el 'tensor de campo' del campo RR, denotado $H_{d+1}$,
definido con una correcci\'on de CS como 
\[
H_{d+1}=dB_d+{\cal Q} _{d+1}(A,F) 
\]
es invariante gauge. SE concluye que m\'odulo formas exactas $\delta _v{\cal %
B}_d=0$ y entonces el t\'ermino de WZW en la acci\'on ec.(105) es invariante
gauge, por lo que toda la acci\'on DDS es invariante gauge.

La acci\'on DDS ha sido relevante en el estudio de dualidades entre objetos
extendidos, ver p.ej. ref.\cite{duff} y referencias ah\'{\i}.\newline

\subsubsection{Teor\'{\i}a de Cuerdas y Acciones de Chern-Simons}

Varios trabajos \cite{moore1,mile,witten1,kogan} se orientaron a interpretar
las acciones de cuerdas en 1+1 dimensiones como teor\'{\i}as de CS en 2+1
dimensiones, buscando aprovechar las buenas propiedades de estas \'ultimas,
como ser teor\'{\i}as de gauge independientes de background y tener un buen
comportamiento cu\'antico.

Moore and Seiberg \cite{moore1} mostraro que una teor\'{\i}a de gauge de CS
en una variedad de dimensi\'on 2+1 con borde de dimensi\'on 1+1 induce una
teor\'{\i}a de campo conforme (CFT) en el borde. Todas las CFT Racionales
pueden construirse de ese modo, y muchos aspectos complejos de estas se
entienden facilmente a partir de la invariancia gauge y la covariancia
general de la CS. Las supercuerdas pueden pensarse como CFTs en 1+1, por lo
que Moore y Seiberg hicieron la conjetura natural de que la misma estrategia
deber\'{\i}a resultar para estas.

En su trabajo en gravedad de CS en 2+1 Witten \cite{witten1} propuso que las
teor\'{\i}as de cuerdas en una superficie de mundo de dimensi\'on 1+1
podr\'{\i}an interpretarse como modelos de CS a trav\'es de un
'engrosamiento de la superficie de mundo'.

M.B. Green prosigui\'o esta l\'{\i}nea de trabajo considerando una
teor\'{\i}a de CS en 2+1 ('volumen de mundo') con una versi\'on no
degenerada del grupo de supertraslaciones en 10D ('espaciotiempo') como
grupo de gauge, en una variedad con borde. Green sugiri\'o que el modelo se
relacionaba con las supercuerdas. Las desventajas de ese modelo son que el
t\'ermino cin\'etico debe agregarse a mano y que la acci\'on no es
invariante gauge sin fijar el gauge en el borde.

\subsection{Formas de Transgresi\'on y Acciones de Branas}

{\it Nothing is more fruitful-all mathematicians know it- than those obscure
analogies, those disturbing reflections of one theory on another, those
furtive caresses, those inexplicable discords; nothing also gives more
pleasure to the researcher}\newline
Andr\'e Weil\newline

Tomando como base los trabajos repasados arriba en refs.\cite{mn,mora} se
contruy\'o una clase de modelos de objetos extendidos en interacci\'on con
los campos de gauge de las gravedades o supergravedades de Chern-Simons , en
la que confluyen las propiedades mas atractivas de los modelos discutidos en
las secciones anteriores. Podr\'{\i}a decirse que estos modelos describen
branas propagandose en un background descrito por una gravedad de CS, pero
esto ser\'{\i}a inexacto, ya que el background no es fijo y predeterminado,
sino que es afectado por las branas, que act\'uan como fuente de los campos
de gauge. Ya se mencionaron en la Introducci\'on las ventajas de los modelos
considerados ac\'a.

La acci\'on se define por 
\begin{equation}
S=\sum_{n=0}^{N}\alpha _n \int_{S^{2n+1}}k_{01}\, \hbox{STr}\, \left(
F^{n+1} \right)
\end{equation}
donde $S^{2N+1}$ es el borde de $M^{2N+2}$, $S^{2N+1}\equiv\partial M^{2N+2}$%
, y las subvariedades $S^{2n+1}$ est\'an inmersas en $S^{2N+1}$. En caso de
que la propia $S^{2N+1}$ tenga borde $\Omega ^{2N}$, entonces $S^{2N+1}$
est\'a estrictamente inclu\'{\i}da en el borde de $M^{2N+2}$. Las
subvariedades $S^{2n+1}$ pueden tener bordes dados por subvariedades $\Omega
^{2n}$, de modo que $\partial S^{2n+1}=\Omega ^{2n}$.

Las variables din\'amicas son lo potenciales de gauge $~A_m^I$, las
coordenadas de inmersi\'on de las subvariedades $S^{2n+1}$, $%
~X^m_{(2n+1)}(\chi _{(2n+1)}^i)$, $m=0,~...,2N+1$ donde las $%
~\chi_{(2n+1)}^i $~ con $i~=~0,~...,~2n+1$~, son coordenadas locales en $%
~S^{2n+1}$ y las coordenadas de inmersi\'on de las subvariedades $\Omega
^{2n}$, $~X^m_{(2n)}(\xi _{(2n+1)}^i)$, $m=0,~...,2N+1$ donde las $%
~\xi_{(2n)}^i$~ con $i~=~0,~...,~2n$~, son coordenadas locales en $%
~\Omega^{2n}$ (por supuesto en el borde $\Omega ^{2n}$ de $S^{2n+1}$ las $X^m
$'s de el mismo punto deben coincidir como funciones de las $\chi$'s, o las $%
\xi$'s correspondientes). Note que de todas estas variedades se asume que
pueden ser no compactas, especialmente en la 'direcci\'on temporal'.

Veremos mas adelante que la consistencia de la teor\'{\i}a cu\'antica
implica que los coeficientes $\alpha _n$ est\'an cuantizados, analogamente a
lo que pasaba en gravedad de CS \cite{zanelli1}. Una elecci\'on consistente
de los coeficientes es la que sale del polinomio $P(F)=h~\, \hbox{STr}%
\,[e^{i F/(2\pi)}]$, donde h es la constante de Planck.

La acci\'on de la ec.(114) es de la forma 
\begin{equation}
S=\sum_{n=0}^N\alpha _n \int_{S^{2n+1}} {\cal T}_{2n+1} =\sum_{n=0}^N\alpha _n
\int_{S^{2n+1}} {\cal L}_{2n+1}
\end{equation}
Puede escribirse de mas explicitamente como 
\begin{equation}
S=\sum_{n=0}^N\alpha _n\left\{ \int_{S^{2n+1}} \left[ {\cal Q}_{2n+1}(F_1,A_1)
-{\cal Q} _{2n+1}(F_0,A_0)\right] -\int_{\Omega ^{2n}}C_{2n} \right\}
\end{equation}
donde $C_{2n}=k_{01} {\cal Q} _{2n+1}(A_t,F_t)$ como en las ec.(25-26).


En ref.\cite{mn} se adicion\'o un 't\'ermino cin\'etico' en el borde $\Omega
^{2n}$ de las subvariedades $S^{2n+1}$ dado por 
\begin{equation}
S_K^{(2n)}=\frac1 2 \int_{\Omega ^{2n}} d^{2n} \xi_{_{(2n)}} \sqrt{%
-\gamma_{_{(2n)}}} \left[ \gamma_{_{(2n)}}^{ij} \, \hbox{STr}\, \left( J_i
J_j \right) - (2n-2) \right]
\end{equation}
donde se introdujo la m\'etrica del volumen de mundo $\gamma_{_{(2n)}}$ .
Otra posibilidad para este t\'ermino cin\'etico es usar una expresi\'on de
la forma de Born-Infeld 
\begin{equation}
\int_{\Omega ^{2n}} d^{2n}\xi_{_{(2n)}} \, \hbox{STr}\, \left[\sqrt{-\,%
\hbox{sdet}\,\big\{ J_iJ_j+(F_0)_{ij}+(F_1)_{ij}\big\} }\right]
\end{equation}
o 
\begin{equation}
\int_{\Omega ^{2n}} d^{2n}\xi_{_{(2n)}} \, \hbox{STr}\, \left[\sqrt{-\,%
\hbox{sdet}\,\big\{ \, \hbox{STr}\, (J_i J_j) +(F_0)_{ij}+(F_1)_{ij}\big\} }%
\right]
\end{equation}
donde el superdeterminante se toma en los \'{\i}ndices curvos $~i~,~j~$ de
los pull-backs en ~$S^d$~ mientras las supertrazas se toman en los
\'{\i}ndices de grupo. En lo que sigue consideraremos estos t\'erminos
cin\'eticos como 'opcionales', y nos concentraremos en el modelo en que
estos no se incluyen.

\subsection{Invariancias de la Acci\'on}


La acci\'on de la ec.(114) y los t\'erminos cin\'eticos opcionales son
invariantes bajo transformaciones generales de coordenadas por
construcci\'on. De la invariancia gauge de las formas de transgresi\'on bajo
transformaciones de gauge, si tanto $A_0$ como $A_1$ transforman con el
mismo elemento del grupo se sigue que la acci\'on dela ec.(114) es
invariante gauge. La invariancia gauge del t\'ermino cin\'etico se deduce de
que $F$ y $J$ son covariantes gauge, la m\'etrica auxiliar en el volumen de $%
\gamma$ es invariante gauge, y de la invariancia y la propiedad c\'{\i}clica
de la traza sim\'etrica.

Si consideramos variaciones de gauge que involucren solo uno de nuestros dos
campos de gauge, manteniendo el otro fijo, la variaci\'on de la
transgresi\'on es una derivada total que puede leerse de las ecuaciones de
descenso ecs.(32-33) y es 
\begin{equation}
\delta _v I^{0}_{2n+1}(A_1,A_0)=- dI^{1}_{2n}(v,A_1,A_0)
\end{equation}
para variaciones involucrando solo $A_1$ (note que $\overline{A}_1\mid
_{\theta =0}=A_1$). La variaci\'on de la acci\'on es una suma de t\'erminos
de borde en los bordes de las branas. Para variaciones involucrando solo $%
A_0 $ vale un resultado similar.

\subsection{Ecuaciones del Movimiento}

En el caso de una teor\'{\i}a CS sin branes o bordes las ecuaciones del
movimiento $\frac{\delta S}{\delta A}=0$ est\'an dadas por ec.(65)\cite
{chamseddine,banados1,troncoso1} 
\[
\, \hbox{STr}\, (T^IF^{n})=0 
\]
En el caso con bordes y branes necesitamos usar que 
\[
\delta _1 J =\delta A_1~~~,~~~\delta _0 J = -\delta A_0 
\]
\[
\delta _r F_t = D_t ( \delta _r A_t ) =d( \delta _r A_t ) +[A_t,( \delta _r
A_t ) ]~~~,~~~r=0,1 
\]
\[
\delta _1 A_t= t \delta A_1~~~,~~~\delta _0 A_t = (1-t)\delta A_0 
\]
Por lo tanto, para variaciones de $A_1$ 
\begin{equation}
\delta _1 {\cal L}_{2n+1}=(n+1)\int_0^1dt \, \hbox{STr}\, (\delta
A_1F_t^{n}) +n(n+1)\int _0^1dt~t\, \hbox{STr}\,( JD_t(\delta A_1)F_t^{n-1})
\end{equation}
pero 
\begin{equation}
d\left[\, \hbox{STr}\, (J\delta A_1F_t^{n})\right] =\, \hbox{STr}\,
(D_tJ~\delta A_1F_t^{n-1}) -\, \hbox{STr}\, (JD_t(\delta A_1)F_t^{n-1})
\end{equation}
donde usamos $d\, \hbox{STr}\, (~~)=\, \hbox{STr}\, (D_t~~)$ y la identidad
de Bianchi $D_tF_t=0$, entonces 
\begin{eqnarray}
\delta _1 {\cal L}_{2n+1}=(n+1)\int_0^1dt \, \hbox{STr}\, (\delta
A_1F_t^{n}) +n(n+1)\int _0^1dt~t\, \hbox{STr}\,(\delta A_1 D_t(J)F_t^{n-1}) 
\nonumber \\
+d\left[n(n+1)\int _0^1dt~t\, \hbox{STr}\,( \delta A_1 J F_t^{n-1})\right]
\end{eqnarray}
El \'ultimo t\'ermino del segundo miembro es un t\'ermino de borde. Bajo
variaciones de $A_0$ tenemos 
\begin{eqnarray}
\delta _0 {\cal L}_{2n+1}=-(n+1)\int_0^1dt \, \hbox{STr}\, (\delta
A_0F_t^{n}) +n(n+1)\int _0^1dt~(1-t)\, \hbox{STr}\,(\delta A_0
D_t(J)F_t^{n-1})  \nonumber \\
+d\left[ n(n+1)\int _0^1dt~(1-t)\, \hbox{STr}\,( \delta A_0 J F_t^{n-1}) %
\right]
\end{eqnarray}
Si escribimos 
\begin{equation}
\delta _r{\cal L}_{2n+1}=\, \hbox{STr}\, (\delta A_r Q^{(r)}_{2n} ) +d\left[%
\, \hbox{STr}\, (\delta A_r R^{(r)}_{2n-1} )\right]
\end{equation}
donde 
\begin{eqnarray}
Q^{(1)}_{2n} =(n+1)\int_0^1dt F_t^{n} +n(n+1)\int _0^1dt~t D_t(J)F_t^{n-1} 
\nonumber \\
Q^{(0)}_{2n}=-(n+1)\int_0^1dt F_t^{n} +n(n+1)\int _0^1dt~(1-t)
D_t(J)F_t^{n-1}  \nonumber \\
R^{(1)}_{2n-1}=n(n+1)\int _0^1dt~t J F_t^{n-1}  \nonumber \\
R^{(0)}_{2n-1}=n(n+1)\int _0^1dt~(1-t) J F_t^{n-1}
\end{eqnarray}

Entonces podemos escribir 
\begin{equation}
\delta _r S=\sum_{n=0}^N\alpha _n \left[ \int_{S^{2n+1}}\, \hbox{STr}\,
(\delta A_r Q^{(r)}_{2n} ) +\int_{\Omega ^{2n}} \, \hbox{STr}\, (\delta A_r
R^{(r)}_{2n-1} )\right]
\end{equation}
o 
\begin{equation}
\delta _r S=\int_{S^{2N+1}}d^{{\small 2N+1}}x~ \delta (A_r)_m^I ~J^{(r)mI}
\end{equation}
donde 
\begin{equation}
J^{(r)mI}(x^m) = \sum_{n=0}^N\alpha _n\bigg[ \int_{S^{2n+1}}d^{2n+1}\chi
_{2n+1} {\cal J}_{(2n+1)}^{(r)mI} +\int_{\Omega ^{2n}}d^{2n}\xi _{2n} {\cal J%
}_{(2n)}^{(r)mI} \bigg]
\end{equation}
con 
\begin{eqnarray}
{\cal J}_{(2n+1)}^{(r)mI}= \delta ^{2N+1}(X_{(2n+1)}(\chi _{2n+1})-x^m) \, %
\hbox{STr}\, \left( T^I (Q^{(r)}_{2n}) _{m_2...m_{2n+1}} \right) \times 
\nonumber \\
\times\partial _{i_1}X^{[m}_{(2n+1)} \partial _{i_2}X^{m_2}_{(2n+1)}...
\partial _{i_{2n+1}}X^{m_{2n+1}]}_{(2n+1)} \epsilon ^{i_1...i_{2n+1}}
\end{eqnarray}
y 
\begin{eqnarray}
{\cal J}_{(2n)}^{(r)mI}= \delta ^{2N+1}(X^m_{(2n)}(\xi _{2n})-x^m) \, %
\hbox{STr}\, \left( T^I (R^{(r)}_{2n-1}) _{m_2...m_{2n}} \right) \times 
\nonumber \\
\times\partial _{i_1}X^{[m}_{(2n)} \partial _{i_2}X^{m_2}_{(2n)}... \partial
_{i_{2n}}X^{m_{2n}]}_{(2n)}\epsilon ^{i_1...i_{2n}}
\end{eqnarray}
Las ecuaciones del movimiento $\frac{\delta S}{\delta A_r}=0$ son entonces 
\begin{equation}
J^{(r)mI} = 0
\end{equation}
Estas ecuaciones pueden interpretarse como las ecuaciones halladas antes en
el caso sin bordes o branas, pero ahora con t\'erminos de fuente dados por
corrientes asociadas a las branas, que act\'uan como fuentes de los campos
de gauge y son afectadas por estos.

Respecto a las ecuaciones del movimiento correspondientes a la
extremizaci\'on de la acci''on bajo variaciones de las funciones que
describen la inmersi\'on de las branas $X$ es conveniente escribir 
\begin{eqnarray}
S=\sum_{n=0}^N\alpha _n \bigg[ \int_{S^{2n+1}} d^{2n+1}\chi _{2n+1} ~(\omega
_{2n+1})_{m_1...m_{2n+1}} \partial _{i_1}X^{[m_1}_{(2n+1)} ... \partial
_{i_{2n+1}}X^{m_{2n+1}]}_{(2n+1)}\epsilon ^{i_1...i_{2n+1}}  \nonumber \\
+\int_{\Omega ^{2n}} d^{2n}\xi _{2n}~ (\omega _{2n})_{m_1...m_{2n}}\partial
_{i_1}X^{[m_1}_{(2n)} ... \partial _{i_{2n}}X^{m_{2n}]}_{(2n)}\epsilon
^{i_1...i_{2n}} \bigg]
\end{eqnarray}
donde separamos las contibuciones del interior ('bulk') y el borde de las
branas a ${\cal L}_{2n+1}$ como en la ec.(116).

En la expresi\'on previa la dependencia de $S$ en las funciones $X$ es a
trav\'es de las $\omega$'s mientras que la dependencia de $S$ en $\partial X$
es a trav\'es de los factores que entran en la construcci\'on de los
pull-backs. Las ecuaciones de Euler-Lagrange para $X_{(p)}^s$ dan entonces 
\begin{equation}
\bigg[p\frac{\partial}{\partial X^r_{(p)}}(\omega _{(p)})_{ sm_2...m_p} -%
\frac{\partial}{\partial X^s_{(p)}}(\omega _{(p)})_{r m_2...m_p}\bigg] %
\partial _{i_1}X^{[r}_{(p)} \partial _{i_2}X^{m_2}_{(p)} ... \partial
_{i_{p}}X^{m_{p}]}_{(p)}\epsilon ^{i_1...i_{p}}=0
\end{equation}
Aplicando las ecuaciones de Euler-Lagrange para el 'bulk' $S^{2n+1}$ nos
sobra un t\'ermino de borde 
\begin{equation}
\partial _{i_1}\bigg[(\omega _{(p)})_{ sm_2...m_p} \partial
_{i_2}X^{[m_2}_{(p)} ... \partial _{i_{p}}X^{m_{p}}_{(p)}\delta X^{s]}_{(p)}
\epsilon ^{i_1...i_{p}} \bigg]
\end{equation}
Podemos requerir que el t\'ermino de borde sea cero, en analog\'{\i}a con
las cuerdas abiertas, 
\begin{equation}
(\omega _{(p)})_{ sm_2...m_p} \partial _{i_2}X^{[m_2}_{(p)} ... \partial
_{i_{p}}X^{m_{p}}_{(p)} \delta X^{s]}_{(p)} \epsilon ^{i_1...i_{p}}=0
\end{equation}
la cual no debe tomarse como condici\'on sobre que puntos pueden ser
recorridos por los bordes de las branas, sobre las velocidades de estos
puntos, o sobre las variaciones $\delta X$ permitidas, sino solo sobre las
derivadas espaciales de las funciones $X$ en el borde (una condici\'on de
coordenadas). Alternativamente se puede agregar ese t\'ermino como una
contribuci\'on extra a las ecuaciones de Euler-Lagrange en el borde. Si
agregamos los t\'erminos cin\'eticos de las ecs.(117-119) habr\'a un
t\'ermino extra en las corrientes localizadas en los bordes $\Omega ^{2n}$
de las branas, y t\'erminos extra en las ecuaciones de Euler-Lagrange. Las
ecuaciones del movimiento de las m\'etricas auxiliares $\gamma$ en los
t\'erminos cin\'eticos de la ec.(117) son algebraicas.\newline

\subsection{Conexiones con la Teor\'{\i}a de Cuerdas y la Teor\'{\i}a M}

\subsubsection{Relaci\'on con la Acci\'on de Green-Schwarz}

En la ref.\cite{mn} consideramos el modelo de ec.(114) mas un t\'ermino
cin\'etico del tipo dado antes para el grupo de la teor\'{\i}a M $~OSp(32|1)$%
~\cite{horava,townsend2,holten} y relacionamos este modelo con las
supercuerdas IIA y IIB. Como vimos este grupo tiene generadores $~P_a$
(traslaciones), $~Q_{\alpha}$ (generadores de supersimetr\'{\i}as), $~M_{ab}$%
~(Lorentz) y $~Z_{a_1...a_5}$, $a=0,...10$. Tomamos la traza sim\'etrica
como la traza est\'andar en la representaci\'on adjunta de $~G$ simetrizada.
Se toma una acci\'on de la clase considerada en la secci\'on 6.1
correspondiente a una membrana con borde, con t\'ermino cin\'etico en el
borde 
\begin{equation}
S=\int _{\Omega _2}d\sigma ^2\sqrt{-\gamma}\gamma ^{ij}STr[J_iJ_j]+\frac{1}{2%
}\int _{\Omega _3} {\cal  T}_3  ( A _0, A _1)
\end{equation}
para $~OSp(32|1)$ \footnote{%
Otras posibilidades menos simples para hacer contacto con la teor\'{\i}a de
cuerdas incluyen considerar un grupo diferente, como $~OSp(64|1)$),
t\'erminos de borde extra como los del tipo de Born-Infeld de secci\'on 6.1,
o una traza sim\'etrica diferente, como alguna extensi\'on apropiada de la
car\'{a}cter\'{\i}stica de Euler.} con $J=A_1-A_0$ y $\gamma $ una m\'etrica
auxiliar en la superficie de mundo correspondiente al borde de la membrana.

Se considera $A_1$ como gauge puro y $A_0=0$ 
\[
A_1=g^{-1}dg 
\]
con el elemento del grupo tomado de la forma 
\[
g=e^{(iX_1^aP_a+\theta ^{\alpha}_1Q_{\alpha})} 
\]
con $a=0,...,9$. Se considera una 'semiespacio' en 11D con borde en un
hiperplano de 10D, esto es una regi\'on en 11D con un borde con la
topolog\'{\i}a de $R^{10}$. Identificamos los par\'ametros de gauge $X^a$, $%
a=0,...,9$, con las coordenadas $x^a$, $a=0,...,9$ $X^a\equiv x^a$. Los
bordes en 10D se eligen en $x^{10}=0$ y $x^{10}=1$ respectivamente.
Consideramos entonces una 2-brana con borde, con este borde contenido en el
hiperplano mencionado en 10D. Las coordenadas del volumen de mundo de la
brana son $\sigma ^0$, $\sigma ^1$ $\sigma ^2$ y $x^a=X^a=X^a(\sigma
^0,\sigma ^1)$ para $a=0,...,9$.

Se requiere que $A_1$ sea gauge puro $A_1=g^{-1}dg$ 
\[
g=e^{(iX_0^aP_a+\theta ^{\alpha}_0Q_{\alpha})} 
\]
con $a=0,...,9$. En el l\'{\i}mite de Inonu-Wigner (ver ap\'endice B)
los conmutadores relevantes son 
\[
[P_a,P_b]=0 
\]
\[
\{ Q_{\alpha} ,Q_{\beta}\}=2i(C\gamma ^a)_{\alpha\beta}P_a 
\]
entonces, considerando que para una matriz $M$ se cumple que 
\[
e^{-M}\delta e^M=\delta M-\frac{1}{2}[M,\delta M]
\]
si $[[M,\delta M],M] = [[M,\delta M],\delta M]=0 $, entonces 
\[
A_1=i(dX_1^a-i\overline{\theta}_1\gamma ^a \theta _1 )P_a+d\theta
^{\alpha}_1Q_{\alpha} 
\]
Vemos que aparecen Los $\Pi ^a= dX_1^a-i\overline{\theta}_1\gamma ^a \theta
_1 $ de la secci\'on 3.2.1. sobre el modelo de Green-Schwarz.

La acci\'on para la 2-brana es 
\[
S_2=\int _{\Omega ^2}C_2 +\int _{S^3}WZW 
\]
como en las ecs.(24-26). Tenemos 
\[
\int _{\Omega ^2}C_2 = \int_{\Omega _2} d\sigma ^2 \sqrt{-\gamma}%
STr[A_{0i}A_{0j}\gamma ^{ij}] = \int_{\Omega _2} d\sigma ^2 \sqrt{-\gamma}%
\eta _{ab} \Pi _i^a \Pi _j ^b\gamma ^{ij}] 
\]
que es el t\'ermino cin\'etico de Green-Schwarz y las $\Pi$'s son las de la
ec.(103). Se requirieron las supertrazas sim\'etricas $\, \hbox{STr}\, (P P) 
$, $\, \hbox{STr}\, (P Q)$~ and $\, \hbox{STr}\, (Q Q )$. Para el t\'ermino
WZW las trazas relevantes son $\, \hbox{STr}\, (P P P) $, $\, \hbox{STr}\,
(P P Q) $, $\, \hbox{STr}\, (Q Q P) $~ y $\, \hbox{STr}\, (Q Q Q)$. Entre
estas, las no nulas despues de la contracci\'on de Inonu-Wigner $%
P_a\rightarrow RP_a$, $Q_{\alpha}\rightarrow \sqrt{R}Q_{\alpha}$ y $%
R\rightarrow\infty$ se normalizan como 
\[
\, \hbox{STr}\, (P_a P_b) = \eta _{a b}~~,~~~~ \, \hbox{STr}\, (P_a Q
_{\alpha}Q_{\beta} ) = \left(\gamma _a~C^{-1}\right) _{\alpha\beta} 
\]
El t\'ermino de WZW surge de que la transgresi\'on se reduce al Chern-Simons
para $A_0=0$, y de ah\'{\i} al WZW para $A_1$ gauge puro.  Para el t\'ermino
de WZW se tiene 
\[
WZW=-\frac{1}{3}\int STr [( g^{-1}_1dg )^3]
\]
que con la forma del $g$ dado y las trazas consideradas es exactamente el
WZW de Green-Schwarz.

En el trabajo original de Green-Schwarz el coeficiente relativo se fijo por
simetr\'{\i}a $\kappa $. Ser\'{\i}a interesante entender como esta
simetr\'{\i}a aparece en nuestro modelo. Una posibilidad es que sea lo que
queda de la parte fermi\'onica de nuestra simetr\'{\i}a de gauge. Mas
concretamente, se podr\'{\i}a pensar en repetir la construcci\'on anterior
para una m\'etrica auxiliar no plana arbitraria en la superficie de mundo,
entonces luego de una transformaci\'on fermi\'onica de gauge recuperar las
condiciones de dualidad o antidualidad haciendo a transformaci\'on de esta
m\'etrica que corresponder\'{\i}a a la simetr\'{\i}a $\kappa $.\newline

Se mostr\'o entonces que la acci\'on propuesta contiene las configuraciones
de la acci\'on de Green-Schwarz (en el gauge en que la m\'etrica de la
superficie de mundo es plana) como parte de su espacio de congiguraciones.
La acci\'on propuesta ac\'a para configuraciones gen\'ericas (no truncadas a
formas especiales de los campos de gauge como en esta subsecci\'on) es sin
embargo invariante gauge e independiente de background.\newline

\subsubsection{Relaci\'on con el Modelo DDS de Branas Heter\'oticas}

Es posible relacionar el modelo DDS para el acoplamiento de campos de
Yang-Mills a branas, repasado en la secci\'on 3.2.2., a trav\'es de los paso
siguientes:

(a) Considerar la acci\'on de la ec.(114) mas un t\'ermino cin\'etico como
el de la ec.(117), con grupo de gauge dado por el producto del grupo de
Poincar\'e y el grupo de gauge 'interno' (con los generadores de uno
conmutando con los del otro).

(b) Tomar el caso particular de un $A_0=K$ gauge puro y un $A_1$ con
componentes de Poincar\'e (los que multiplican los generadores de
Poincar\'e) iguales a cero, y componentes en los generadors del grupo de
gauge interno arbitrarias.

(c) Hacer una reducci\'on dimensional doble (en el background y en el bulk
de la brana) asumiendo que todos los campos son independientes de una
coordenada del background, la cual se identifica con una coordenada en el
bulk de la brana. El borde de la brana se asume contenido en las dimensiones
no compactificadas.

(d) Si $K={\cal G}^{-1}d{\cal G}$ con ${\cal G}=g_{spacetime}g_{internal}$ y 
$g_{spacetime}=exp[i\hat{X}^rP_r]$, $r\neq D-1$ ($D-1$ es la dimensi\'on
compactificada), identificar las coordenadas espaciotemporales $X^r$ en las
dimensiones no compactificadas con los par\'ametros de gauge $\hat{X}^r$
(una elecci\'on de coordenadas).

Mirando las ecs.(116-117) se ve que la acci\'on DDS con la brana DDS siendo
el borde de la brana de la que partimos se recupera luego del proceso
descrito. La forma de WZW para $K$ tiene solo componentes seg\'un el grupo
de gauge interno (las trazas que podr\'{\i}an dar un WZW espaciotemporal
para $K$ son cero) y el campo-RR esta dado como un campo compuesto en
t\'erminos de las formas de CS para $A_1$ dimensionalmente reducidas. Se
us\'o $\, \hbox{STr}\, [P_rP_s]\approx\eta _{rs}$ , la que no es realmente
una traza en una representaci\'on matricial del grupo de Poincar\'e, ya que
este tiene representaciones de dimensi\'on infinita, pero podemos eludir
este problema bien tomando el grupo dS o AdS con un 'radio' (el par\'ametro
en la contracci\'on de Inonu-Wigner) muy grande y viendo la acci\'on DDS
como una aproximaci\'on, o bien simplemente definiendo el tensor invariante
de ese modo.\newline

\subsubsection{D-branas y Teor\'{\i}a K}

Algunos de los modelos de 'D-branas' tienen acciones con alguna similitud
con la nuestra con un t\'ermino cin\'etico, por ejemplo los modelos de
Douglas \cite{douglas} y de Green, Hull y Townsend \cite{green3} (aunque en
ese trabajo el t\'ermino cin\'eticose agrega en el bulk). En esos trabajos
sin embargo los grupos espaciotemporal e interno se mantienen separados, el
background es fijo (al contrario que en nuestro modelo donde el background
es din\'amico e interact\'ua con las branas), y existen campos-RR para
asegurar la invariancia gauge (como en el modelo DDS).

Varios trabajos relativamente recientes tratan de la relaci\'on de las
D-branas con la teor\'{\i}a \cite{minasian,witten3}. La situaci\'on descrita
en esos art\'{\i}culos puede resumirse en el enunciado de que "la
teor\'{\i}a K debe preferirse a la cohomologi\'{\i}a" o equivalentemente
"formulaciones en t\'erminos de campos de gauge deben preferirse a
formulaciones en t\'erminos de p-formas, o campos-RR". En nuestro modelo
podemos reproducir campos-RR, como se menciona en la subsecci\'on previa,
como compuestos con las formas CS de uno de los potenciales de gauge (p.ej. $%
A_1$) el cual se acopla (con lo que usualmente se llama 'acoplamiento
an\'omalo') al otro ($A_0$). La 'regla de transformaci\'on an\'omala' de ese
campo-RR compuesto es entonces autom\'atica.

Notese que el mapa est\'andar entre las clases de teor\'{\i}a K de fibrados
sobre una variedad diferenciable y las clases de cohomolog\'{\i}a sobre la
variedad est\'a dada por los caracteres de Chern, los cuales aparecen en la
acci\'on de ec.(137). Tambi\'en la duplicaci\'on de campos que aparece en
nuestro modelo es una propiedad de teor\'{\i}a K.

Estas observaciones parecen sugerir que nuestra clase de modelos es mas
fundamental que los modelos de D-branas mencionados, al tener en forma
expl\'{\i}cita propiedades sugeridas impl\'{\i}citamente en estos.

\subsection{Propiedades Cu\'anticas}

\subsubsection{ Acci\'on Efectiva Cu\'antica}

La teor\'{\i}a cu\'antica se define formalmente a trav\'es de la integral de
caminos 
\[
Z=\sum_{topologies}\sum_p\int{\cal D}A~{\cal D}X_{(p)}~e^{iS/\hbar} 
\]

donde se entiende que se debe sumar sobre todas las configuraciones de los
campos de gauge y sobre todas lasgeometr\'{\i}as y topolog\'{\i}as de las
branas y edel espacio base.

Est\'a claro que la teor\'{\i}a incluye configuraciones con muchas branas,
ya que la suma incluye cnfiguraciones con cualquier n\'umerode partes no
conectadas para branas de cualquier dimensionalidad posible. Esto tamb\'en
es cierto a nivel cl\'asico, donde pueden considerarse soluciones con
cualquier n\'umero de branas.\newline

Como en teor\'{\i}as de gauge y gravedad de Chern-Simons la consistencia de
la teor\'{\i}a cu\'antica lleva al requerimiento de que las constantes en la
acci\'on est\'an cuantizadas. Podemos considerar branas sin bordes, con uno
solo de los campos de gauge $A_0$ o $A_1$ no nulo. Entonces se puede usar un
argumento similar al de la secci\'on 3.1.4 para probar que las constantes
est\'an cuantizadas. Se puede considerar la brana recorriendo un camino
cerrado en el espaciotiempo,y entonces contraer este camino a cero, de modo
que la brana no se mueve. El camino es una subvariedad del espaciotiempo de
dimensi\'on p+1 para una p-brana (con p par), la cual es el borde de
infinitas subvariedades de dimensi\'on p+2. Cuando el camino se contrae a
cero la variedad de dimensi\'on p+2 se vuelve cerrada. La amplitud
cu\'antica para ese camino contra\'{\i}do debe ser 1, ya que la brana no se
mueve. Debe ser tambi\'en $exp(iS/\hbar)$. Se sigue que $S=2\pi m$ con m
entero. Pero 
\[
S=k\int _{S^{p+1}}{\cal Q } _{p+1} = k\int _{\Omega ^{p+2}} STr[F ^{p/2+1}] 
\]
Entonces 
\[
k\int _{\Omega ^{p+2}} STr[F ^{p/2+1}]=2\pi m\hbar 
\]
Si $\int _{\Omega ^{p+2}} STr[F ^{p/2+1}]$ es proporcional a un entero
debido al teorema de \'{\i}ndice (ver secci\'on 2.1.6. p.ej. el n\'umeo de
Chern es $i^n/(2\pi )^n\int _{\Omega ^{p+2}} STr[F ^{p/2+1}]=l$) se sigue
entonces que k debe estar cuantizado.\newline

Un corolario interesante es que si $\int _{\Omega ^{p+2}} STr[F ^{p/2+1}]$
es no nulo hay un flujo no trivial a trav\'es de la subvariedad $\Omega
^{p+2}$, se\~nalando la presencia de una d-p-4-brana solit\'onica
'magn\'etica'(al contrario que las branas originales fundamentales
'el\'ectricas') rodeada por $\Omega ^{p+2}$. Esto implica que a nuestras
branas fundamentales con volumen de mundo de dimensi\'on impar corresponden
branas solit\'onicas con volumenes de mundo de dimensi\'on par. P.ej. hay
una 5-brana solit\'onica asociada a nuestra 2-brana fundamental en 11D .%
\newline

La integraci\'on de caminos dada arriba es claramente solo formal a este
nivel. Ser\'{\i}a deseable desarrollar los detalles t\'ecnicos como
procedimientos para evitar redundancias al sumar sobre configuraciones de
los campos de gauge y m\'etodos de regularizaci\'on. Sin embargo como en el
caso de teor\'{\i}as de Chern-Simons podemos conjeturar que la acci\'on
cl\'asica es ya la acci\'on efectiva cu\'antica, basandonos en argumentos
similares, relacionados con el teorema de Adler-Bardeen de no
renormalizaci\'on de las anomal\'{\i}as.

Como se mencion\'o en el caso de las teor\'{\i}as de CS, tambi\'en escierto
para nuestro modelo de branas que el formalismo matem\'atico es
estrechamente an\'alogo al que se usa en el estudio de las anomal\'{\i}as.
Se podr\'{\i}a decir en cierto modo que la acci\'on es en si misma una pura
anomal\'{\i}a.

Otro aspecto de la relaci\'on entre las anomal\'{\i}as y nuestros modelos en
el contexto de la discusi\'on al final de la secci\'on 3.1.4. y el hecho de
que las anomal\'{\i}as 'tienen el mismo aspecto' en todas las escalas (no
renormalizaci\'on). Esto \'ultimo implica que la estructura de
anomal\'{\i}as de una teor\'{\i}a efectiva dada debe ser la misma que la de
la teor\'{\i}a microsc\'opica correspondiente, una propiedad conocida como
'condici\'on de compatibilidad de anomal\'{\i}as de 't Hooft'\cite{'t hooft}%
. El hecho de que la teor\'{\i}a efectiva 'recuerda' estas caracteristicas
de la teor\'{\i}a microsc\'opica debe dar indicios sobre la forma de esta
\'ultima, si se conoce la primera, como sugiri\'o Stelle \cite{stelle}
(quien llam\'o a esta propiedad 'atavismo' ). Si el modelo de ec.(137)
corresponde a la teor\'{\i}a M, deber\'{\i}a haber una correspondencia entre
este y sus propiedades de transformaci\'on y los t\'erminos an\'omalos y
reglas de transformaci\'on an\'omalas del l\'{\i}mite de bajas energ\'{\i}as
de esta teor\'{\i}a, que es la supergravedad est\'andar en 11D.\newline

\subsubsection{Relaci\'on con la Teor\'{\i}a M a Nivel Cu\'antico}

Si nuestro modelo en once dimensiones con grupo $~OSp(32|1)$ tiene como
casos l\'{\i}mite las cinco teor\'{\i}as consistentes de supercuerdas,
entonces las consideraciones de consistencia y cancelaci\'on de
anomal\'{\i}as de estas \'ultimas deber\'{\i}an reflejarse en que la primera
es la \'unica de nuestra clase de modelos que esconsistente. Podemos hacer
un argumento heur\'{\i}stico en el sentido de que una teor\'{\i}a
completamente consistente de la naturaleza deber\'{\i}a tener un desarrollo
perturbativo 'suave' en torno a todo 'punto' de su espacio de fases, en el
sentido de que cada orden sea finito, a\'un si el punto no es un
'vac\'{\i}o' de la teor\'{\i}a y en ese caso la serie completa no converge.

La versi\'on cu\'antica de nuestro modelo ofrece otro modo de relacionar
modelos de CS con supergravedad est\'andar. Las idea es que como tenemos
configuraciones correspondientes a supercuerdas en el espacio plano, podemos
sumar las contribuciones a la integral de caminos de estas configuraciones y
tomar prestado el argumento de teor\'{\i}a de cuerdas acerca de que
talcondensado de cuerdas corresponde a bajas energ\'{\i}as a diferentes
supergravedades est\'andar. Aunque uno no puede esperar que el resultado de
esta suma parcial de lugar a un backround que sea un verdadero vac\'{\i}o
(soluci\'on de las ecuaciones del movimiento de la acci\'on efectiva
cu\'antica) de la teor\'{\i}a completa.

Al hacer esta suma parcial podemos agregar una contribuci\'on que no sea
gauge puro en la placa de secci\'on 4.2.1. y reducirla dimensionalmente
asumiendo que el potencial es independiente de las coordenadas a trav\'es de
la placa. El resultado \cite{mn} es un t\'ermino en la acci\'on de cuerdas
que se ve como el dilat\'on por el escalar de curvatura de la superficie de
mundo,con el dilaton dado por la und\'ecima componente del vielbein y la
curvatura dada por el pull-back de la espaciotemporal. La integral de la
curvatura en 2D es el n\'umero de Euler, por lo que podemos ordenar la suma
parcial de estas configuraciones en la integral de caminos seg\'un el
g\'enero de la superficie de mundo y la potencia correspondiente del
dilat\'on, el cual se interpreta como la constante de acoplamiento. Como el
valor del dilat\'on corresponde a la und\'ecima componente del vielbein,
podemos decir que la constante de acoplamiento en el desarrollo perturbativo
de las cuerdas est''a directamente relacionado con el tama\~no de la
und\'ecima dimensi\'on, comose hab\'{\i}a encontrado en el estudio de las
dualidades en la teor\'{\i}a M (ver \cite{townsend1}).\newline

\subsubsection{Vac\'{\i}o y Fenomenolog\'{\i}a}

Si la acci\'on es ya la acci\'on efectiva cu\'antica, como se sugiri\'o, el
problema de hallar un 'vac\'{\i}o' se reduce a encontrar una soluci\'on de
las ecuaciones del movimiento cl\'asicas (sin necesidad de buscar
correcciones cu\'anticas a esta). Hacer fenomenolog\'{\i}a requiere
encontraruna soluci\'on realista, en el sentido de tener cuatro dimensiones
espaciotemporales 'grandes' aproximadamente planas con signatura 3+1 (al
menos en cierta etapa de la evoluci\'on c\'osmica). Las masas y constantes
de acoplamiento de la f\'{\i}sica de part\'{\i}culas podr\'{\i}an leerse de
los coeficientes de los t\'erminos de menor orden de una expansi\'on
perturbativa alrededor de este vac\'{\i}o (o background). Ver ref.\cite
{brandenberger} y referencias ah\'{\i} por trabajos recientes en modelos
cosmol\'ogicos para teor\'{\i}as con t\'erminos de mayor orden en la
curvatura y como problemas de modelos mas est\'andar se resuelven en este
contexto. Un punto importante es que la acci\'on original no solo contiene
constantes adimensionadas, las cuales adem\'as est\'an cuantizadas, por lo
que todas las constantes dimensionadas que aparezcan deben aparecer
din\'amicamente, asociadas a un vac\'{\i}o determinado, por una suerte de
'transmutaci\'on dimensional'. Por ejemplo el tama\~no de una dimensi\'on
compactificada proporcionar\'{\i}a una constante dimensionada, la cual
aparecer\'{\i}a en las expresiones para masas y constantes de acoplamiento
de peque\~nas perturbaciones alrededor de ese background. La
renormalizaci\'on (o dependencia con la escala) de estas constantes
podr\'{\i}a leerse en la dependencia expl\'{\i}cita de estas en la escala (o
longitud de onda) de las perturbaciones.

La dependencia de las constantes en el vac\'{\i}0 que se considere es
an\'aloga a la dependencia de las frecuencias normales de las peque\~nas
oscilaciones de un sistema alrededor de un m\'{\i}nimo de un potencial con
muchos m\'{\i}nimos en este m\'{\i}nimo.

Se debe subrayar que estamos usando dos nociones diferentes, pero
compatibles, de cuantizaci\'on. Por un lado tenemos la teor\'{\i}a
cu\'antica completa, la cual da amplitudes entre dos configuraciones
diferentes en t\'erminos de una integral de caminos entre configuraciones
arbitrarias como una suma sobre todas las geometr\'{\i}as y topolog\'{\i}as
interpolando entre estas, la cual es an\'aloga a la 'Teor\'{\i}a de Campos
de Cuerdas', pero a\'un m\'as dif\'{\i}cil debido a las dimensiones mas
altas de las branas. Por otro lado tenemos las peque\~nas perturbaciones
alrededor de un vac\'{\i}o, lo cual tiene sentido porque nosotros 'vivimos
en este' y podemos ver solo peque\~nas perturbaciones alrededor de este
vac\'{\i}o. Deber\'{\i}a haber una cierta amplitud (idealmente peque\~na) de
trancisi\'on a otros vac\'{\i}os o configuraciones lejanas a nuestro
vac\'{\i}o, pero no podemos detectarlos porque no 'existimos' en estas
'ramas cu\'anticas'. Fenomenologicamente solo procesos de muy alta
energ\'{\i}a, como los dados en agujeros negros y cosmolog\'{\i}a
requerir\'{\i}an un apartamiento de la aproximaci\'on de peque\~nas
perturbaciones alrededor de un background. Por supuesto el requerimiento de
finitud debe valer para la teor\'{\i}a completa, y no solo un sector de esta.

Para modelos 'de juguete' como modelos de CS en 2+1 tiene sentido, y es
posible, considerar la teor\'{\i}a completa como en \cite{witten1}, pero
esto ser\'{\i}a imposible para teor\'{\i}as mas complejas.

Otro problema interesante en el estudio de soluciones de nuestra clase de
teor\'{\i}as tiene que ver con el trabajo de Aros et al. \cite{aros}. En
este se consideraro soluciones tipo 'branas negras' a las ecuaciones de las
supergravedades de Chern-Simons. Uno puede conjeturar que membranas
fundamentales de CS como las que se consideran ac\'a pueden proporcionar una
teor\'{\i}a efectiva para perturbaciones de longitudes de onda largas de
estas 'branas negras', de modo an\'alogo a como las supermembranas son
teor\'{\i}as efectivas para soluciones tipo branas negras ('solit\'onicas')
de las supergravedades est\'andar \cite{supermembrane}.

Finalmente debe decirse que el significado de la aparici\'on de dos campos
de gauge cuando se usan formas de transgresi\'on en vez de formas de CS es
oscuro a nivel fenomenol\'ogico. Es tentador comparar la situaci\'on con la
de la teor\'{\i}a de cuerdas heter\'oticas con grupo de gauge $E_8\times E_8$%
, con su sector de 'materia normal' y su 'sector oculto'. Sin embargo la
situaci\'on es bastante diferente, ya que los sectores de las cuerdas
heter\'oticas interact\'uan gravitacionalmente, mientras que en nuestro
modelo todas las interacciones vienen de t\'erminos de borde, de modo que
silas branas no tienen bordes, entonces los dos sectores no interact\'uan en
absoluto. El estudio de soluciones concretas de branas con bordes, o quiz\'a
de la teor\'{\i}a inducida en el borde cuando ambos campos son gauge puro,
deber\'{\i}a ayudar a esclarecer este punto.

\newpage

\section{Discusi\'on y Conclusiones}

{\it  Si el Se\~nor me hubiera preguntado, le hubiera sugerido algo mucho mas simple.}\\
 Comentario del Rey Alfonso X  el Sabio, al conocer el sistema astron\'omico de Tolomeo\\

Como ya dijimos antes, la evoluci\'on de la teor\'{\i}a de campos y la
f\'{\i}sica de part\'{\i}culas durante las \'ultimas d\'ecadas nos ha
ense\~nado que la invariancia gauge es el principio subyacente a las
teor\'{\i}as que describen tres de las cuatro interacciones fundamentales.
Este principio tiene un alcance que va mucho mas all\'a de su origen en la
teor\'{\i}a cl\'asica del campo electromagn\'etico, y es esencial para la
consistencia cu\'antica del modelo est\'andar.

Por otro lado, el desarrollo de la teor\'{\i}a cu\'antica de campos dio
lugar a tres grandes sorpresas:

(i) Renormalizaci\'on: la necesidad de renormalizar la teor\'{\i}a para
obtener predicciones finitas y la  renormalizaci\'on de las constantes
f\'{\i}sicas, las cuales dependen de la escala (de longitud, tiempo o
energ\'{\i}a, equivalentes en unidades naturales), 

(ii) Ruptura espont\'anea de simetr\'{\i}a, 

(ii) Anomal\'{\i}as: asociadas a la violaci\'on a nivel cu\'antico de leyes
de conservaci\'on cl\'asicas, las cuales han tenido valor predictivo en
casos concretos tanto para anomal\'{\i}as quirales como de gauge. Su
estructura matem\'atica est\'a profundamente relacionada con las clases
caracter\'{\i}sticas de fibrados (ver p. ej. ref.\cite{alvarez}).\\ 

Mientras que la ruptura espont\'anea de simetr\'{\i}a probablemente no es
fundamental y el campo de Higgs posiblemente solo un campo efectivo , como
sucede en superconductividad, donde el campo de Higgs corresponde a los
pares de Cooper, creo que los otros puntos son pistas importantes en la
b\'usqueda de una eventual teor\'{\i}a unificada. Estas pistas se recogen en
la construcci\'on de los modelos discutidos en este trabajo, los cuales
consisten en teor\'{\i}as de gauge independientes de background incluyendo
la gravitaci\'on. Las anomal\'{\i}as aparecen en que la forma matem\'atica
de las teor\'{\i}as consideradas, la \'unica consistente con la invariancia
gauge y la independencia de background es tal que uno podr\'{\i}a decir,
ir\'onicamente, que las teor\'{\i}as son una pura anomal\'{\i}a. Respecto a
la renormalizaci\'on, como consecuencia de la ausencia de contrat\'erminos y
la cuantizaci\'on de las constantes se espera que estas teor\'{\i}as no
reciban correcciones cu\'anticas, propiedad correspondiente al teorema de
Adler-Bardeen para las anomal\'{\i}as.\newline

Hemos visto como el pasar de formas de Chern-Simons a transgresiones,
reemplazando acciones cuasi-invariantes gauge por acciones estrictamente
invariantes, proporciona adem\'as (en virtud del principio de
gauge) una prescripci\'on general para los t\'erminos de borde y
regularizaci\'on de la acci\'on que da las cargas conservadas y la
entrop\'{\i}a correctas.\newline

Una importante cuesti\'on, que deber\'{\i}a analizarse en el futuro, tiene
que ver con el significado f\'{\i}sico y la importancia del segundo campo de
gauge. En el caso de gravitaci\'on la prescripci\'on de variedad cobordante
para el segundo campo es la que permite apartarse lo menos posible,
agregando un m\'{\i}nimo de estructura adicional, de las acciones de
Chern-Simons puras. Uno podr\'{\i}a sentirse tentado a detenerse ah\'{\i}, y
considerar solamente configuraciones de este tipo, considerando el segundo
campo como no f\'{\i}sico. Sin embargo esto no parece natural, ya que nada
hay en el formalismo que marque un campo como f\'{\i}sico y el otro como
auxiliar. Considerar solo estas configuraciones parece an\'alogo a disponer
del formalismo del an\'alisis vectorial pero limitarse a considerar campos
seg\'un la direcci\'on z, por ejemplo. Creo que la configuraci\'on de
variedad cobordante deber\'{\i}a verse  elecci\'on especialmente conveniente,
 entre muchas posibles.

Dado que las ecuaciones del movimiento, en el caso de teor\'{\i}as de campos
(sin objetos extendidos) son las mismas de Chern-Simons, una situaci\'on
interesante ser\'{\i}a estudiar un caso en que ambos campos interact\'uan,
por ejemplo el caso en que ambos son gauge puro y viven en una variedad con
borde, ya que en el borde se induce una acci\'on correspondiente a dos
acciones de WZW restadas y un t\'ermino de interacci\'on (que viene de $C_{2n}$). 
En el caso de objetos extendidos los bordes de las branas se acoplan a
ambos campos.\newline

Las acciones de branas construidas, adem\'as de ser invariantes bajo transformaciones de
gauge e
independientes de background tienen una forma sugestiva en el siguiente
sentido: los niveles de estructura que se pueden dar a una variedad son en
orden de precedencia topolog\'{\i}a, estructura diferencial y m\'etrica. Uno
de los principales logros de la Relatividad General fue hacer la m\'etrica
parte de la din\'amica, en vez de ser un escenario fijo para los dem\'as
fen\'omenos f\'{\i}sicos. Hacer lo mismo con la estructura diferencial,
considerando los diferenciales $dx ^{m}$ objetos f\'{\i}sicos que
anticonmutan, como en la ref.\cite{witten4}, considerando funciones
gen\'ericas de estos objetos, que son polinomios truncados, como
lagrangianas llevar\'{\i}a a acciones con objetos extendidos de diversas
dimensiones, como las consideradas en este trabajo surgiendo de la suma
formal  de formas diferenciales involucrada en los polinomios 
caracter\'{\i}sticos \footnote{%
Esta imagen se parece algo a la idea de Thorn sobre teor\'{\i}a de cuerdas
conocida como 'string bits approach' y recuerda la frase de Witten sobre
esta teor\'{\i}a "...to do justice to such a theory, one needs building
blocks more graceful than big, floppy strings"\cite{witten5}.}. Requerir que
el modelo sea invariante gauge restringe mucho las formas posibles de la
acci\'on, como hemos visto.\newline

Entre los problemas interesantes a estudiar en el futuro est\'an la
b\'usqueda de otras conexiones con la teor\'{\i}a de supercuerdas, en
particular como se traducen en el contexto de nuestros modelos los
argumentos que en las supercuerdas llevan a un n\'umero muy peque\~no de
teor\'{\i}as consistentes ('traducir' la cancelaci\'on de anomal\'{\i}as
parece prometedor); y la b\'usqueda de posibles conexiones con el efecto
Hall cu\'antico (QHE) en dimensiones mas altas (que dos) \cite{zhang} , dado
que la teor\'{\i}a de campos efectiva de este fen\'omeno esta relacionada
con la de Chern-Simons \cite{zhang1, zhang11}, siendo en el caso de
dimensiones m\'as altas muy similar a los modelos discutidos aca, ya que
consiste en branas de Chern-Simons en un background de Chern-Simons.\newline

{\it "Nunca persegu\'{\i} la gloria}

{\it ni dejar en la memoria}

{\it de los hombres mi canci\'on;}

{\it yo amo los mundos sutiles,}

{\it ingr\'avidos y gentiles}

{\it como pompas de jab\'on.}

{\it Me gusta verlos pintarse}

{\it de sol y grana, volar}

{\it bajo el cielo azul, temblar}

{\it s\'ubitamente quebrarse."}

-Antonio Machado\footnote{
Siendo esta la \'ultima de varias citas incluidas en este trabajo, debo
referir al lector al entretenido art\'{\i}culo por Peter Rodgers titulado
'Who said that', publicado en el n\'umero de junio del 2002 de la revista Physics
World, el cual puede encontrarse en el sitio web www.physicsweb.org. El
se\~nor Rodgers tiene algunas duras palabras para el h\'abito de los
f\'{\i}sicos de abusar de las citas literarias.}\newline

{\bf Agradecimientos}\newline
Estoy agradecido a mi colaborador en uno de los art\'{\i}culos en que este
trabajo se basa, Hitoshi Nishino, por muchas discusiones.\\
Estoy muy agradecido a
Rodrigo Olea, Ricardo Troncoso y Jorge Zanelli, por muchas discusiones en las cuales
aprend\'{\i} un mont\'on, y una muy disfrutable y fruct\'{\i}fera
colaboraci\'on. A Jorge debo agradecerle por su paciente y sabio desempe\~no
como mi Orientador y por su permanente apoyo.\\
Estoy muy agradecido a mi Co-Orientador Rodolfo Gambini por su permanente apoyo.\\
Agradezco la c\'alida hospitalidad de los miembros del Centro de Estudios Cient\'{\i}ficos
CECS de Valdivia, Chile,  en mis  varias visitas durante las cuales se concret\'o esta tesis.\\
Agradezco el apoyo econ\'omico del CONICYT-Uruguay y de la Universidad de la Rep\'ublica
durante el per\'{\i}odo de mis estudios de doctorado en el que estube en
los Estados Unidos.\\
Agradezco el apoyo econ\'omico recibido de la Iniciativa Cient\'{\i}fica Milenio-Chile, del Proyecto
FONDENCYT-Chile N$^o$ 7010450 de R. Troncoso, y del International Center of Theoretical Physics 
de Trieste ICTP, apoyo que hizo posibles mis vistas a Valdivia. 

\newpage

\section{\bf Ap\'endices}

\subsection{Ap\'endice A. Complemento en fibrados y campos de gauge.}

\subsubsection{Propiedades generales}

 
Las trazas de productos de formas diferenciales matriciales satisfacen la
propiedad c\'{\i}clica 
$$
tr[ \Sigma _q\Lambda _p ]=(-1)^{pq}tr[\Lambda _p \Sigma _q]
$$
Una importante propiedad es 
$$
d\, \hbox{STr}\, (\Omega )=\, \hbox{STr}\, (D~\Omega )
$$
donde $\Omega $ es una forma cualquiera 
con \'{\i}ndices en el grupo.
De esta y de la identidad de Bianchi resulta 
$$
dP(F)=0~~~,~~~d\, \hbox{STr}\, \big( F^{n+1} \big)=0
$$
de donde $P(F)$ y $\, \hbox{STr}\, \big( F^{n+1} \big)$ son formas
localmente exactas.\newline


\subsubsection{Transformationes de Gauge}


Bajo transformaciones de gauge los potenciales cambian como 
$$
A_r^g=g^{-1}(A_r+d)g~~,~~r=0,1
$$
donde $g(x)\equiv exp[v_I(x)T^I]$ es un elemento del grupo. Se sigue que $%
~J\equiv A_1-A_0$~ transforma covariantemente si tanto $A_1$ como $A_0$
transforman con el mismo $g$ 
$$
J^g=g^{-1}Jg
$$
Tambi\'en 
$$
F^g=g^{-1}Fg
$$
Para un polinomio invariante, por definici\'on 
$$
P(F^g)=P(g^{-1}Fg)=P(F)
$$

La variedad diferencial de base esta descrita en general por un conjunto de
cartas locales de coordenadas $U_i$. Los campos de gauge en dos cartas
locales con intersecci\'on no vac\'{\i}a est\'an relacionados en la
intersecci\'on de las cartas por una transformaci\'on de gauge 
$$
A_{U_j}=t^{-1}_{ij}(A_{U_i}+d)t_{ij}
$$
La informaci\'on sobre la topolog\'{\i}a del fibrado est\'a contenida en las
'funciones de transici\'on' $t_{ij}$.

Un campo de gauge se dice que es {\it gauge puro} si se puede hacer cero en una
carta local cualquiera por una transformaci\'on de gauge (a\'un cuando esto
puede no ser posible en general en todas las cartas locales
simult\'aneamente). Un campo gauge puro es de la forma 
$$
A_{gauge~puro}\equiv V=g^{-1}dg
$$
La forma $V$ se llama {\it forma de Maurer-Cartan (MC) invariante por la
izquierda}. Es invariante por la izquierda en el sentido de que si
reemplazamos $g\rightarrow g_0g$, donde $g_0$ es un elemento del grupo de
gauge independiente de $x$, entonces $V$ no cambia. Las formas de MC
satisfacen la {\it ecuaci\'on de Maurer-Cartan} 
$$
dV+V^2=0
$$
Se sigue que $F_{gauge~puro}=0$. El rec\'{\i}proco tambi\'en es verdadero,
esto es: si $F=0$ entonces $A$ es gauge puro.\newline


\subsubsection{Operador y F\'ormula de Homotop\'{\i}a de Cartan}


Sea  $A_t$ la interpolaci\'on entre dos potenciales de gauge $%
A_0$ y $A_1$, 
$$
A_t=t A_1+(1-t)A_0~~,~~F_t=d A_t+A_t^2 ~~.
$$
El {\it operador de Homotop\'{\i}a de Cartan} $k_{01}$ act\'ua sobre polinomios $%
{\cal P}(F_t,A_t)$ y se define como 
$$
k_{01}{\cal P}(F_t,A_t) =\int_0^1 dt~l_t{\cal P}(F_t,A_t) ~~,
$$
donde la acci\'on del operador~$l_t$~ en polinomios arbitrarios de $A_t$ y $%
F_t $ es definida a trav\'es de 
$$
l_t A_t=0~~,~~~~ l_t F_t = A_1 - A_0\equiv J ~~,
$$
y la convenci\'on de que ~$l_t$ act\'ua como una antiderivaci\'on 
$l_t(\Lambda _p\Sigma _q) = (l_t\Lambda
_p)\Sigma _q +(-1)^p \Lambda _p(l_t\Sigma _q) $, donde $\Lambda _p$ y $\Sigma _q$ son 
p y q-formas (funciones de A y de F) respectivamente.

Se puede verificar directamente la relaci\'on
$$
\big( l_td+dl_t\big) {\cal P}(F_t,A_t) =\frac{\partial}{\partial t}{\cal P}%
(F_t,A_t)
$$
la cual se puede integrar entre 0 y 1 en $t$ para obtener la
{\it f\'ormula de homotop\'{\i}a de Cartan} 
$$
\big( k_{01}d+dk_{01}\big) {\cal P}(F_t,A_t) ={\cal P}(F_1,A_1)-{\cal P}%
(F_0,A_0) ~~.
$$
 
Para variaciones arbitrarias $\delta A$ uno puede definir la
antiderivaci\'on $l$ (correspondiente a $dt~l_t)$ actuando como $lA=0$ y $%
lF=\delta A$. Entonces $ld+dl=\delta$ en polinomios en $A$ y $F$.\newline


\subsubsection{Formas de Transgresi\'on y de Chern-Simons}


La {\it forma de transgresi\'on} ${\cal T}_{2n+1}(A_1,F_1,A_0,F_0)$ se define como 
$$
{\cal T}_{2n+1}(A_1,A_0)\equiv k_{01}\, \hbox{STr}\, \left(F_t^{n+1}\right)=
(n+1)\int _0^1 dt~\, \hbox{STr}\, \left((A_1-A_0)F_t^{n}\right)
$$
La {\it forma de Chern-Simons} ${\cal Q} _{2n+1}(A,F)$ es la forma de Transgresi\'on
en el caso $A_1=A$ y $A_0=0$. 
$$
{\cal Q} _{2n+1}(A,F)\equiv {\cal T}_{2n+1}(A,F,0,0) =(n+1)\int_0^1 d s~\, \hbox{STr}\,
\left(AF_s^{n}\right)
$$
con $A_s=sA$ y $F_s=dA_s+A_s^2=sdA+s^2A^2=sF+s(s-1)A^2$.

De la f\'ormula de homotop\'{\i}a de Cartan ${\cal P}(F_t,A_t)=\, \hbox{STr}%
\, \left(F_t^{n+1}\right)$ se sigue la {\it f\'ormula de transgresi\'on}
$$
\, \hbox{STr}\, \left(F_1^{n+1}\right)-\, \hbox{STr}\,
\left(F_0^{n+1}\right)=d{\cal T}_{2n+1}(A_1,A_0) ~~.
$$
la cual es v\'alida globalmente. Por lo tanto las integrales $\int
_{M^{2n+2}}\, \hbox{STr}\, \left(F^{n+1}\right)$ en una variedad sin borde,
conocidas como 'n\'umeros de Chern' son invariantes topol\'ogicos en el
sentido de que solo dependen de la topolog\'{\i}a del fibrado (esto es, de
las funciones de transici\'on) y no cambian bajo difeomorfismos.

Para las formas de Chern-Simons tenemos 
$$
\, \hbox{STr}\, \left(F^{n+1}\right)=d{\cal Q} _{2n+1}(A,F)
$$
Esta ecuaci\'on vale solo localmente, ya que si tomamos $A_0$ cero en una
carta local, no ser\'a cero en general en otras, debido a funciones de
transici\'on no triviales. Una consecuencia de esto es que en caso de campos
gauge puro ($F=0$) la forma de Chern-Simons es localmente exacta, y esta dada
explicitamente por la 'forma de Wess-Zumino-Witten' (WZW) 
$$
{\cal Q}_{2n+1} ( g^{-1}dg ,0)= (-1)^n\frac{(n+1)!n!}{(2n+1)!} \, \hbox{STr}\, %
\left[ (g^{-1}dg )^{2n+1}\right]
$$
La forma de transgresi\'on se puede escribir como la diferencia de dos
formas de Chern-Simons mas un t\'ermino de borde usando la f\'ormula de
homotop\'{\i}a de Cartan aplicada a ${\cal P}(F_t,A_t)={\cal Q} _{2n+1}(F_t,A_t)$,
Resulta 
$$
{\cal T}_{2n+1}(A_1,F_1,A_0,F_0)= {\cal Q} _{2n+1}(A_1,F_1) -{\cal Q} _{2n+1}(A_0,F_0) -d 
\left[ k_{01}{\cal Q} _{2n+1}(A_t,F_t) \right]
$$
donde se uso que $d{\cal Q} _{2n+1}(A,F)=\, \hbox{STr}\, (F^{n+1})$. El \'ultimo
t\'ermino es un t\'ermino de borde $C_{2n}= k_{01}{\cal Q} _{2n+1}(A_t,F_t) $ dado
mas explicitamente por 
$$
C_{2n}(F_1,A_1;A_0,F_0)\equiv -n(n+1)\int_0^1 d s~\int_0^1 d t~s~ \, %
\hbox{STr}\, \left(A_tJF_{st}^{n-1}\right)
$$
con $F_{st}=sF_t+s(s-1)A_t^2$ y $A_t=tA_1+(1-t)A_0$.

La invariancia de la forma de Transgresi\'on bajo transformaciones de gauge
involucrando ambos potenciales $A_0$ y $A_1$ se sigue de la covariancia de $%
J=A_1-A_0$ y $F_t$ bajo esas transformaciones, la definici\'on de ${\cal T}_{2n+1}$
y la invariancia de la traza.\newline


\subsubsection{Teoremas de Indice}


Finalmente repasaremos algunos resultados sobre Teoremas de Indice que
usaremos mas adelante. Consideramos el operador 
$$
\rlap{\hbox{$\mskip 3 mu /$}}D = e^{\mu}_a \gamma ^a\left(\partial
_{\mu}+A_{\mu} \right)
$$
definido en una variedad $M^{2n}$ de dimensi\'on par. En la expresi\'on
previa $e^{\mu}_a$ es el inverso del vielbein, $a$ es un \'{\i}ndice 'plano'
en el espacio tangente y $\mu$ es un \'{\i}ndice 'curvo' de la variedad. Las
matrices gamma de Dirac $\gamma ^a$ satisfacen el \'algebra de Clifford $%
\{\gamma ^a,\gamma ^b\}=2\eta ^{ab}$. Definimos $\gamma ^{2n+1}=\eta \gamma
^0~.~.~.\gamma ^{2n-1}$, donde $\eta$ es una constante num\'erica elegida de
modo que $(\gamma ^{2n+1})^2=1$. Entonces los autovalores de $\gamma ^{2n+1}$
son mas o menos uno, y sus autoestados se dicen espinores de quiralidad
positiva o negativa respectivamente. Si definimos el 'hamiltoniano' $H=(i%
\rlap{\hbox{$\mskip 3 mu /$}}D)^2$ entonces se puede mostrar que $\{i%
\rlap{\hbox{$\mskip 3 mu /$}}D,\gamma ^{2n+1}\}=0$ y $[H,\gamma ^{2n+1}]=0$.
Se sigue que podemos diagonalizar simult\'aneamente $H$ y $\gamma ^{2n+1}$,
o sea que podemos elegir una base de autoestados de that is $H$ de
quiralidad definida. Si tenemos un autoestado $\psi$ de $H$ con autovalor $E$%
, $H\psi=E\psi$, entonces $\phi=i\rlap{\hbox{$\mskip 3 mu /$}}D\psi $ es
tambie\'en un autoestado de $H$ con el mismo autovalor $E$, $H\phi=H(i%
\rlap{\hbox{$\mskip 3 mu /$}}D\psi)= i\rlap{\hbox{$\mskip 3 mu /$}}%
D(H\psi)=E\phi$. Por otro lado $\phi$ y $\psi$ tienen quiralidades opuestas,
porque $\gamma ^{2n+1}\phi =\gamma ^{2n+1}(i\rlap{\hbox{$\mskip 3 mu /$}}%
D\psi )= -i\rlap{\hbox{$\mskip 3 mu /$}}D\gamma ^{2n+1}\psi $, entonces si $%
\gamma ^{2n+1}\psi =\pm \psi$ obtenemos $\gamma ^{2n+1}\phi =\mp \phi$. Se
sigue que los autoestados de $H$ con un autovalor dado existen en parejas de
quiralidad opuesta. Sin embargo el razonamiento previo no vale si $\phi =i%
\rlap{\hbox{$\mskip 3 mu /$}}D\psi =0$, entonces $H\psi =0$ y $E=0$, y los
estados con autovalor cero ({\it modos cero}) no aparecen en parejas.

El {\it Indice} del operador de Dirac se define como la diferencia entre el
n\'umero de modos cero linealmente independientes de quiralidad positiva
menos el n\'umero de modos cero linealmente independientes de quiralidad
negativa, 
\[
ind ~i\rlap{\hbox{$\mskip 3 mu /$}}D=n_{+}-n_{-} 
\]
Para ver que el \'{\i}ndice es un invariante topol\'ogico se observa que
bajo deformaciones continuas de la variedad algunos modos cero pueden
convertirse en modos con autovalor no nulo y viceversa, pero deben hacerlo
en pares de quiralidad opuesta, de modo que la diferencia es constante.

El {\it teorema de indice de Atiyah-Singer} da el \'{\i}ndice en t\'erminos de la
integral de un polinomio invariante sobre la variedad. Un caso particular
del teorema que utilizaremos es 
$$
ind~i\rlap{\hbox{$\mskip 3 mu /$}}D=\int _{M^{2n}}\left[ch(F)\right]
$$
donde el {\i car\'{a}cter de Chern} se define como la suma formal de formas
diferenciales dada por 
$$
ch(F)=\, \hbox{STr}\, \left(e^{i\frac{F}{2\pi}}\right)
$$
y esta \'ultima integral se entiende que selecciona la forma del orden
correcto en la suma formal que define el car\'{a}cter de Chern.

\subsubsection{Variaci\'on general de la transgresi\'on}

El contenido de esta subsecci\'on es la \'unica parte de esta secci\'on que
es nuevo \cite{motz1}, hasta donde yo se.

La forma de transgresi\'{o}n es 
$$
{\cal  T}_{2n+1}=(n+1)\int_{0}^{1}dt<JF_{t}^{n}>
$$
con $J=A_{1}-A_{0}$. Adem\'{a}s 
$$
A_{t}=tJ+A_{0}=t A_1+(1-t)A_0
$$
y 
$$
F_{t}=dA_{t}+A_{t}^{2}=F_{0}+tD_{0}J+t^{2}J^{2}
$$
con $F_{0}=dA_{0}+A_{0}^{2}$ y $D_{0}J=dJ+A_{0}J+JA_{0}$ Notese que la
derivada de $F_{t}$ con respecto al par\'{a}metro $t$ satisface 
$$
\frac{d~}{dt}F_{t}=D_{t}J=dJ+A_{t}J+JA_{t}=dJ+2tJ^{2}+A_{0}J+JA_{0}
$$
Para la variaci\'{o}n general de la forma de transgresi\'{o}n tenemos 
$$
\delta {\cal  T}_{2n+1}=(n+1)\int_{0}^{1}dt\{<F_{t}^{n}\delta
J>+<nJF_{t}^{n-1}D_{t}[\delta A_{t}]>\}
$$
pero 
\[
D_{t}[JF_{t}^{n-1}\delta A_{t}]=D_{t}JF_{t}^{n-1}\delta
A_{t}-JF_{t}^{n-1}D_{t}[\delta A_{t}]=\frac{d~}{dt}F_{t}F_{t}^{n-1}\delta
A_{t}-JF_{t}^{n-1}D_{t}[\delta A_{t}] 
\]
y usando $\delta A_{t}=t\delta J+\delta A_{0}$ 
\[
\delta {\cal  T}_{2n+1}=(n+1)\int_{0}^{1}dt\{<[F_{t}^{n}+tn\frac{d~}{dt}%
F_{t}F_{t}^{n-1}]\delta J>+<n\frac{d~}{dt}F_{t}F_{t}^{n-1}\delta A_{0}>\}
\]
\[
-n(n+1)~d~\int_{0}^{1}dt<JF_{t}^{n-1}\delta A_{t}>
\]
pero, dentro del bracket, $F_{t}^{n}+tn\frac{d~}{dt}F_{t}F_{t}^{n-1}=\frac{d~%
}{dt}[tF_{t}^{n}]$ y $n\frac{d~}{dt}F_{t}F_{t}^{n-1}=\frac{d~}{dt}F_{t}^{n}$
\ entonces las dos primeras integrales en $t$ pueden calcularse dando 
$$
\delta {\cal  T}_{2n+1}=(n+1)<F_{1}^{n}\delta
J>+(n+1)<(F_{1}^{n}-F_{0}^{n})\delta
A_{0}>-n(n+1)~d~\int_{0}^{1}dt<JF_{t}^{n-1}\delta A_{t}>
$$
y finalmente tenemos para variaciones gen\'{e}ricas de las transgresiones 
$$
\delta {\cal  T}_{2n+1}=(n+1)<F_{1}^{n}\delta A_{1}>-(n+1)<F_{0}^{n}\delta
A_{0}>-n(n+1)~d~\int_{0}^{1}dt<JF_{t}^{n-1}\delta A_{t}>
$$

Bajo transformaciones de gauge involucrando solo $A_1$ tenemos $\delta
A_1=D_1\lambda $, $\delta A_t=tD_1\lambda $ y entonces 
$$
\delta {\cal T}_{2n+1} =d[(n+1)<F_1^n\lambda>
-n(n+1)\int_0^1dt~t~<JF_t^{n-1}D_1\lambda >]
$$
La expresi\'on previa con $A_1=A$ y $A_0=0$ da la variaci\'on de gauge de la
forma de Chern-Simons.

Bajo transformaciones de gauge involucrando solo $A_0$ tenemos $\delta
A_0=D_0\lambda $, $\delta A_t=(1-t)D_0\lambda $ y entonces 
$$
\delta {\cal T}_{2n+1} =d[-(n+1)<F_0^n\lambda>
-n(n+1)\int_0^1dt~(1-t)~<JF_t^{n-1}D_0\lambda >]
$$

\newpage

\subsection{ Ap\'endice B. Supergrupos}

En este trabajo consideraremos teor\'{\i}as de gauge con grupos de gauge
dados por extensiones supersim\'etricas de grupos espaciotemporales. En esta
secci\'on repasar\'e brevemente las propiedades b\'asicas de los grupos
espaciotemporales y sus extensiones supersim\'etricas. Hay muchas
referencias muy buenas sobre supersimetr\'{\i}a en general, una de ellas es
ref.\cite{westbook}. Mis principales referencias sobre extensiones
supersim\'etricas de los grupos de Sitter y Conforme son\cite
{holten,troncoso2}, mientras que trabajos \'utiles en este tema son \cite
{zanelli2,bergshoeff}. Una lista extensiva de referencias puede encontrarse
en estos trabajos.\newline


\subsubsection{Generalidades}


El {\it grupo ortogonal} $O(M,N)$ se define como el grupo de $U^n_{~m}$ que
dejan invariante la forma cuadr\'atica 
$$
x^m\eta _{mn}x^n=~constant
$$
donde $\eta _{mn}$ es una matriz diagonal de $(M+N)\times (M+N)$ con $M$
entradas igual a +1 y $N$ entradas igual a -1. El {\it grupo ortogonal especial} 
$SO(M,N)$ es el grupo de matrices de $O(M,N)$ con determinante igual a uno, $%
det[M]=1$. Considerando transformaciones infinitesimales $U^n_{~m}=\delta
^n_m+\omega ^{rs} (M_{rs})^n_{~m}$, con $\omega ^{rs}=-\omega ^{sr}$ real, $%
\omega ^{rs}\ll 1$, entonces los generadores $( M_{rs} )^n_{~m}$ satisfacen
el \'algebra 
$$
[M_{rs} ,M_{pq} ]=+ \eta _{rq} M_{sp}- \eta _{rp} M_{sq} +\eta _{sp} M_{rq}
-\eta _{sq} M_{rp}
$$
El {\it grupo simpl\'ectico} $Sp(N)$ se definecomo el grupo de matrices que
dejan invariante la forma cuadr\'atica 
$$
\theta ^{\alpha }C_{\alpha\beta}\theta ^{\beta}=~constant
$$
donde $C_{\alpha\beta}$ es una matriz antisim\'etrica de $N\times N$ y los
par\'ametros $\theta ^{\alpha}$ son variables de Grassman que anticonmutan .
El {\it grupo ortosimpl\'ectico} $OSp(N\mid M)$ es el grupo de matrices que
dejan invariante la forma cuadr\'atica 
$$
x^m\delta _{mn}x^n+\theta ^{\alpha }C_{\alpha\beta}\theta ^{\beta}=~constant
$$
con $m,n=1,~.~.~.,~N$ and $\alpha ,\beta =1,~.~.~.,~M$. Claramente tanto $%
O(N)$ como $Sp(M)$ son subgrupos de $OSp(N\mid M)$.\newline


\subsubsection{Los Grupos Espaciotemporales}


Los grupos espaciotemporales en dimensi\'on $D$ son el {\it grupo de Lorentz} 
$SO(D-1,1)$, el {\it  grupo de Poincar\'e} $ISO(D-1,1)$ (que consiste de
transformaciones de Lorentz y traslaciones espaciotemporales), el {\it grupo de
de Sitter} (dS) $SO(D,1)$ y el {\it grupo de anti-de Sitter Group} (AdS) 
$SO(D-1,2)$.

El grupo de Poincar\'e es el grupo de isometr\'{\i}as (transformaciones que
dejan la m\'etrica invariante) del espacio de Minkowski.

Los espacios de de Sitter se definen como el hiperboloides 
$$
-x_0^2+x_1^2+~.~.~.~+x_{D-1}^2+\epsilon \left( \frac{w}{R}\right) ^2=R^2
$$
inmersos en un espacio plano de dimensi\'on $D+1$ con m\'etrica 
$$
ds^2=-dx_0^2+dx_1^2+~.~.~.~+dx_{D-1}^2 +\epsilon \frac{1}{R^2}dw^2
$$
donde $R$ es el radio de curvatura del hiperboloide. La m\'etrica en el
hiperboloide es la inducida por la inmmersi\'on. El caso $\epsilon =1$
corresponde al {\it espacio de de Sitter}, $\epsilon =-1$ corresponde al
{\it espacio de anti-de Sitter} y $\epsilon =0$ corresponde al espacio de
Minkowski. Los grupos de de Sitter son los grupos de isometrias de los
espacios de de Sitter correspondientes. Los espacios de Minkowski y de de
Sitter son los espacios con mayor n\'umero de isometr\'{\i}as en una
dimensi\'on dada $D$ (tienen $D(D+1)/2$ generadores), por lo que se
denominan {\it espacios maximalmente sim\'etricos}. En el l\'{\i}mite $%
R\rightarrow\infty$ los espacios de de Sitter se reducen al espacio de
Minkowski. En el mismo l\'{\i}mite los grupos de de Sitter se reducen al
grupo de Poincar\'e, a trav\'es de lo que se conoce como {\it contracci\'on de
Inonu-Wigner }. Esta contracci\'on consiste en definir los momentos 
$\overline{P}_s$ 

$$
\overline{P}_s=R^{-1}~M_{sD}
$$
y tomar el l\'{\i}mite $R\rightarrow\infty$ en el \'algebra de los grupos in
the dS o AdS. Es f\'acil verificar que el \'algebra se reduce a la del grupo
de Poincar\'e.\newline


\subsubsection{Supersimetr\'{\i}a}


Supersimetr\'{\i}a \cite{westbook} es una extensi\'on de las simetr\'{\i}as
espaciotemporales que mezcla bosones y fermiones (materia e interacci\'on)
Hay teoremas de imposibilidad, como el teorema de Coleman-Mandula y el
teorema de Sohnius-Haag-Lopuzsanski que afirman que las \'unicas extensiones
no triviales de las simetr\'{\i}as espaciotemporales (que no sean producto
directo o extensiones centrales) est\'an dadas por {\it supergrupos} los cuales
son grupos con algunos par\'ametros dados por variables de Grassmann, o lo
que es lo mismo,con un \'algebra de generadores consistente de conmutadores
y anticonmutadores. Si denotamos por $B$ y $F$ los generadores bos\'onicos y
fermi\'onicos respectivamente, tenemos un \'algebra que se lee
esquem\'aticamente como 
$$
[B,B]\approx B~~,~~[B,F]\approx F~~,~~\{ F,F\}\approx B
$$
Para agregar un conjunto de generadores fermi\'onicos a un \'algebra
bos\'onica se debe satisfacer las condiciones de consistencia dadas por la
'identidad de Jacobi' (la cual vale autom\'aticamente para cualquier
representaci\'on matricial del \'algebra, si esta existe) 
$$
[A,[B,C\}\} = [A,B\},C\}+(-1)^{bc} [A,C\},B\}
$$
donde $(-1)^{bc}=-1$ si tanto $B$ como $C$ son fermi\'onicos, y $%
(-1)^{bc}=+1 $ en cualquier otro caso.

En general la manera de obtener superidentidades matriciales a partir de
las identidades v\'alidas para matrices bos\'onicas usuales es considerar
los generadores fermi\'onicos multiplicados porvariables de Grassmann,
tratando la combinaci\'on como una matriz usual, y entonces factorizar y
reordenar los par\'ametros de Grassmann correspondientes, lo que producir\'a
signos relativos entre los diferentes t\'erminos. La generalizaci\'on de la
identidad de Jacobi dada arriba es un ejemplo de esto, y esta es la regla
que usaremos para definir 'Supertrazas'. Esta regla debe aplicarse tambi\'en
a nuestra definici\'on de {\it traza sim\'etrica}, la que ser\'a de hecho
antisim\'etrica en los \'{\i}ndices fermi\'onicos.

Resulta que los generadores fermi\'onicos $F$ tienen que ser espinores $%
Q_{\alpha}^i$ donde $\alpha$ es el \'{\i}ndice espinorial espaciotemporal y $%
i$ es un \'{\i}ndice en la representaci\'on vectorial de alg\'un grupo de
simetr\'{\i}as internas.

El \'algebra de Super-Poincar\'e se sabe que tiene, en adici\'on al
\'algebra de Poincar\'e, las siguientes relaciones de conmutaci\'on (modulo
constantes multiplicativas) 
$$
\{Q^i_{\alpha},Q^j_{\beta}\}=\delta ^{ij}\gamma ^s_{\alpha\beta}\overline{P}%
_s~~, ~~[M_{rs},Q^i_{\alpha}]=\left(\gamma _{rs}\right)
^{\beta}_{\alpha}Q^i_{\beta}
$$
donde $\gamma ^s$ son las matrices de Dirac de la dimensi\'on
correspondiente, y denotamos su producto antisimetrizado como haremos en lo
que sigue por 
$$
\gamma _{[k]}\equiv \gamma ^{[r_1} ~.~.~.~\gamma ^{r_k]}\equiv \gamma
^{r_1...r_k}
$$
El grupo de de Sitter no tiene extensiones supersim\'etricas, debido a que
no es posible agregarle generadores fermi\'onicos de modo consistente con la
identidad de Jacobi.

Para el grupo de anti-de Sitter es conveniente\cite{troncoso2,zanelli2}
considerar una representaci\'on concreta dada por 
$$
P_s\equiv M_{sD}=\left( 
\begin{array}{ll}
\frac{1}{2}(\gamma _s)_{\alpha\beta} & 0 \\ 
0 & 0
\end{array}
\right)
$$

$$
M_{rs}=\left( 
\begin{array}{ll}
\frac{1}{2}(\gamma _{rs})_{\alpha\beta} & 0 \\ 
0 & 0
\end{array}
\right)
$$

Resulta que, excepto para $D=5~mod~4$ (no necesitaremos este caso, pero se
discute en \cite{troncoso2,zanelli2}) se puede extender el \'algebra a una
superalgebra adicionando el espinor de pseudo-Majorana $Q^k_{\alpha}$ tal
que $\overline{Q}_k^{\alpha}=C^{\alpha\beta}u_{kj}Q^j_{\alpha}$ donde $%
C^{\alpha\beta}$ es la matriz de conjugaci\'on de carga y $u_{ij}$ es un
Casimir cuadr\'atico del grupo interno. $Q^k_{\alpha}$ esta dado
explicitamente por 
$$
\left(Q^k_{\gamma}\right)_{\beta j}^{\alpha i} = \left( 
\begin{array}{ll}
0 & \delta ^{\alpha}_{\gamma}\delta ^k_j \\ 
-C_{\gamma\beta}u^{ki} & 0
\end{array}
\right)
$$
Los generadores del grupo de simetr\'{\i}a interna son 
$$
\left(R^{kl}\right)_{j}^{i} = \left( 
\begin{array}{ll}
0 & 0 \\ 
0 & ({\cal R}^{kl})_{j}^{i}
\end{array}
\right)
$$
donde $({\cal R}^{kl})_{j}^{i}$ son los generadores en la representaci\'on
adjunta.

Para satisfacer la identidad de Jacobi tambien se necesita a\~nadir nuevos
generadores bos\'onicos a $P_s$ y $M_{rs}$. Los nuevos generadores son de la
forma 
$$
Z_{[k]}=\left( 
\begin{array}{ll}
\frac{1}{2}(\gamma _{r_1...r_k})_{\alpha\beta} & 0 \\ 
0 & 0
\end{array}
\right)
$$
Podemos definir $P_s \equiv (Z_{[1]})_s$ y $M_{rs} \equiv (Z_{[2]})_{rs}$.
Los $Z_{[k]}$ requeridos son aquellos tales que $\left(C \gamma _{[k]}\right)
^T=+C \gamma _{[k]}$ si $D=~2,~6,~7,~8~mod~8$ o aquellos tales que $\left(C
\gamma _{[k]}\right) ^T=-C\gamma _{[k]}$ si $D=~2,~3,~4,~6~mod~8$. $%
D=~2,~6~mod~8$ aparece en ambas listas debido a que en esa dimensi\'on hay
dos elecciones no equivalentes de la matriz de conjugaci\'on de carga $%
C^T=\pm C$. Excepto en $D=5~mod~4$ las superalgebras obtenidas son $OSp(2^{
[D/2] }\mid N)$ para d=2,3,4 mod 8 y $OSp(N\mid 2^{ [D/2] })$ para d=6,7,8
mod 8, donde $[D/2]$ denota la parte entera de $D/2$.

Un importante ejemplo es el {\it supergrupo de la Teor\'{\i}a M} \cite{townsend3}
$OSp(1\mid 32)$, el cual es la extensi\'on supersim\'etrica minimal del
grupo AdS en $D=11$, $SO(10,2)$. El grupo de la teor\'{\i}a M tiene
generadores $P_s \equiv (Z_{[1]})_s$(translations), $~Q_{\alpha}$ (espinores
de Majorana generadores se supersimetr\'{\i}as), $M_{rs} \equiv
(Z_{[2]})_{rs}$ (Lorentz) and $(Z_{[5]})_{r_1...r_5} $, con $\alpha\beta
=1,...,32$ y $r,s=0,...,10$. El \'algebra es 
$$
[Z_{[i]},Z_{[j]}]=2y \sum_{k=1,2~mod~4} \left\{ ^{i~j}_{~k~}\right\}Z_{[k]}
$$
$$
[ Q , Z_{[k]} ] = (-1)^k y \gamma _{[k]}Q \\
$$
$$
\left\{Q_{\alpha},Q_{\beta}\right\}= \sum_{k=1,2,5}\frac{1}{k!} \left(\gamma
^{r_1...r_k}C^{-1}\right)_{\alpha\beta} (Z_{[k]})_{r_1...r_k}
$$
donde $y$ es un par\'ametro de normalizaci\'on arbitrario, y debemos definir 
$$
(Z_{[D-k]})_{s_1...s_{D-k}} =\frac{i}{k!} \epsilon ^{r_k...r_1}_
{~~~~~~s_1...s_{D-k}} (Z_{[k]})_{r_1...r_k}
$$
Los coeficientes de Clebsch-Gordan son 
$$
\left\{ ^{i~j}_{~k~}\right\}=\frac{i!~j!}{s!~t!~u!}
$$
donde $s= \frac{1}{2}(i+j-k)$, $t= \frac{1}{2}(i-j+k)$ y $u= \frac{1}{2}%
(-i+j+k)$. Los conmutadores de los generadores fermi\'onicos del supergrupo
de la teor\'{\i}a M son explicitamente 
\[
\left\{Q_{\alpha},Q_{\beta}\right\}= \left(\gamma
^{r}C^{-1}\right)_{\alpha\beta} P_{r}+ \frac{1}{2} \left(\gamma
^{rs}C^{-1}\right)_{\alpha\beta} M_{rs} + \frac{1}{5!} \left(\gamma
^{r_1...r_5}C^{-1}\right)_{\alpha\beta} (Z_{[5]})_{r_1...r_5} 
\]
Podemos definir contracciones de las \'algebras super-AdS, an\'alogasa las
contracciones de Inonu-Wigner para AdS. Sea 
$$
\overline{P}_s \equiv \frac{1}{2y~R} (Z_{[1]})_s~,~ M_{rs} \equiv \frac{1}{2y%
} (Z_{[2]})_{rs}~,~ \overline{Q}_{\alpha} =\frac{1}{\sqrt{R}}Q_{\alpha}~,~ 
\overline{Z_{[5]}} \equiv \frac{1}{R} Z_{[5]}
$$
En el l\'{\i}mite $R\rightarrow \infty$ recobramos el \'algebra de
super-Poincar\'e mas los generadores $\overline{Z_{[5]}}$ que son
extensiones centrales con respecto a las supertraslaciones (lo que significa
que conmutan con $\overline{P}_s$ y $\overline{Q}_{\alpha}$) y tienen los
conmutadores con los generadores de Lorentz que corresponden a un tensor de
cinco \'{\i}ndices. A veces se llama \'algebra de la teor\'{\i}a M a esta
contracci\'on, pero en este trabajo se reservar\'a ese t\'ermino para el
\'algebra de $OSp(1\mid 32)$. Claramente los generadores $\overline{Z_{[5]}}$
no son necesarios para la clausura del \'algebra de super-Poincar\'e.

\subsubsection{ Trazas Invariantes}


Un tensor invariante es un objeto con \'{\i}ndices en una o mas
representaciones de un grupo, tal que todos sus componentes son constantes
(n\'umeros puros) y que bajo transformaciones arbitrarias en el grupo
transforma en si mismo. Por ejemplo $\eta _{rs}$ es un tensor invariante de $%
O(N,M)$ por definici\'on (y por supuesto tambi\'en de $SO(N,M)$), el
s\'{\i}mbolo de Levi-Civita $\epsilon _{r_1...r_D}$ es un tensor invariante
de $SO(N,M)$, $N+M=D$ y tambi\'en lo son las matrices de Dirac para esa
dimensi\'on y signatura $\gamma _{\alpha\beta}^r$ (con \'{\i}ndices en las
representaciones fundamental y adjunta). Los tensores invariantes pueden
usarse para producir invariantes, saturando los \'{\i}ndices de objetos con
\'{\i}ndices en la representaci\'on correspondiente. Se puede obtener
tensores invariantes tomando la traza de un producto de generadores del
grupo. La traza sim\'etrica se obtiene simetrizando 
\[
\, \hbox{STr}\, \big( T^{I_1}\dots T^{I_{n+1}} \big)
= \sum_P Tr \big( T^{(I_1}\dots T^{I_{n+1})} \big)
\]
donde la suma es sobre todas las permutaciones de \'{\i}ndices. Llamamos en
general 'traza invariante' al resultado de la contracci\'on de todos los
\'{\i}ndices de un objeto dadocon un tensor invariante, y una 'traza
sim\'etrica invariante' si el tensor invariante es sim\'etrico. Como se dijo
antes estas trazas son antisim\'etricas en los \'{\i}ndices fermi\'onicos.
Las propiedades de simetr\'{\i}a correctas pueden obtenerse considerando los
generadores fermi\'onicos mulriplicados por par\'ametros de Grassmann,
tomando las definiciones usuales para el caso bos\'onico, y entonces
factorizando y reordenando los par\'ametros de Grassmann.

Los tensores invariantes con \'{\i}ndices en la representaci\'on adjunta
para los grupos $SO(D)$ con cualquier signatura (lo que incluye dS y AdS, y
su contracci\'on Poincar\'e) son esencialmente $\eta _{rs}$ y $\epsilon
_{r_1...r_D}$ (el tensor de Levi-Civita), y productos tensoriales y
contracciones de estos. Productos de $\eta$'s son equivalentes a productos
de trazas de productos de generadores $Tr~[T^{r_1}...T^{r_{n_1}}]
...Tr~[T^{r_1}...T^{r_{n_k}}]$ con el rango del tensor invariante igual a $%
N=n_1+...+n_k$.

En este trabajo se usa tambi\'en la notaci\'on 
$$
<T^{r_1}...T^{r_{k}}> \equiv g ^{r_1...r_k}
$$
para denotar una traza sim\'etrica invariante. Los \'{\i}ndices de grupo se
suben y bajan con la 'm\'etrica del grupo', que para $SO(D)$ es $\eta _{rs}$.

\newpage


\begin{thebibliography}{99}

\bibitem{zanelli2}  {\small J. Zanelli, Braz. Jour. Phys. 30(2000)251,
hep-th/0010049 . }

\bibitem{aschwarz}  {\small A. Schwarz, Lett. Math. Phys. 2(1978)247 }

\bibitem{deser}  {\small S. Deser, R. Jackiw and S. Templeton, Phys. Rev.
Lett. 48(1983)975,\newline
Ann. Phys. NY 140(1984)372. }

\bibitem{witten-cs}  {\small Witten, Comm. Math. Phys. 121(1989)351 }

\bibitem{vannieuwen}  {\small P. Van Nieuwenhuizen, Phys.~Rev.~ {\bf D32}(19{%
85}){872} }

\bibitem{achu}  {\small A. Achucarro and P.K. Townsend, Phys.~Lett.~{\bf B180%
}(19{86}){89}. }

\bibitem{witten1}  {\small E. Witten, Nucl.~Phys.~{\bf 311B}(19{88}){46};
Nucl.~Phys.~{\bf 323B}(19{89}){113}. }

\bibitem{chamseddine}  {\small A.H. Chamseddine, Phys.~Lett.~{\bf B233}(19{89%
}){291}; Nucl.~Phys.~{\bf 346B}(19{90}){213}. }

\bibitem{superstring}  {\small M.B. Green, J.H. Schwarz and E. Witten, {\it %
Superstring Theory}, Vols. 1 \& 2, (Cambridge Univ. Press, 1987). }

\bibitem{banados1}  {\small M. Ba\~nados, R. Troncoso and J. Zanelli,
Phys.~Rev.~ {\bf D54}(19{96}){2605}. }

\bibitem{zanelli1}  {\small J. Zanelli, Phys.~Rev.~ {\bf D51}(19{95}){490},
hep-th/9406202. }

\bibitem{troncoso1}  {\small R. Troncoso and J. Zanelli, Phys.~Rev.~ {\bf D58%
}(19{98}){101703}, hep-th/9710180. }

\bibitem{troncoso2}  {\small R. Troncoso and J. Zanelli, Int. Jour. Theor.
Phys. 38(1999)1181, hep-th/9807029. }

\bibitem{troncoso3}  {\small R. Troncoso and J. Zanelli, Class. Quan. Grav.
17(2000)4451, hep-th/9907109. }

\bibitem{banados2}  {\small M. Ba\~ nados, Nucl. Phys. Proc. Suppl.
88(2000)17, hep-th/9911150 }

\bibitem{hassaine}  {\small M. Hassaine, R. Troncoso and J. Zanelli, {\it %
Eleven-dimensional supergravity as a gauge theory for the M-algebra},
hep-th/0306258. }

\bibitem{townsend1}  {\small P.K. Townsend, Phys.~Lett.~{\bf B350}(19{95}){%
184}, hep-th/9501068 }

\bibitem{hull1}  {\small C.~Hull and P.K.~Townsend, Nucl.~Phys.~{\bf 348B}(19%
{95}){109}. }

\bibitem{witten2}  {\small E. Witten, Nucl.~Phys.~{\bf 443B}(19{95}){85}. }

\bibitem{townsend2}  {\small P.K. Townsend, {\it Four Lectures on M-Theory},
in {\it `Proceedings of ICTP Summer School on High Energy Physics and
Cosmology'}, Trieste (June 1996), hep-th/9612121; }

\bibitem{kogan}  {\small I. Kogan, Phys.~Lett.~{\bf B231}(19{89}){377} }

\bibitem{moore1}  {\small G. Moore and N. Seiberg, Phys.~Lett.~{\bf B220}(19{%
89}){422} }

\bibitem{horava}  {\small P. Horava, Phys. Rev. {\bf D59}(1999){046004}%
, hep-th/9712130. }

\bibitem{nastase}  {\small H. Nastase, {\it Towards a Chern-Simons M theory
of } $OSp(1\mid 32)\times OSp(1\mid 32)$, hep-th/0306269.}

\bibitem{stora}  {\small R. Stora, {\it Algebraic Structure of Chiral
Anomalies}, in {\it 'Recent Progress in Gauge Theories '}, H. Lehmann ed.,
NATO ASI Series, (Plenum, NY, 1984). }

\bibitem{zumino}  {\small B. Zumino, {\it Chiral Anomalies and Differential
Geometry}, in {\it `Relativity, Groups and Topology II'}, B.S. De Witt and
R. Stora eds., (North Holland, Amsterdam, 1984). }

\bibitem{manes}  {\small J. Ma\~nes, R. Stora and B. Zumino, Comm. Math.
Phys. 102(1985)157 }

\bibitem{alvarez}  {\small L. Alvarez-Gaum\' e and P. Ginsparg,
Ann.~of~Phys.~{\bf 161}(19{85}){423}. }

\bibitem{chern}  {\small S. S. Chern, {\it Complex Manifolds without
Potential Theory}, 2nd Ed., (Springer, Berlin, 1979) }

\bibitem{naka} M. Nakahara, {\it "Geometry, Topology and Physics"}, Adam Hilger, (1991)

\bibitem{motz1}  {\small P. Mora, R. Olea, R. Troncoso and J. Zanelli, {\it %
Covariant Conserved Charges and Black Hole Thermodynamics for Chern-Simons
Gauge Theories via Transgression Forms}, hep-th/03xxxxx }

\bibitem{motz2}  {\small P. Mora, R. Olea, R. Troncoso and J. Zanelli, {\it %
Boundary Terms for Chern-Simons AdS Gravity}, hep-th/03xxxxx}

\bibitem{mile}  {\small M.B. Green, Phys.~Lett.~{\bf B223}(19{89}){157} }


\bibitem{mn}  {\small P. Mora and H. Nishino, Phys. Lett. {\bf B482}(2000){%
222}, hep-th/0002077. }

\bibitem{mora}  {\small P. Mora, Nucl. Phys. {\bf 594B}(2001)229,
hep-th/0008180. }

\bibitem{woit}  {\small P. Woit, {\it String Theory: An Evaluation},
Physics/0102051 }

\bibitem{bertlmann}  {\small R. Bertlmann, {\it Anomalies in Quantum Field
Theory}, (Oxford U.P., Oxford,1996) }

\bibitem{westbook}  {\small P. C. West, {\it Introduction to Supersymmetry
and Supergravity}, 2nd Ed., (World Scientific, Singapore, 1990),\newline
{\it Supergravity, Brane Dynamics and String Duality}, hep-th/9811101 }

\bibitem{holten}  {\small J.W. Van Holten and A. Van Proeyen,
Jour.~Math.~Phys.~{\bf 15}(19{82}){3763}. }

\bibitem{bergshoeff}  {\small E. Bergshoeff and A. Van Proeyen, {\it The
many faces of OSp(1,32)}, hep-th/0003261 }

\bibitem{townsend3}  {\small P.K. Townsend, {\it M-theory from its
Superalgebra}, hep-th/9712004. }

 \bibitem{Adler}  {\small S. Adler and W.A. Bardeen, Phys. Rev. 182(1969)1517 
}

\bibitem{piguet}  {\small O. Piguet and S. Sorella , Nucl.~Phys.~{\bf 381B}%
(19{92}){373},\newline
Nucl.~Phys.~{\bf 395B}(19{93}){661}. }

\bibitem{gross11}  {\small D. Gross, Nucl.~Phys.~{\bf 74B}(19{99}){426}
Proc. Suppl., hep-th/9809060 }

\bibitem{weinberg11}  {\small S. Weinberg, {\it What is Quantum Field Theory
and What Did We Think it is}, hep-th/9702027 }

\bibitem{gates}  {\small S.J. Gates Jr., M. Grisaru, M. Rocek and W. Siegel, 
{\it Superspace: Or One Thousand and One Lessons in Supersymmetry},
(Reading: Benjamin/Cummings, 1983) }

\bibitem{greenschwarz}  {\small M.B. Green and J.H. Schwarz, Phys.~Lett.~%
{\bf B136}(19{84}){367},\newline
Nucl.~Phys.~{\bf 243B}(19{84}){285} }

\bibitem{henneaux}  {\small M. Henneaux and L. Mezincescu, Phys.~Lett.~{\bf %
B152}(19{85}){340}. }

\bibitem{supermembrane}  {\small E. Bergshoeff, E. Sezgin and P.K. Townsend,
Ann. of Phys. 185(1988)330 }

 \bibitem{dixon} J.A. Dixon, M. Duff and E. Sezgin, Phys. Lett. {\bf B279}(1992)265.
 
\bibitem{gross}  {\small D.J. Gross, J.A. Harvey, E. Martinec and R. Rohm,
Nucl.~Phys.~{\bf 256B}(19{85}){253} }

\bibitem{duff}  {\small M. Duff, J. Liu and R. Minasian, Nucl.~Phys.~{\bf %
452B}(19{95}){261}, hep-th/9506126 }

\bibitem{postdam}  {\small R. Aros, M. Contreras, R. Olea, R. Troncoso and
J. Zanelli,{\it Charges in 2+1 Dimensional Gravity and Supergravity},
presented at the {\it Strings'99 Conference}, Postdam, Germany, July 1999.}

\bibitem{francaviglia1}  {\small A. Borowiec, M. Ferraris and M.
Francaviglia, J. Phys. {\bf A36}(2003)2589.}

\bibitem{francaviglia2}  {\small G. Allemandi, M. Francaviglia, M. Raitieri,
Class. Quant. Grav. 20(2003)483,\newline
{\it Charges and Energy in Chern-Simons Theory and Lovelock Gravity},
hep-th/0308019.}

\bibitem{brown}  {\small J.C. Brown and M. Henneaux, J. Math. Phys.
27(1986)489,\newline
Comm. Math. Phys. 104(1986)207.}

\bibitem{regge}  {\small T. Regge and C. Teitelboim, Ann. Phys. (NY)
88(1974)286.}

\bibitem{sarda}  {\small G. Sardanashvily, {\it Gauge conservation laws in
higher-dimensional Chern-Simons models }, hep-th/0303059,\newline
{\it Energy-momentum conservation in higher-dimensional Chern-Simons models}%
, hep-th/0303148. }

\bibitem{aroscargas}  {\small R. Aros, M. Contreras, R. Olea, R. Troncoso
and J. Zanelli, Phys. Rev. Lett. 84(2000)1647,\newline
Phys. Rev. {\bf D62}(2000)044002.}

\bibitem{dimensionally}  {\small M. Ba\~nados, C. Teitelboim and J. Zanelli,
Phys. Rev. Lett. 69(1992)1849,\newline
Phys. Rev. {\bf D49}(1994)975.}

\bibitem{scan}  {\small J. Cris\'ostomo, R. Troncoso and J. Zanelli, Phys.
Rev. {\bf D62}(2000)084013.}

\bibitem{barnich}  {\small G. Barnich and F. Brandt, Nucl. Phys. {\bf B633}%
(2002)3, hep-th/0111246}

\bibitem{bekenstein}  {\small J.D. Bekenstein, Phys. Rev. {\bf D7}(1973)2333,%
\newline
Phys. Rev. {\bf D9}(1974)3292.}

\bibitem{hawking}  {\small S.W. Hawking, Nature 248(1974)30,\newline
Comm. Math. Phys. 43(1975)199.}

\bibitem{gibbonshawking}  {\small G.W. Gibbons and S.W. Hawking, Phys. Rev. 
{\bf D15}(1977)2753.}

\bibitem{douglas}  {\small M.R. Douglas, {\it Branes within Branes},
hep-th/9512077. }

\bibitem{green3}  {\small M.B. Green, C.M. Hull and P.K. Townsend,
Phys.~Lett.~{\bf B382}(19{96}){65}, hep-th/9604119. }

\bibitem{minasian}  {\small R. Minasian and G. Moore, JHEP 9711(1997)002,
hep-th/9710230 }

\bibitem{witten3}  {\small E. Witten, JHEP 9812(1998)019, hep-th/9810188 }

\bibitem{'t hooft}  {\small G. 't Hooft, {\it Naturalness, Chiral Symmetry,
and Spontaneous Chiral Symmetry Breaking}, in {\it Recent Developments in
Gauge Theory}, ed. G. 't Hooft et al., (Plenum, New York, 1980) }

\bibitem{stelle}  {\small K. Stelle, {\it The Unification of Quantum Gravity}%
, Nucl. Phys. {\bf B} (Proc. Suppl.) 88(2000)3 }

\bibitem{brandenberger}  {\small R. Brandenberger, {\it A Status Review in
Inflationary Cosmology}, hep-ph/0101119. }

\bibitem{aros}  {\small R. Aros, C. Mart\'{\i}nez, R. Troncoso and J.
Zanelli, {\it Supersymmetry of gravitational ground states}, hep-th/0204029 }

\bibitem{witten4}  {\small E. Witten, Mod. Phys. Lett. A 5(1990)487 }

\bibitem{witten5}  {\small E. Witten, {\it M theory and Quantum Mechanics},
Nucl. Phys. {\bf B} (Proc. Suppl.) 62A-C(1998)463 }

\bibitem{zhang}  {\small S.C. Zhang and J. Hu, Science 294(2001)823,
cond-mat/0110572. }

\bibitem{zhang1}  {\small S.C. Zhang, Int. J. Mod. Phys. 25(1992). }

\bibitem{zhang11}  {\small B.A. Bernevig, C-H Chern, J-P Hu, N. Toumbas and
S-C Zhang, {\it Effective field theory description of the higher dimensional
quantum hall liquid}, cond-mat/0206164. }


\end{thebibliography}
\end{document}